\documentclass[11pt,fleqn]{article}

\usepackage{bm}
\usepackage{amsmath}
\usepackage{graphicx}
\usepackage{cite}
\usepackage[margin=1.25in]{geometry}
\usepackage{booktabs}
\usepackage{caption}
\usepackage{subcaption}
\usepackage{float}

\usepackage{color}
\usepackage[bottom]{footmisc}

\usepackage{multicol}
\usepackage{framed}
\usepackage{nomencl}
\makenomenclature
\newcommand{\tensr}[1]{\bm{\mathsf{#1}}}
\newcommand{\ms}{\scriptscriptstyle}
\newcommand{\um}{\scalebox{0.75}[1.0]{\( - \)}}

\newcommand{\K}{\kappa}

\newcommand{\PP}{\tensr{P}}
\newcommand{\F}{\tensr{F}}

\newcommand{\subscript}[1]{_{\ms #1}}

\usepackage{color}
\definecolor{orange}{rgb}{1,0.5,0}
\definecolor{darkorchid}{rgb}{0.6,0.196,0.8}
\definecolor{olivedrab}{rgb}{0.42,0.56,0.14}

\makeatletter
\newcommand{\mathleft}{\@fleqntrue\@mathmargin0pt}
\newcommand{\mathcenter}{\@fleqnfalse}
\makeatother

\newcommand{\Keq}[1]{\kappa\subscript{#1}^{eq}}

\newcommand{\wphi}[1]{\omega\subscript{#1}^{\phi}}

\definecolor{olivedrab}{rgb}{0.42,0.56,0.14}
\definecolor{oxfordblue}{rgb}{0.0, 0.13, 0.28}
\definecolor{deepsaffron}{rgb}{1.0, 0.6, 0.2}

\DeclareMathOperator{\sech}{sech}

\title{\vspace{-2.0cm} Investigation of Surfactant-Laden Bubble Migration Dynamics in Self-Rewetting Fluids using Lattice Boltzmann Method}

\author{{Bashir Elbousefi\footnotemark, William Schupbach, Kannan N. Premnath, Samuel W.J. Welch}\\Department of Mechanical Engineering\\ University of Colorado Denver\\ 1200 Larimer Street, Denver, CO 80204, U.S.A.}

\begin{document}

\maketitle
\begin{abstract}
Self-rewetting fluids (SRFs), such as aqueous solutions of long-chain alcohols, show anomalous nonlinear (quadratic) variations of surface tension with temperature involving a positive gradient in certain ranges, leading to different thermocapillary convection compared to normal fluids (NFs). They have recently been used for enhancing thermal transport, especially in microfluidics and microgravity applications. Moreover, surface-active materials or surfactants can significantly alter interfacial dynamics by their adsorption on fluid interfaces. The coupled effects of temperature- and surfactant-induced Marangoni stresses, which arise due to surface tension gradients, on migration bubbles in SRFs remain unexplored. We use a robust lattice Boltzmann (LB) method based on central moments to simulate the two-fluid motions, capture interfaces, and compute the transport of energy and surfactant concentration fields, and systematically study the surfactant-laden bubble dynamics in SRFs. When compared to motion of bubbles in normal fluids, in which they continuously migrate without a stationary behavior, our results show that they exhibit dramatically different characteristics in SRFs in many different ways. Not only is the bubble motion directed towards the minimum temperature location in SRFs, but, more importantly, the bubble attains an equilibrium position. In the absence of surfactants, such an equilibrium position arises at the minimum reference temperature occurring at the center of the domain. The addition of surfactants moves the equilibrium location further upstream, which is controlled by the magnitude of the Gibbs elasticity parameter that determines the magnitude of the surface tension variation with surfactant concentration. The parabolic dependence of surface tension in SRF is parameterized by a quadratic sensitivity coefficient, which modulates this behavior. The lower this quantity, the greater is the role of surfactants modifying the equilibrium position of the bubble in SRF. Furthermore, the streamwise gradient in the surfactant concentration field influences the transient characteristics in approaching the terminal state of the bubble. These findings provide new means to potentially manipulate the bubble dynamics, and especially to tune its equilibrium states, in microchannels and other applications by exploiting the interplay between surfactants and SRFs.
\end{abstract}

\let\thefootnote\relax\footnote{*Corresponding author (Email: Bashir.Elbousefi@ucdenver.edu)}

\section{Introduction}
Dispersed two-phase flows, such as those involving bubbles or drops, are common in nature, everyday life, and industrial applications, and represent a topic of major research interests. Because surface tension forces can alter how fluids behave at the interface, they are important in multiphase and heat transport processes~\cite{de2004capillarity} and, in particular, variable surface tension flows have found various applications related to microfluidics and microgravity transport phenomena. Such technological applications rely on Marangoni stresses~\cite{scriven1960marangoni}, which are caused by variations in the surface tension (i.e., the surface tension gradients) at the interface between two immiscible fluids and arise due to variations in the local interfacial temperature or due to the use of surfactants or due to the coupling between the two. These stresses cause convective motions to occur close to the interfaces due to the viscous effects of the fluids~\cite{probstein2005physicochemical}. Surface tension driven flows due to temperature, or the thermocapillary convection, and the associated migration of bubbles or drops in fluids have been the subject of many studies over the years (see e.g.,~\cite{young1959motion,subramanian1981slow, subramanian1983thermocapillary,welch1998transient,ma2011direct}). In the case of micro-electro-mechanical-systems, the thermocapillary convection is utilized to manipulate fluid streams and thermal transport phenomena, as well as the motion of bubbles and drops in microchannels as interfacial forces become dominant at the small device scales (see, e.g., \cite{darhuber2005principles,karbalaei2016thermocapillarity}).

Surface tension is a common property of fluids that decreases with increasing temperatures in most fluids. However, some fluids exhibit anomalous behavior in which the surface tension deviates from this linear relationship and instead exhibits a non-linear parabolic dependence on temperature with a range involving a positive gradient. Aqueous solutions of high carbon content-based alcohols (e.g., 1-butanol, 1-pentanol, and 1-heptanol) exhibit an increase in surface tension as the temperature increases in certain operating ranges. Over the years, various experiments on such special classes of fluids conducted by different researchers have measured and demonstrated this behavior~\cite{vochten1973study,petre1984experimental,limbourg1986thermocapillary,villers1988temperature}. It may be noted that in addition to mixtures of certain alcohols with water, some liquid metallic alloys and nematic liquid crystals also exhibit anomalous behavior in which the surface tension decreases with temperature up to a certain point, after which it begins to increase. Abe~\emph{et al.}~\cite{abe2004microgravity} named these fluids "self-rewetting" fluids (SRFs). They experimentally revealed that SRFs have much potential for transferring heat because of how easily they can change phases, which, compared to common or normal fluids (NFs), exhibit a significantly altered thermocapillary convection that promotes a desired wetting effect especially over hotspots.

\begin{figure}[H]
 \centering
  \begin{subfigure}{0.48\textwidth}
    \centering
    \includegraphics[width=\textwidth]{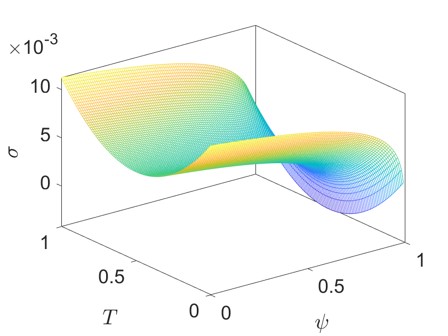}
    \caption{} \label{T_psi_surface_SRF}
\end{subfigure}
\begin{subfigure}{0.5\textwidth}
    \centering
    \includegraphics[width=\textwidth]{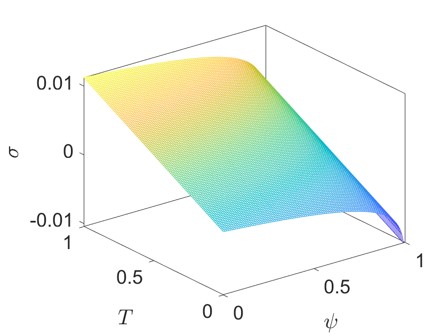}
    \caption{} \label{T_psi_surface_NF}
\end{subfigure}
\begin{subfigure}{0.5\textwidth}
    \centering
    \includegraphics[width=\textwidth]{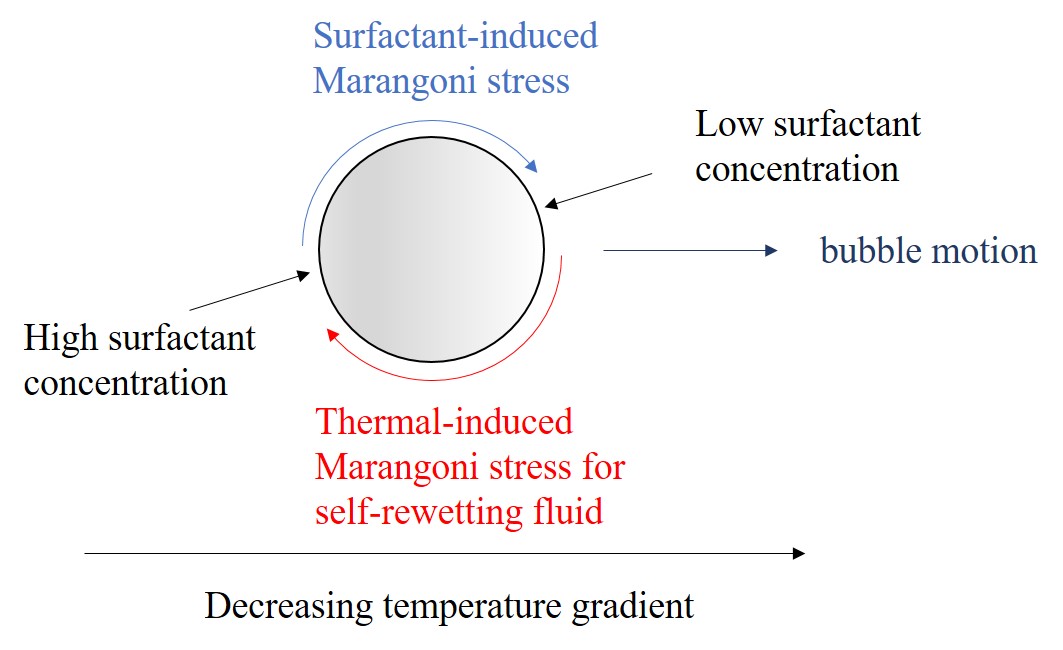}
    \caption{} \label{surfactant_thermal_Marangoni_stress}
\end{subfigure}
\caption{$(a)$ Surface tension variation with respect to temperature $T$ and surfactant concentration $\psi$ for a SRF $(M_2=5)$ and  $(b)$ for a NF $(M_1=-2.5)$. $(c)$ Schematic diagram shows the directions of both surfactant-induced and thermal-induced Marangoni stresses for SRF acting on the bubble surface. See Eq.~(\ref{three-prime}) for definitions of $M_1$ and $M_2$.}
\label{fig1_NF_vs_SRF}
\end{figure}

Marangoni stresses, in particular, cause fluids near the interfaces to move towards higher temperatures (where the fluids have higher surface tensile strength) in SRFs (as opposed to NFs). For example, the schematic diagram in Fig.~\ref{fig1_NF_vs_SRF} shows how the two kinds of fluids behave differently with temperature and the resulting Marangoni stress in SRF around a bubble in motion. In self-rewetting fluids (SRFs), the surface tension increases with temperature.
As a result, when a temperature gradient is present at the interface of two fluids, the SRF flows from the cold to the hot region, in contrast to what occurs in normal fluids (NFs), where the fluid flows from hot to cold regions due to the decrease in surface tension with increasing temperature~\cite{schupbach2022lattice,elbousefi2023lattice,elbousefi2024lattice}. Effectively, this causes fluid currents towards higher temperature zones in SRF, which can be exploited to enhance transport. As such, SRFs have been used as working fluids in a variety of thermal management applications in both terrestrial and microgravity environments~\cite{abe2004microgravity,abe2007terrestrial}, such as in heat pipes~\cite{savino2009surface,savino2013some,hu2014heat,wu2017study,cecere2018experimental,hu2018review,zhu2020thermal,zhang2022heat,gao2022study,su2023study,liang2023comparisons}, flow boiling~\cite{sitar2015heat} and evaporation~\cite{sefiane2020heat} in microchannels, pool boiling processes~\cite{hu2015heat,hu2018marangoni,hu2019marangoni,movze2021pool,kim2022pool}, and two-phase heat transfer devices~\cite{zaaroura2021thermal}. Moreover, some experiments~\cite{shanahan2014recalcitrant,mamalis2017bubble} have recently studied how bubble migration in SRFs has certain unique characteristics unlike what has been observed in common fluids.

Apart from temperature variations, surfactants can also significantly modify the local surface tension in fluid interfaces. Surfactants, also known as surface-active substances, are molecules that preferentially absorb at the interface between two phases. They can be soluble or insoluble in the bulk fluids and their net adsorption/desorption characteristics on interfaces effectively cause a local reduction in surface tension as well generating its tangential gradients or the Marangoni stresses. The latter arises due to a nonuniform concentration of surfactants on the interfaces. For example, a bubble in motion involves an adsorption at its trailing edge, while a part of its front region remains free of surfactants and the resulting Marangoni stress causes the bubble to move towards areas with lower surface tension. See the schematic Fig.~\ref{surfactant_thermal_Marangoni_stress} illustrating this effect. Surfactants are used widely in many areas related to technical, chemical, and biological applications~\cite{rosen2012surfactants}, including in microfluidic systems~\cite{baroud2010dynamics,seemann2011droplet,kovalchuk2019mass,luo2020effect,kovalchuk2023review}.

Multiphase flows arise in a wide range of complex situations (see e.g.,~\cite{issakhani2023geometrically,qenawy2024intermittent}). To provide insights into the underlying physics associated with the transport processes in such cases, computational studies that are based on appropriate physical models are essential~\cite{mirhoseini2024novel,yan2024thermocapillary}. It would be interesting to investigate the coupled effects of both thermal- and surfactant-induced Marangoni stresses on bubble migration in self-rewetting fluids (SRFs).
However, much of the prior research on SRFs in this regard involving either analytical~\cite{slavtchev1998thermocapillary,oron1994nonlinear,batson2017thermocapillary,yu2018thermocapillary,zubair2023dynamics} or numerical~\cite{tripathi2015non,balla2019non,majidi2020single,mitchell2021computational,xu2021motion} approaches do not include considerations of surfactants, including our recent work on analytical and numerical study on thermocapillary convection of superimposed SRF layers~\cite{elbousefi2023thermocapillary}. One notable exception is that of~\cite{su2022enhanced}, which studied the SRF droplet spreading over a heated film in the presence of surfactants. Nevertheless, most of existing studies that combine thermocapillary effects with surfactant behavior are all directed towards normal fluids (NFs), where the surface tension varies linearly with temperature (see e.g.,~\cite{luo2020effect,sharanya2019surfactant,guo2023nonlinear}), and have not explored them for SRFs, especially involving bubble dynamics.

Thus, the aim of this work is to systematically investigate the coupled effects of both thermal- and surfactant-induced Marangoni stresses on bubble migration in self-rewetting fluids (SRFs) in order to provide a more complete physical understanding of the behavior of the dispersed phases in such configurations, which can, in turn, inform novel ways to potentially manipulate them. More specifically, previous studies have examined the effects of thermal-induced Marangoni stresses on bubble migration in SRFs, but have neglected the additional influence of surfactant-induced Marangoni stresses which, as this work will show, leads to certain interesting outcomes for the bubble dynamics. One primary objective is to present and apply a numerical simulation approach based on the lattice Boltzmann method (LBM) in conjunction with a phase field model for capturing of interfaces along with the attendant transport models for energy and surfactant dynamics. The LBM is a numerical approach inspired by the kinetic theory of gases~\cite{benzi1992lattice,yu2003viscous,lallemand2021lattice}, and its natural parallelization features, efficient and simpler implementation algorithms, and ability to handle complex geometries and boundary conditions make it a powerful and versatile technique for simulating fluid motions under a variety of situations.
As such, the LBM has been found to be a versatile approach to tackle a variety of complex fluid flows and beyond. The recent advances of this method have been discussed in a number of reviews for various applications, such as thermal and aerodynamic flows~\cite{sharma2020current}, turbulent and multiphase systems of complex fluids~\cite{petersen2021lattice}, phase change problems including melting and solidification~\cite{samanta2022review}, non-ideal fluids~\cite{hosseini2023lattice} and reacting flows~\cite{hosseini2024lattice}. Over the years, the LBM attracted attention especially for simulation of multiphase flows~\cite{he1999lattice,he2002thermodynamic,lee2005stable,premnath2007three,hajabdollahi2021central,alsadik2024lattice}. As a result, the LBM has become an important numerical simulation tool for multiphase flows alongside more conventional techniques such as volume of fluid, level set method, and front tracking method in multiphase flows. LB methods have also been used to model thermocapillary flows (see e.g.,~\cite{liu2012modeling,majidi2020single,mitchell2021computational,elbousefi2023thermocapillary,scherr2023volume}). In recent years, new collision models based on central moments have been developed that enhance the robustness of the LBM~\cite{geier2006cascaded} and allow for the application of the LBM method to simulate high-density ratio multiphase flows that are subject to Marangoni stress~\cite{hajabdollahi2021central}. Such collision models have been used in the LBM for simulation of thermocapillary flows in SRFs in our recent work~\cite{elbousefi2023thermocapillary}, which will be further extended by incorporating surfactant dynamics, including its adsorption/desorption process and bulk solubility effects, and its contribution to the surface tension equation of state.

Thus, in this work, we will use a robust central moment LB approach that involves the computing of four different distribution functions. One is used to compute the two-fluid motion and includes the effect of Marangoni stresses based on surface tension gradients, another one to capture the interfacial motions represented by the conservative Allen-Cahn equation (ACE), and, the third one to compute the energy transport and, finally, the fourth one for the surfactant concentration field. Local surface tension is then modulated by using a parabolic dependence on temperature for self-rewetting fluids coupled with the effect of the surfactant concentration via the Langmuir isotherm. As another key objective, we present the simulation results of surfactant-laden bubble dynamics and its terminal states in SRFs by studying the effect of variations of different characteristic parameters associated with this flow configuration. We will compare and contrast the behavior of bubbles in SRFs with those in NFs with the addition of surfactants.

The remaining sections of the paper are structured as follows. In the upcoming section (Section~\ref{Sec.1}), we will explain the problem setup of the migration of bubbles laden with surfactants through thermocapillary effects in a SRF. We will also discuss the governing equations that describe the incompressible two-fluid motion, the transport of energy and surfactant concentration, and the equation of state for interfacial surface tension. The diffuse-interface computational model equations for the LBM of surfactant-laden self-rewetting fluids are provided in Section~\ref{Sec.3}. We summarize the discretized central moment LB algorithms for simulating multiphase flows of SRFs in Section~\ref{Sec.LBschemes}, while Appendices~\ref{Sec.4.1},\ref{Sec.4.2},\ref{Sec.4.3}, and~\ref{Sec.4.4} discuss the LB schemes for interfacial dynamics, two-fluid motion, energy transport, and surfactant concentration transport, respectively. Section~\ref{Sec.5} presents a numerical validation of our computational approach by comparing it to established benchmark problems. The results and discussion regarding the influence of various characteristic parameters on the physics of surfactant-laden thermocapillary bubble migration in SRF using LB schemes are presented in Section~\ref{Results and Discussion}. Finally, we provide an overview of the main conclusions and contributions of our research in Section~\ref{Sec.9}, and additional supporting details can be found in the appendix sections.

\section{Problem setup, governing equations, and modeling of surface tension effects} \label{Sec.1}
\begin{figure}[H]
\centering
\includegraphics[trim = 0 0 0 0,clip, width = 100mm]{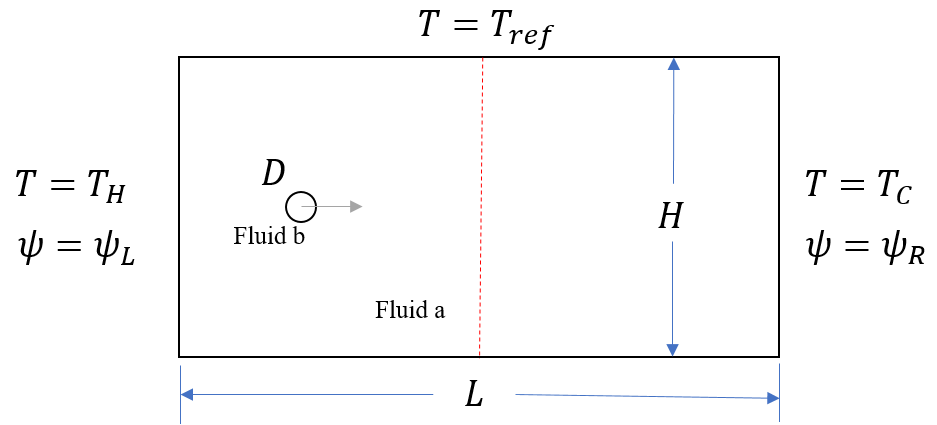}
\caption{\label{fig: Model_setup} Schematic of the initial configuration of the surfactant-laden thermocapillary bubble migration in self-rewetting fluid (SRF).}
\label{Model_setup}
\end{figure}

The objective of the study is to simulate the surfactant-laden thermocapillary bubble migration in self-rewetting fluid (SRF). The simulations are conducted within a rectangular computational domain with dimensions $(L \times H)$, where $L=600$ and $H=0.5L$ are the lengths of the domain in the $x$ and $y$ directions, respectively. The problem is schematically represented in Fig.~\ref{Model_setup}, which shows the computational domain and the bubble with a diameter of $D=0.1H$ is initially located at $(0.25L,0.5H)$ where the origin of the coordinate system is chosen to be at the bottom left corner of the domain. The lattice units are used for all the values as is typical for the implementation of the LB method (see e.g. Ref.~\cite{kruger2017lattice}). No slip boundary conditions are imposed on the left and right walls, while periodic boundary conditions are used on the top and bottom sides. As noted below, the side walls are imposed with different thermal conditions to drive the bubble motion via thermocapillary effects. Since they are generally solid surfaces, the no-slip boundary conditions are appropriate. Moreover, periodic conditions are imposed on the sides to ensure accurate representation of bubble dynamics without boundary interference, which is a common approach used in such dispersed flow simulations.
We initialize this problem with a constant temperature gradient applied in the $x$ direction, with a value of $\lvert \bm{\nabla} T \rvert = (T_H-T_C)/L$ such that a fixed hot temperature on the left side $(T_H=1)$ and a constant cold temperature, $(T_C=0)$ on the right side, and then we maintain the reference temperature $(T_{ref}=0.5)$ to occur at the middle of the domain. Moreover, we introduce a background surfactant gradient such that the gradient in the dimensionless surfactant concentration $\psi$ decreases from the left side ($\psi = \psi_L$) to the right side ($\psi = \psi_R$) (see the paragraph above Eq.~(\ref{SUR_mu_psi}) for the definition of $\psi$).
The initial bubble placement and surfactant gradients were chosen to adequately study the interesting features of dynamics of the bubble in the SRF. Given the temperature boundary conditions noted above with the temperature having a minimum in the middle of the domain and increasing on both sides, we placed the bubble halfway between the left side and the middle so that it accommodates the peculiar features of its motion in the SRF under thermocapillary effects. This ensures it sufficiently captures the transient effects before the bubble is expected to equilibrate in the vicinity of the center of the domain. A surfactant gradient is introduced in the background to generate additional Marangoni stresses to manipulate the dynamics and the final equilibrium position of the bubble. For convenience, the various symbols used in this paper are listed in Appendix~\ref{sec:nomenclature}.

Here, and in what follows, we use a subscript notation with 'A' for the ambient fluid and 'B' for bubble.
The thermocapillary convection in the SRFs including surfactant obeys the equations of mass and momentum (i.e., the Navier–Stokes equations (NSE)), energy transport, and the equation of surfactant concentration transport. They can be respectively written as follows:
\mathleft
\begin{subequations}
\begin{equation}
\qquad \bm{\nabla} \cdot {\bm{u}}=0,
\end{equation}
\begin{equation}
\qquad \rho \left( \frac{\partial \bm{u}}{\partial t} + \bm{u}\cdot\bm{\nabla}\bm{u} \right) = - \bm{\nabla} p + \bm{\nabla} \cdot \left[ \mu (\bm{\nabla} \bm{u} + \bm{\nabla} \bm{u}^{\dagger})\right]+ \bm{F}_{ext},
\end{equation}
\begin{equation}    \label{energy eqn}
\qquad  \frac{\partial {T}}{\partial t} + \bm{u} \cdot \bm{\nabla}T   =  \bm{\nabla} \cdot \left(\alpha \bm{\nabla}T \right),
\end{equation}
\begin{equation}    \label{surfactant concentration eqn}
\qquad  \frac{\partial {\psi}}{\partial t} + \bm{u} \cdot \bm{\nabla}\psi   =  \bm{\nabla} \cdot \left(M_{\psi} \bm{\nabla}\mu_{\psi} \right),
\end{equation}
\end{subequations}
where $\rho$, $\mu$, and $\alpha$ are the fluid density, dynamic viscosity, and thermal diffusivity of the fluid, respectively, with $\alpha=k/(\rho c_p)$ based on the thermal conductivity $k$ and specific heat $c_p$. In the above, $\bm{u}$, $p$, and $T$ denote the velocity, pressure, and temperature fields of the fluids, respectively, and the superscript symbol $\dagger$ represents taking the transpose of the dyadic velocity gradient $\bm{\nabla} \bm{u}$. Also, $\bm{F}_{ext}$ is any external body force. While the mass, momentum, and energy equations are standard, the modeling of surfactants as represented by Eq.~(\ref{surfactant concentration eqn}) calls for further discussion. Few different approaches to model surfactants within the phase field framework for its use with the LBM have been proposed (see e.g.,~\cite{van2006diffuse,liu2010phase,van2014mesoscale,van2016analysis,shi2019improved}).

In Eq.~(\ref{surfactant concentration eqn}), the surfactant concentration $\psi$ is non-dimensionalized by the maximum surfactant concentration on the interface $\psi_{max}$, which is a material property that determines its interfacial adsorption capacity, such that $\psi=\psi/\psi_{max}$ and $M_{\psi} = m_{\psi} \psi (1-\psi)$ is the local surfactant mobility, with $m_{\psi}$ being the scale for the mobility parameter. Hence, the non-dimensional local surfactant concentration will be restricted between $0$ and $1$ $(0 \le \psi \le 1)$; $\mu_{\psi}$ is the chemical potential whose gradients drive the diffusion-adsorption dynamics of the advecting surfactant concentration field and can be written as \cite{van2006diffuse,liu2010phase}
\mathleft
\begin{equation}\label{SUR_mu_psi}
\qquad \mu_\psi  = \lambda \ln \left(\frac{\psi}{1-\psi}\right) - \frac{s}{2} |\bm{\nabla} \phi|^2 + \frac{w}{2} (\phi-\phi_m)^2.
\end{equation}
Here, $\phi$ is the order parameter of the phase field variable (see Eq.~(\ref{eqn1}) for its governing equation based on a diffuse interface model) that is used to capture the dynamics of interfaces nominally located at $\phi_m = (\phi_A+\phi_B)/2$, where $\phi_A$ and $\phi_B$ are the values of $\phi$ in the bulk fluids A and B. The preferred tendency of the surfactant adsorption on the interface is represented by the second term on the right hand side (RHS) of the above chemical potential Eq.~(\ref{SUR_mu_psi}) represented in terms of a delta function through a square gradient of the order parameter. However, some recent studies~\cite{engblom2013diffuse,soligo2019coalescence} suggested replacing it with a gradient-free regularized delta function formulation obtained using the hyperbolic tangent profile across the interface in the normal direction for the phase field variable $\phi$ in order to achieve well-posedness with better numerical properties for a wider range of parameter choices. By using the latter approach and taking $\phi_o = (\phi_A-\phi_B)/2$, it is possible to rewrite the chemical potential utilized in this work as
\mathleft
\begin{equation}\label{SUR_mu_psi_1}
\qquad \mu_\psi  = \lambda \ln \left(\frac{\psi}{1-\psi}\right) - \frac{s}{2} \frac{4}{\phi_o^2 W^2} \left[ \phi_o^2 - (\phi-\phi_m)^2 \right]^2 + \frac{w}{2} (\phi-\phi_m)^2.
\end{equation}
The model parameters $\lambda$, $s$, and $w$ here represent the relative strengths of the several competing processes as explained below.

Each term in Eq.~(\ref{SUR_mu_psi_1}) on the RHS of the chemical potential has the following possible interpretation~\cite{soligo2019coalescence,engblom2013diffuse}: The first term, known as the entropy term, bounds the surfactant concentration between $0$ and $1$ and indicates an increase in the system's entropy when the substance is evenly distributed everywhere. Higher concentrations $\lambda$ drive stronger diffusion, which tends to redistribute the substance more evenly throughout the domain. The parameter $s$ may be used to adjust the strength of the second term, which is also referred to as the adsorption term. It indicates the surfactant's energetic propensity to get adsorbed on interfaces. The final term, also known as the bulk term, is a penalty term that represents the solubility effect and penalizes the presence of surfactants in bulk fluids. Its magnitude is controlled by the parameter $w$. The surfactant concentration profile surrounding interfaces is sharpened by these latter two terms, which oppose the diffusive process (first term). The surfactant concentration field $\psi$ evolution equation may be further reformulated by substituting Eq.~(\ref{SUR_mu_psi_1}) into Eq.~(\ref{surfactant concentration eqn}) and rearranging to get
\mathleft
\begin{equation}\label{SUR_mu_psi_2}
\qquad \frac{\partial {\psi}}{\partial t} + \bm{u} \cdot \bm{\nabla}\psi = \bm{\nabla} \cdot \left(\lambda m_{\psi} \bm{\nabla} {\psi} \right) +  \bm{\nabla} \cdot \left(m_{\psi} \psi (1-\psi) \bm{R} \right),
\end{equation}
where the second term in the RHS of the above equation can be considered as a flux term related to the surfactant's adsorption and solubility, and $\bm{R}$ is given by the following equation~\cite{ouderji2019phdthesis}
\mathleft
\begin{equation}\label{SUR_P}
\qquad \bm{R} = \bm{\nabla} \left( - \frac{s}{2} \frac{4}{\phi_o^2 W^2} \left[ \phi_o^2 - (\phi-\phi_m)^2 \right]^2 + \frac{w}{2} (\phi-\phi_m)^2 \right).
\end{equation}

\subsection{Surface tension equation of state including surfactant and thermocapillary effects for self-rewetting fluids} \label{Subsec.1.3}
In the case of normal fluids (NFs), the effect of inclusion of surfactants was modeled by Luo~\emph{et al.}~\cite{luo2020effect} who reported a form for the surface tension equation of state based on a linear term related to its variation with temperature which is combined with an additional term associated with surfactant effects. Our work extends this for self-rewetting fluids (SRFs) with surfactants. To impose an equation for the surface tension at the interface, we need to relate it to the variations in temperature $T$ and surfactant concentration $\psi$. To combine both effects, we use a nonlinear (quadratic) dependence of surface tension on temperature for SRF and Langmuir isotherm for surfactant effects which are given in the following equation:
\mathleft
\begin{equation}\label{ST_SRF}
\qquad \sigma (T,\psi) = \sigma_{0} + \sigma_{0} \beta \left(\frac{T}{T_{ref}}\right) \ln(1-\psi) +  \sigma_T (T-T_{ref})+ \sigma_{TT} (T-T_{ref})^2,
\end{equation}
where $\sigma_{0}$ denotes the value of the surface tension at a reference temperature $T_{ref}$ which is in the middle of the domain according to the quadratic functional distribution given above; $\sigma_{T}=\frac{d\sigma}{dT}\big\vert_{T_{ref}}$ and $\sigma_{TT}=\frac{1}{2}\frac{d^2\sigma}{dT^2}\big\vert_{T_{ref}}$  are the sensitivity coefficients of the surface tension with linear and quadratic dependencies on temperature, respectively. In Eq.~(\ref{ST_SRF}), $\beta$ is the Gibbs elasticity parameter, which is a dimensionless quantity that determines the influence of the surfactant on the surface tension variations and is given by
\mathleft
\begin{equation} \label{Gibbs_beta_Eg}
\qquad \beta = \frac{R_g T_{ref} \psi_{max}} {\sigma_{0}},
\end{equation}
where $R_g$ is the ideal gas constant and $\psi_{max}$ is the maximum surfactant concentration on the interface. Thus, $\beta$ models the magnitude of the variations in the surface tension due to the addition of surfactants which is a material property of the latter.

When the above governing equations are nondimensionalized using a reference velocity scale $U$ and a length scale $L$ corresponding to the radius of the bubble $R$, the additional dimensionless groups that arise in the study of the thermocapillary motion of a bubble in SRF are as follows: The Reynolds number $\mbox{Re}$, Marangoni number $\mbox{Ma}$, and capillary number $\mbox{Ca}$ are defined by
\mathleft
\begin{equation}
\qquad \mbox{Re} = \frac{U R} {\nu_A},    \quad  \quad     \mbox{Ma} = \frac{U R} {\alpha_A} ,     \quad \textrm{and} \quad     \mbox{Ca}= \frac{U \mu_A} {\sigma_{0}},                   \label{ReMaCa}
\end{equation}
respectively, where $\nu=\mu/\rho$ is the kinematic viscosity of the fluid, and the rest of the symbols are already defined below Eq.~(\ref{surfactant concentration eqn}). Also, the following ratios of the material properties govern the thermocapillary-driven motion of a bubble in a SRF:
\mathleft
\begin{equation}
\qquad \tilde{\rho} =  \frac{\rho_A}{\rho_B},   \quad \quad   \tilde{\mu} =  \frac{\mu_A}{\mu_B},  \quad \quad  \tilde{k} = \frac{ k_A}{k_B}, \quad \textrm{and} \quad  \tilde{c}_p =  \frac{c_{p_A}}{c_{p_B} }.  \label{three}
\end{equation}
Moreover, the dimensionless forms of the linear and quadratic sensitivity coefficients of the surface tension $\sigma_{T}$ and $\sigma_{TT}$ (see Eq.~(\ref{ST_SRF})) can be, respectively, represented as
\begin{equation}
\qquad M_1 =\left(\frac{\Delta T}{\sigma_0}\right)\sigma_T,\quad M_2 =\left(\frac{\Delta T^2}{\sigma_0}\right)\sigma_{TT}, \label{three-prime}
\end{equation}
where $\Delta T$ is the characteristic temperature difference in the domain. Based on the above, we can also rewrite the surface tension equation of state given in Eq.~(\ref{ST_SRF}) in the following dimensionless form:
\begin{equation}\label{ST_SRF_dimensionless}
\qquad \frac{\sigma (T,\psi)}{\sigma_{0}} = 1 + \beta \left(\frac{T}{T_{ref}}\right) \ln(1-\psi) +  M_1 \frac{(T-T_{ref})}{\Delta T}+ M_2 \frac{(T-T_{ref})^2}{\Delta T^2}.
\end{equation}
In addition, the non-dimensional numbers involving the surfactant dynamics based on Eqs.~(\ref{SUR_mu_psi_2}) and (\ref{SUR_P}) include $\mbox{Pi} = \lambda W^2 / (8\sigma_{0})$, which indicates the proportionate contribution of surfactant diffusion, and $\mbox{Ex} = 4s/(w W^2)$, which represents the relative intensity of adsorption and solubility effects where the parameters $\lambda$, $s$, and $w$, represent the relative strengths of the various competing processes. Finally, We can estimate the scale for the reference velocity $U$ of thermocapillary bubble migration in a fluid used in the above via $U_o \sim (\sigma_{0}/\mu_a) (R/L) (M_1+M_2)$ by balancing the scale for the viscous shear stress or $\mu_aU/R$ with that of the primary Marangoni stress due to the surface tension gradient arising from temperature variations or $|d\sigma/dT|(\Delta T/L)+(d^2\sigma/dT^2)(\Delta T^2/L)$.

\section{Diffuse-interface computational modeling of surfactant-laden self-rewetting fluids} \label{Sec.3}
In the upcoming section, we will explore a modeling approach that utilizes the LBM to simulate thermocapillary convection in SRFs. To accurately capture interfacial dynamics and maintain the segregation of two immiscible fluids, we will employ a phase-field lattice Boltzmann approach based on the conservative Allen-Cahn equation (ACE)~\cite{chiu2011conservative}. This approach is an improvement over an earlier model~\cite{sun2007sharp} that utilized a counter-term approach \cite{folch1999phase}. In this model, the binary fluids are distinguished by an order parameter or the phase field variable $\phi$, with fluid $A$ being identified by $\phi = \phi_{A}$ and fluid $B$ by $\phi = \phi_{B}$. Ultimately, we will utilize the interface-tracking equation based on the conservative ACE to solve for the phase field variable which is given as
\mathleft
\begin{equation}\label{eqn1}
\qquad \frac{\partial \phi}{\partial t} + \bm{\nabla} \cdot (\phi \bm{u}) = \bm{\nabla} \cdot [M_\phi(\bm{\nabla} \phi - \theta \bm{n})].
\end{equation}
The fluid velocity is denoted by $\bm{u}$, the mobility by $M_{\phi}$, and the unit normal vector by $\bm{n}$. The unit normal vector may be computed using the order parameter $\phi$ as follows: $\bm{n} = \bm{\nabla}{\phi}/|\bm{\nabla}{\phi}|$. In this case, $\theta = -4\left(\phi - \phi_{A} \right) \left(\phi - \phi_{B} \right)/[W \left(\phi_{A} - \phi_{B} \right)]$ may be used to define the parameter $\theta$, where $W$ is the interface width. In Eq.~(\ref{eqn1}), the term $M_\phi\theta \bm{n}$ functions as the interface sharpening term which counteracts the diffusive flux $-M_\phi\bm{\nabla} \phi$ that occurs after $\phi$ is advected by the fluid velocity. A hyperbolic tangent profile across the diffuse interface is the order parameter that the conservative ACE reduces to at equilibrium. It can be found as follows: $\phi\left(\zeta \right)= \frac {1}{2}\left(\phi_{A} + \phi_{B} \right)+ \frac {1}{2}\left(\phi_{A} - \phi_{B} \right)\tanh\left(2\zeta/W\right)$, where $\zeta$ is a spatial coordinate along the normal with the origin at the interface.

In a single-field formulation depicting the motion of binary fluids, the interfacial surface tension effects may now be easily implemented inside a diffuse interface by the use of a distributed or smoothed volumetric force term. Thus, the equivalent Navier-Stokes equations for binary fluids as a single-field formulation for incompressible flow may be expressed as
\mathleft
\begin{equation}\label{eqn4}
\qquad \bm{\nabla} \cdot {\bm{u}}=0,
\end{equation}
\mathleft
\begin{equation}\label{eqn5}
\qquad \rho \left( \frac{\partial \bm{u}}{\partial t} + \bm{u}\cdot\bm{\nabla}\bm{u}\right) = - \bm{\nabla} p + \bm{\nabla} \cdot \left[ \mu (\bm{\nabla} \bm{u} + \bm{\nabla} \bm{u}^{\dagger})\right] + \bm{F}_s + \bm{F}_{ext},
\end{equation}
where any exterior body force is denoted by $\bm{F}_{ext}$ and the surface tension force by $\bm{F}_{s}$. Here, as surface tension varies with temperature and surfactant concentration, and the surface tension force effectively acts in both the normal and tangential directions to the interface. The continuous surface force approach~\cite{brackbill1992continuum} is a geometric technique that can be used to model these features. It is represented by the following equation, which uses the Dirac delta function $\delta_s$:
\mathleft
\begin{equation}\label{eqn6}
\qquad \bm{F}_{s} = \left( \sigma \kappa \bm{n} + \bm{\nabla}_s \sigma \right)\delta_{s}.
\end{equation}
Here, $\bm{n} = \bm{\nabla}{\phi}/|\bm{\nabla}{\phi}|$ and ${\kappa} = \bm{\nabla} \cdot \bm{n}$ are the unit normal vector and interface curvature, respectively. The normal or capillary force operating on the interface is represented by the first term on the right side of Eq.~(\ref{eqn6}), while the tangential or Marangoni force caused by surface tension gradients is represented by the second term, which involves the surface gradient operator $\bm{\nabla}_{s}$. The delta function $\delta_{s}$ is necessary to satisfy the property that $\int_{-\infty}^{+\infty} \delta_{s} dy = 1$ since surface tension only affects the interface. Here, $\delta_{s} = 1.5 W |\bm{\nabla} \phi|^2$ is a formulation of $\delta_{s}$ that localizes the smoothed surface tension force suitable for use within the phase-field modeling framework. In addition,
$\bm{\nabla}_s = \bm{\nabla} - \bm{n}(\bm{n} \cdot \bm{\nabla})$ is the surface gradient $\bm{\nabla}_{s}$ in Eq.~(\ref{eqn6}). Consequently, it is possible to define the Cartesian components of the surface tension force in Eq.~(\ref{eqn6}) as
\mathleft
\begin{eqnarray}
\qquad F_{sx} &=& -\sigma(T,\psi) |\bm{\nabla} \phi|^2   (\bm{\nabla} \cdot \bm{n}) {n}_x  +|\bm{\nabla} \phi|^2 \left[ ( 1-{n}_x^2 ) \partial_x \sigma(T,\psi)  -  {n}_x {n}_y  \partial_y \sigma(T,\psi)  \right],\nonumber\\
\qquad F_{sy} &=& -\sigma(T,\psi) |\bm{\nabla} \phi|^2   (\bm{\nabla} \cdot \bm{n}) {n}_y  +|\bm{\nabla} \phi|^2 \left[ ( 1-{n}_y^2 ) \partial_y \sigma(T,\psi)  -  {n}_x {n}_y  \partial_x \sigma(T,\psi)  \right]. \label{eq:surfacetensionforcecomponents}
\end{eqnarray}
Here, the combined effect of the nonlinear (parabolic) component and the Langmuir isotherm term, respectively, yield the functional dependence of the surface tension on temperature for the SRF and on surfactant concentration (see Eq.~(\ref{ST_SRF}) for $\sigma=\sigma(T,\psi)$). This study uses an isotropic finite differencing approach \cite{kumar2004isotropic} to determine the needed spatial gradients $\partial_x \sigma(T,\psi) $ and $\partial_y \sigma(T,\psi)$ in Eq.~(\ref{eq:surfacetensionforcecomponents}) for numerical implementations. The temperature field $T$ and the surfactant concentration field $\psi$ are determined by solving the equations for the energy transport and surfactant concentration, which are previously provided in Eq.~(\ref{energy eqn}) and  Eq.~(\ref{surfactant concentration eqn}), respectively. Finally, it is possible to represent the variations in fluid properties across the interface, including density and viscosity, as a continuous function of the phase field variable. This expression is then utilized in Eq.~(\ref{eqn5}). In this work (see e.g., \cite{ding2007diffuse}), we employ the following linear interpolation to account for the interfacial variations of fluid properties:
\mathleft
\begin{equation}\label{eqn11}
\qquad \rho  = \rho_{B} + \frac {\phi - \phi_{A}}{\phi_{A} - \phi_{B}} \left(\rho_{A} - \rho_{B} \right), \quad
\mu  = \mu_{B} + \frac {\phi - \phi_{A}}{\phi_{A} - \phi_{B}} \left(\mu_{A} - \mu_{B} \right),
\end{equation}
where $\rho_{A}$, $\rho_{B}$ and $\mu_{A}$, $\mu_{B}$ are the densities and the dynamic viscosities in the fluid phases, respectively and denoted by $\phi_{A}$ and $\phi_{B}$. An equation similar to Eq.~(\ref{eqn11}) will also be utilized for distributing the interfacial jump in the thermal conductivity in solving the energy equation. In this study, we use $\phi_{A}=0$ and $\phi_{B}=1$.

\section{Central moment lattice Boltzmann schemes for simulation of surfactant-laden self-rewetting fluid motions} \label {Sec.LBschemes}
This section will introduce a numerical LB approach based on more robust collision models involving central moments~\cite{geier2006cascaded,premnath2009incorporating,premnath2011three,hajabdollahi2021central,elbousefi2023thermocapillary} for solving the equations of the phase-field model for tracking the interface (Eq.~(\ref{eqn1})) and the binary fluid motions (Eqs.~(\ref{eqn4})-(\ref{eq:surfacetensionforcecomponents})) given in the previous section, along with the transport of energy and the surfactant concentration presented in Eqs.~(\ref{energy eqn}) and Eqs.~(\ref{SUR_mu_psi_2}) and (\ref{SUR_P}), respectively, earlier. In order to solve these four equations, four different distribution functions need be evolved on the standard two-dimensional square lattice (D2Q9). This is done by performing a \emph{collision step} that is based on the relaxation of the different central moments of such distribution functions to their respective equilibria. The distribution functions then undergo lock-step advection to their adjacent nodes along the characteristic directions in the \emph{streaming step}. The moments of the corresponding distribution functions are then used to derive the macroscopic variables, which include the order parameter, the fluid pressure and velocity, as well as the temperature and the surfactant concentration fields. Note that this requires the use of appropriate mappings that transform between these quantities pre- and post-collision steps since the streaming step is carried out by means of the distribution functions, while the collision step is carried out using central moments. When compared to the other collision models in the LB framework, the central moment LB approaches are demonstrated to be more robust (e.g., enhanced numerical stability) (see~\cite{hajabdollahi2021central,yahia2021central,yahia2021three,elbousefi2023thermocapillary} for recent examples). Although an orthogonal moment basis was used to develop the original central moment LB scheme for two-fluid interfacial flows~\cite{hajabdollahi2021central}, we will adopt an improved structure employing the non-orthogonal moment basis in the following.

Brief descriptions of the central moment LB schemes and attendant details fof the capturing of interfaces based on the conservative Allen-Cahn equation, two-fluid motion, energy transport, and surfactant concentration transport, respectively, are presented Appendices~\ref{Sec.4.1},~\ref{Sec.4.2},~\ref{Sec.4.3}, and~\ref{Sec.4.4}, respectively. While these methods are applicable for general class variable surface-tension driven flows, in this work, they will be mainly applied to study the effect of various characteristic parameters on the flow patterns and the intensity of thermocapillary convention in migrating surfactant-laden bubble in self-rewetting fluids (SRFs). Before proceeding with this, we will first validate our numerical approach given above for some standard thermocapillary benchmark problems for normal fluids for which analytical solutions are available in the literature in the next section.

\section{Model validation} \label{Sec.5}

\subsection{Laplace-Young law of a static drop}

\begin{figure}[H]
\centering
\includegraphics[trim = 0 0 0 0,clip, width = 100mm]{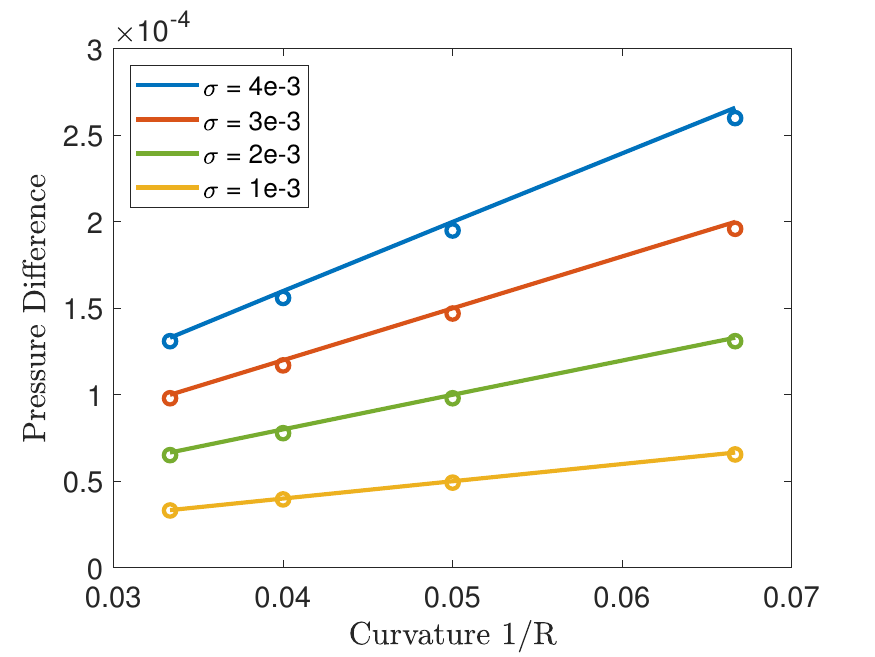}
\caption{Comparison of the computed pressure differences (symbols) against the analytical predictions using the Laplace-Young relation for various values of the drop curvature 1/R with four different values of surface tension.}
\label{dpress_Curvature}
\end{figure}

The static drop test case is first used to validate the capability of this model in the absence of body force effects or surfactants. In this test, the analytical solution from the Laplace-Young's relation for a 2D drop at rest relates the pressure difference inside and outside of the drop $(\triangle P)$ of radius $R$ to the surface tension $\sigma$ and its radius of curvature $1/R$ via $\triangle P = \sigma / R$, which will be used as a comparison. To do the simulation, we consider a drop of density $\rho_{A}$ is placed at the center of a square domain that is divided into 200 $\times$ 200 lattice nodes and filled with an ambient fluid of density $\rho_{B}$. Periodic boundary conditions are specified for all walls. For four different surface tensions of $\sigma = 1\times 10^{-3}$, $2\times 10^{-3}$, $3\times 10^{-3}$, and $4\times 10^{-3}$, we ran the simulations at a density ratio of $1000$. The estimated pressure differences between the drop and ambient fluid are displayed in Fig.~\ref{dpress_Curvature} in comparison to the Laplace-Young relation equation. It is evident that there is a predicted linear relationship between $\triangle P$ and $1/R$; a good quantitative agreement between the analytical solution and the computed results is seen.

\subsection{Equilibrium distribution of surfactant concentration on a flat interface}
To further validate the model quantitatively in the presence of surfactants, we will next compare the numerical results with an analytical solution representing the equilibrium surfactant concentration profile at a planar interface. A schematic figure of this problem is shown in Fig.~\ref{fig: Model_setup}.
\newpage
\begin{figure}[H]
\centering
\includegraphics[trim = 0 0 0 0,clip, width = 100mm]{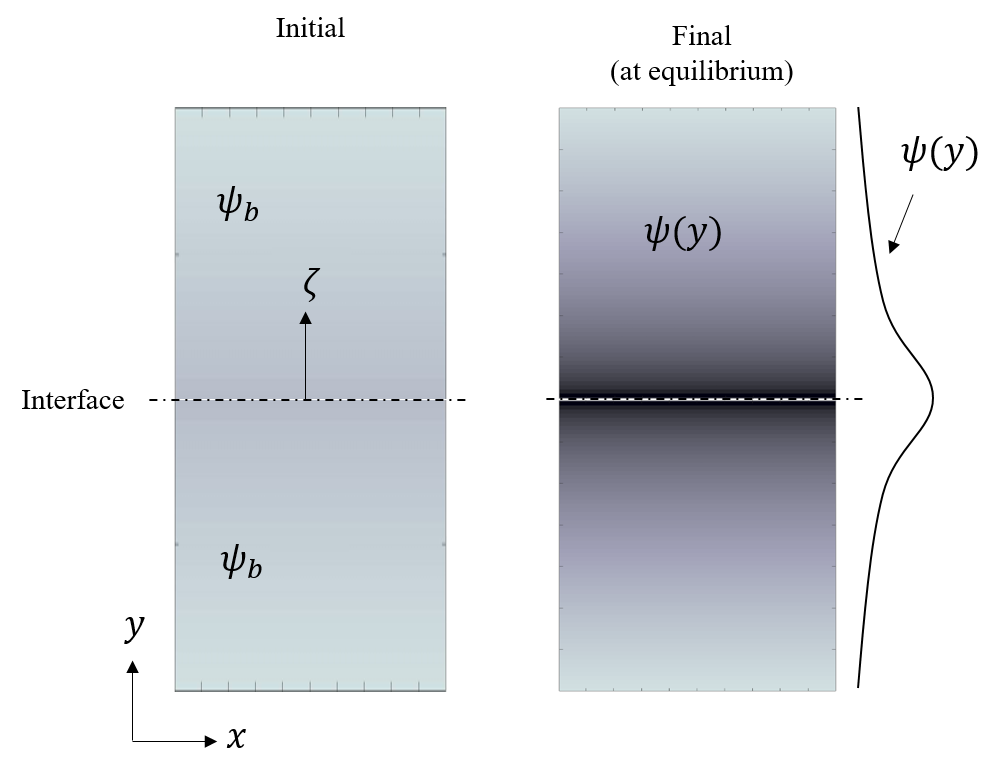}
\caption{\label{fig: Model_setup} Schematic of the initial and final distributions of the surfactant concentration field around a flat interface illustrating the tendency of interfacial surfactant adsorption.}
\end{figure}
This analytical solution can be derived when the chemical potential $\mu_\psi$ (see Eq.~(\ref{SUR_mu_psi_1})) is uniformly distributed throughout the domain. By setting $\mu_\psi = \mu_{\psi,b}$, where $\mu_{\psi,b}$ corresponds to the value in the bulk region that can be determined by $\mu_{\psi,b}=\lambda \ln \psi_b + w (\phi_A+\phi_B)^2 /8$, we obtain the analytical solution for the equilibrium surfactant concentration profile as~\cite{ouderji2019phdthesis}
\begin{equation}
\psi^{eq}(\zeta)=  \frac{\psi_b}{\psi_b + \psi_c(\zeta)}.   
\end{equation}
Here,  $\psi_b$ is the fixed bulk surfactant concentration loading, and  $\psi_c(\zeta)$ in an auxiliary function determined by
\begin{equation}
\psi_c(\zeta)= \exp \Bigg\{ -\frac{s}{2\lambda} \bigg[\frac{2}{W}\phi_m \sech^2 \left( \frac{2\zeta}{W} \right) \bigg]^2 + \frac{w}{2\lambda} \big[\phi(\zeta)-\phi_m\big]^2 - \frac{w}{8}(\phi_A+\phi_B)^2 \Bigg\}.  
\end{equation}
Here, $\phi(\zeta)=\phi_m + \phi_o \tanh(2\zeta/W)$ is the corresponding equilibrium hyperbolic tangent profile of the phase variable, $\phi_m = (\phi_A+\phi_B)/2$, $\phi_o = (\phi_A-\phi_B)/2$ and $\zeta$ is a coordinate along the normal direction that originates at the interface in the above equation. For this test, we select a computational domain with a grid resolution of $5 \times 101$ and an interface width of $W = 5$. The values of the parameters $s$, $w$, and $\lambda$ may be computed using the non-dimensional numbers $\mbox{Ex} = 0.025$ and $\mbox{Pi} = 50$. Other parameters are set as follows: $\sigma_{0}=0.01, M_{\phi}=M_{\psi}=0.02$. The profiles of equilibrium surfactant concentration are obtained through numerical simulations for three different bulk surfactant concentration values: $\psi_b = 0.05, \psi_b = 0.15, \text{and}\ \psi_b = 0.225$. These profiles are displayed alongside the analytical solution for comparison in Fig.~\ref{equilibrium_surf_conc_profile_flat_interface}.
\begin{figure}[H]
\centering
\includegraphics[trim = 0 0 0 0,clip, width = 100mm]{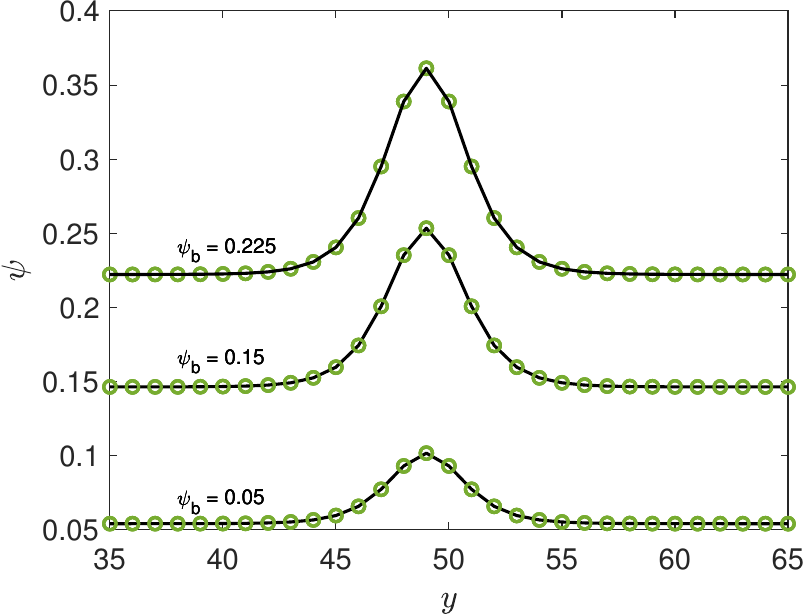}
\caption{Profiles of the surfactant concentration $\psi$ for a planar interface for various bulk surfactant concentrations  $\psi_b = 0.05, 0.15, \text{and}\ 0.225$. The numerical solution and the analytical prediction are represented by symbols and solid lines, respectively.}
\label{equilibrium_surf_conc_profile_flat_interface}
\end{figure}
It is evident that the interface is where the surfactant concentration is highest. This represents the preferred surfactant adsorption at the interface, which is resisted by diffusion. Furthermore, the estimated equilibrium profiles and the analytical solution match up nicely, further confirming our choice to solve the surfactant concentration problem using the central moment LB method accurately.
\subsection{Buoyancy-driven rising bubble with large density ratio}
Finally, we will simulate a rising air bubble in water with a high density ratio and a complex interface change that includes the surface tension force, buoyancy, and viscous effect in order to further test our computational approach; this case study serves as a precursor to studying the physics of the surfactant-laden bubble migration in SRFs that will be discussed in the next section. We study, under various parametric conditions, a bubble with a diameter of $D$ and a density of $\rho_B$ rising due to buoyancy forces in an ambient fluid with a density of $\rho_A$, where the density ratio $\rho_A/\rho_B$ equals $1000$ and the dynamic viscosity ratio is given as $\mu_A/\mu_B=100$. Our objective is to evaluate the capability of the central moment LB approach to reproduce the temporal history of the bubble route quantitatively and to represent the various shape changes the bubble experiences due to the balance between competing forces.

The non-dimensional parameters specified in this problem are the Reynolds number $\mbox{Re}$ and the Eotvos number $\mbox{Eo}$, which can be defined as follows:
\begin{equation}
\qquad \mbox{Re} = \frac{\rho_A U_0 D } {\mu_A}, \quad \quad \mbox{Eo} = \frac{\rho_A U_0^2 D} {\sigma }, \label{}
\end{equation}
where $g$ is the gravitational acceleration and $\sigma$ denotes the surface tension coefficient.
The length scale $L = D$, the velocity scale $U_0 =\sqrt {g D}$, which denotes a characteristic gravitational velocity, and the time scale $T = L/U_0$ are taken into consideration in this problem. In this benchmark problem, a body force of $\boldsymbol{F}_{text}=-(\rho - \rho_A) g \boldsymbol{j}$ is applied to set the bubble in motion. In a rectangular domain of $161\times 481$ lattice nodes, a gas bubble with a diameter of $D=80$ is centered at a location $(80,120)$. Periodic boundary conditions are used on the left and right sides and no-slip boundary conditions are imposed on the upper and lower walls. For reporting time histories, we use the dimensionless time which is denoted by $t^*= t/ T$. Depending on the magnitudes of these dimensionless groups, the bubble experiences intricate interfacial shape changes, leading to a variety of forms, including spherical-cap, dimpled ellipsoidal-cap, and skirted (see \cite{clift2005bubbles} for more details). In this work, we perform buoyancy-driven bubble rise simulations at four Eotvos numbers $(i.e., \mbox{Eo} = 0.1, 10, 50, \text{and}\ 125)$ to illustrate the interface shapes of rising bubbles in different flow regimes. We set $\rho_A/\rho_B=1000$ and $\mu_A/\mu_B=100$ at a fixed Reynolds number $\mbox{Re} = 35$.
\begin{figure}[H]
\centering
\begin{subfigure}{0.15\textwidth}
\includegraphics[trim = 100 10 20 15,clip, width = 40mm]{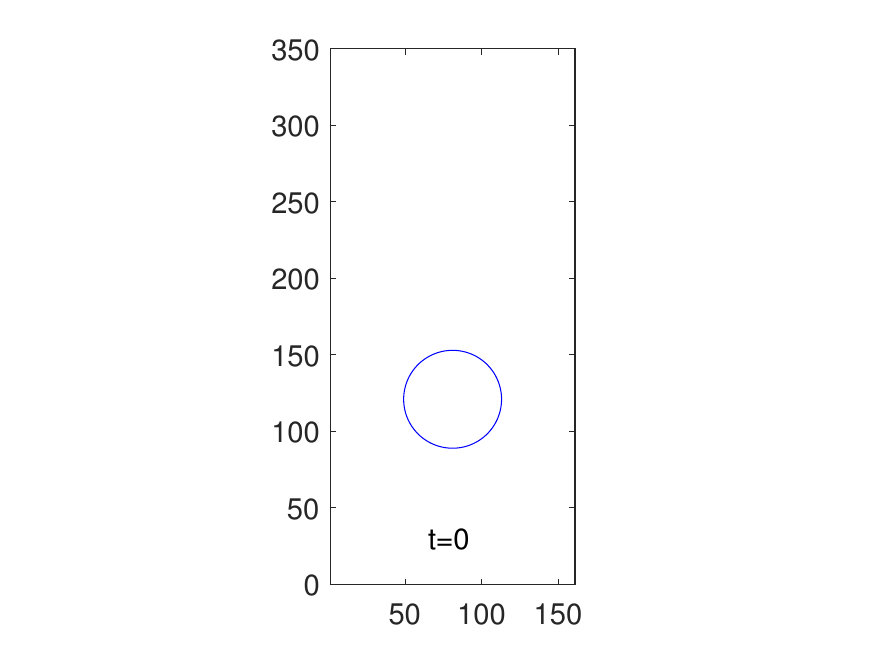}
\end{subfigure}
\begin{subfigure}{0.15\textwidth}
\includegraphics[trim = 100 10 20 15,clip, width = 40mm]{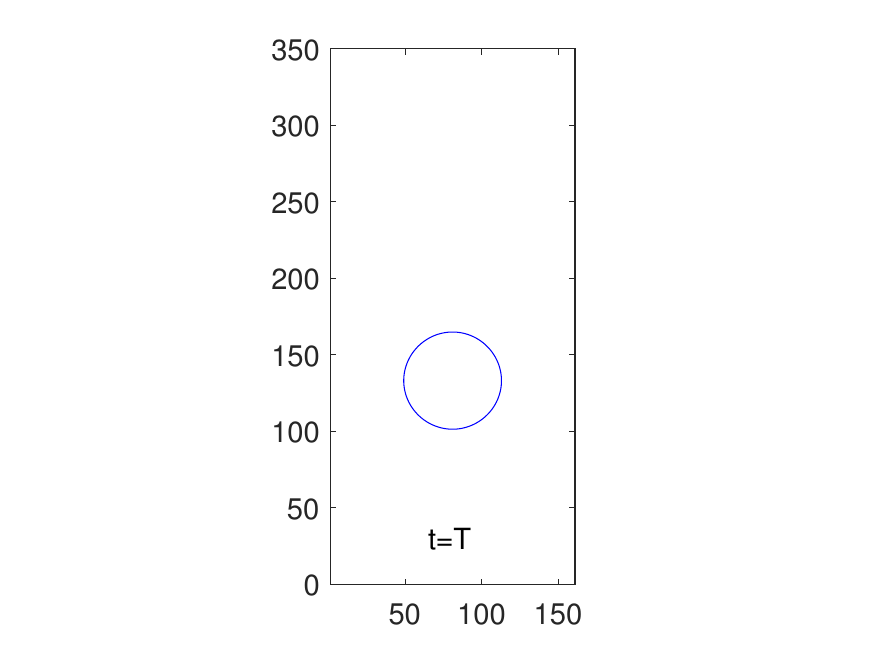}
\end{subfigure}
\begin{subfigure}{0.15\textwidth}
\includegraphics[trim = 100 10 20 15,clip, width = 40mm]{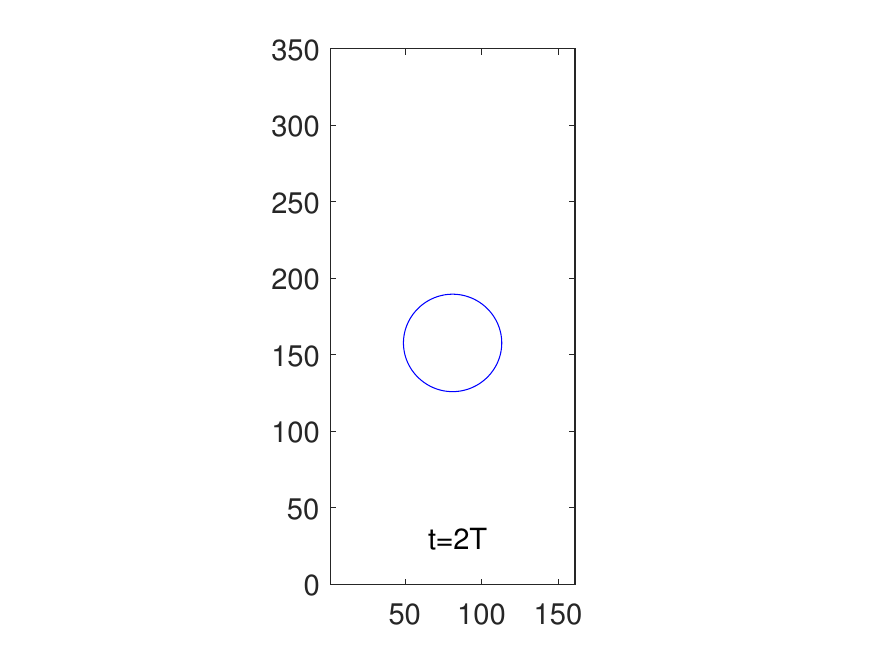}
\end{subfigure}
\begin{subfigure}{0.15\textwidth}
\includegraphics[trim = 100 10 20 15,clip, width = 40mm]{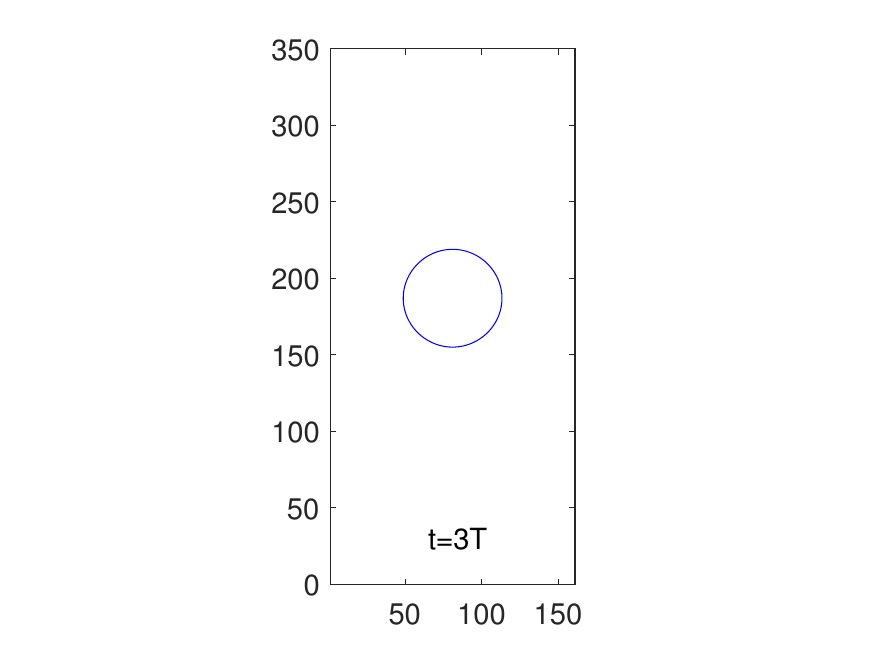}
\end{subfigure}
\begin{subfigure}{0.15\textwidth}
\includegraphics[trim = 100 10 20 15,clip, width = 40mm]{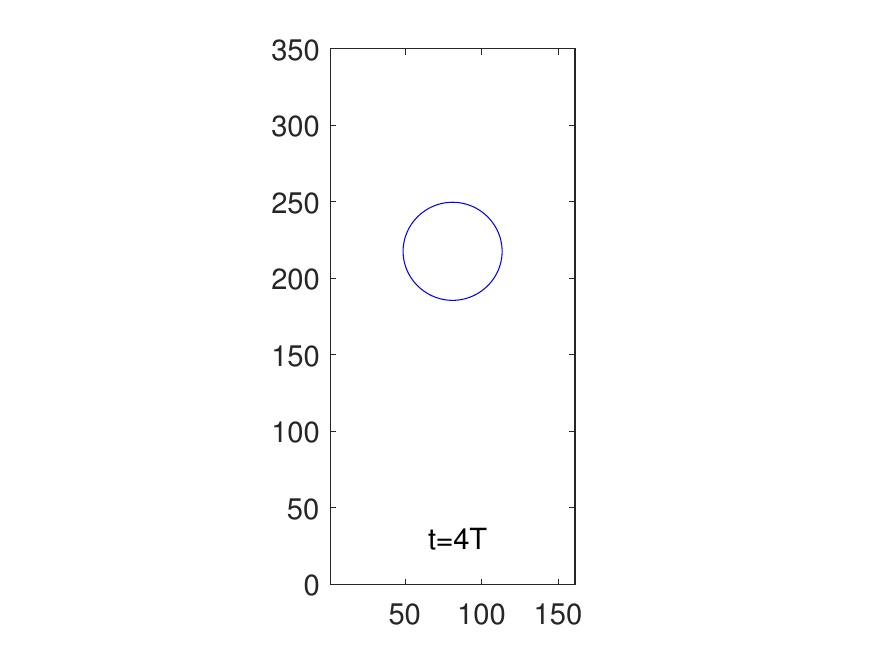}
\end{subfigure}
\begin{subfigure}{0.15\textwidth}
\includegraphics[trim = 100 10 20 15,clip, width = 40mm]{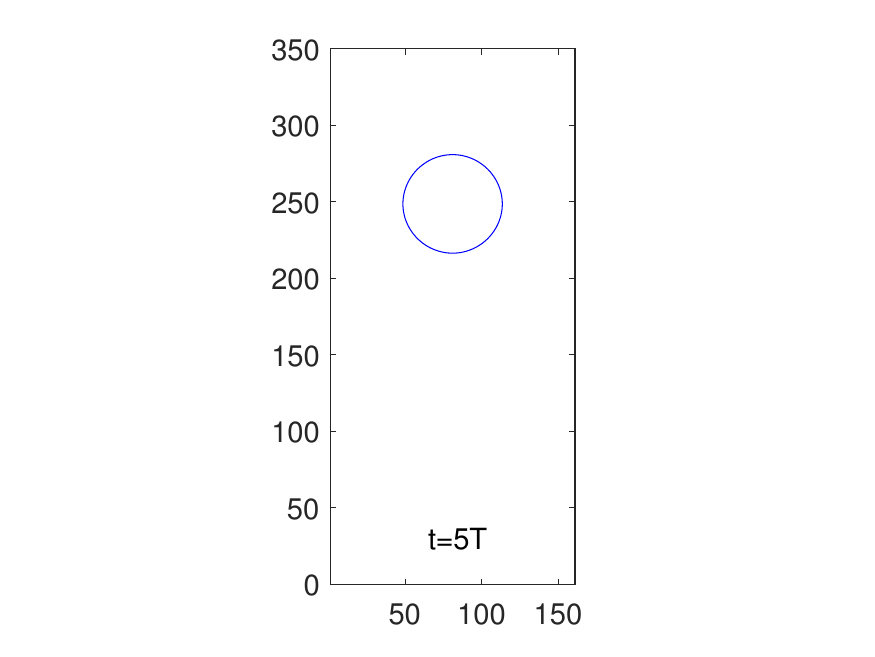}
\end{subfigure}
\text{(a) $\mbox{Eo} = 0.1$}
\par\bigskip

\centering
\begin{subfigure}{0.15\textwidth}
\includegraphics[trim = 100 10 20 15,clip, width = 40mm]{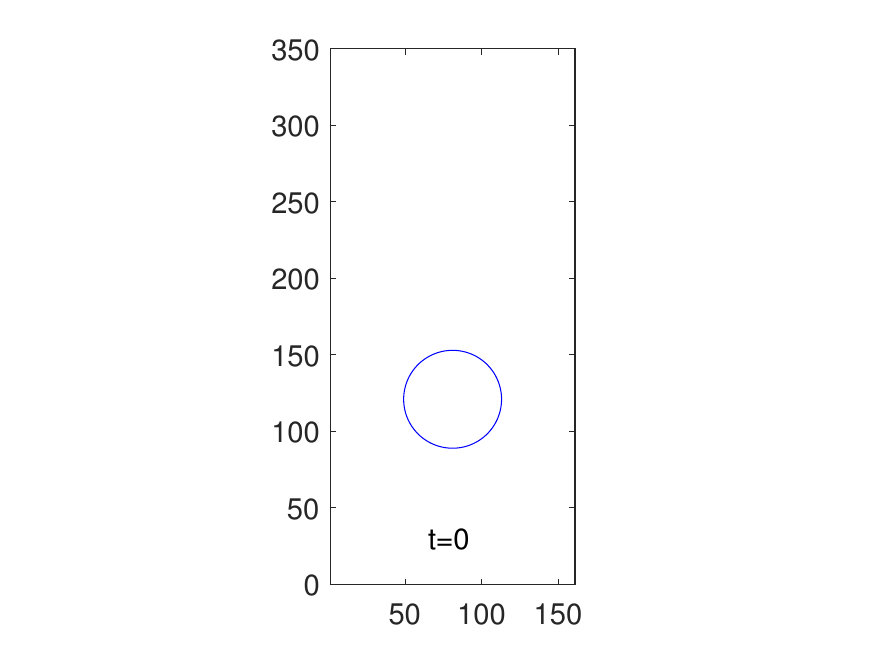}
\end{subfigure}
\begin{subfigure}{0.15\textwidth}
\includegraphics[trim = 100 10 20 15,clip, width = 40mm]{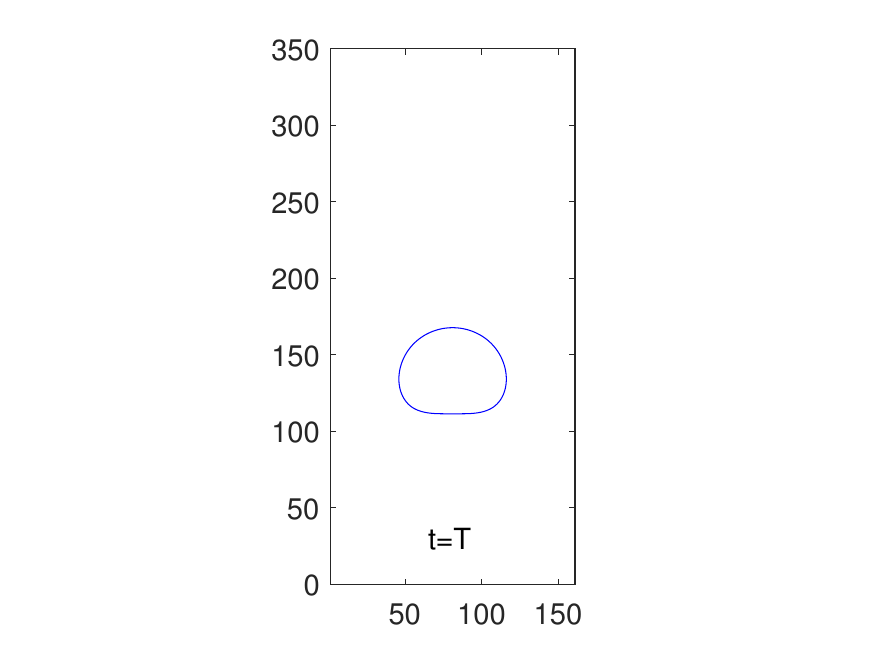}
\end{subfigure}
\begin{subfigure}{0.15\textwidth}
\includegraphics[trim = 100 10 20 15,clip, width = 40mm]{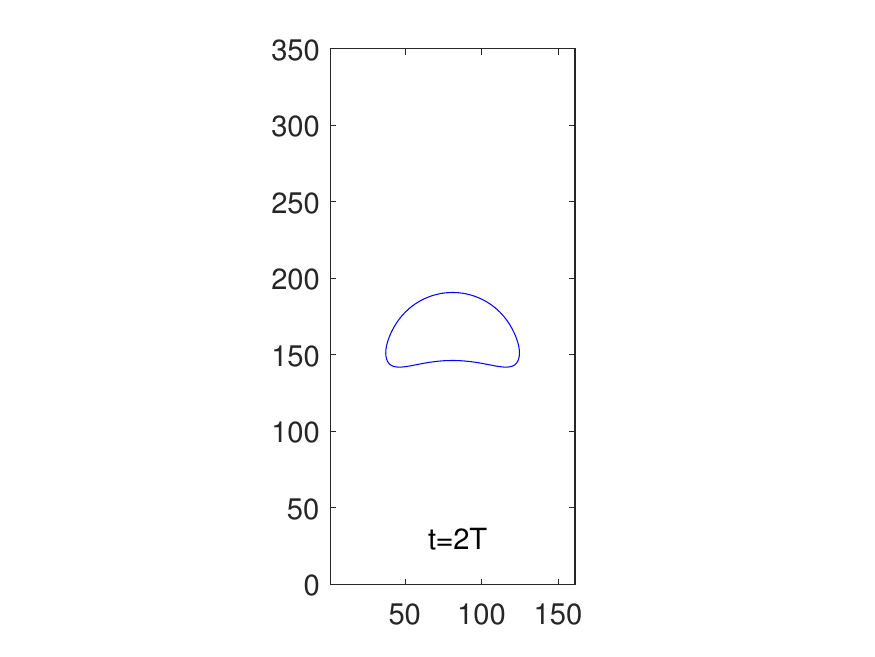}
\end{subfigure}
\begin{subfigure}{0.15\textwidth}
\includegraphics[trim = 100 10 20 15,clip, width = 40mm]{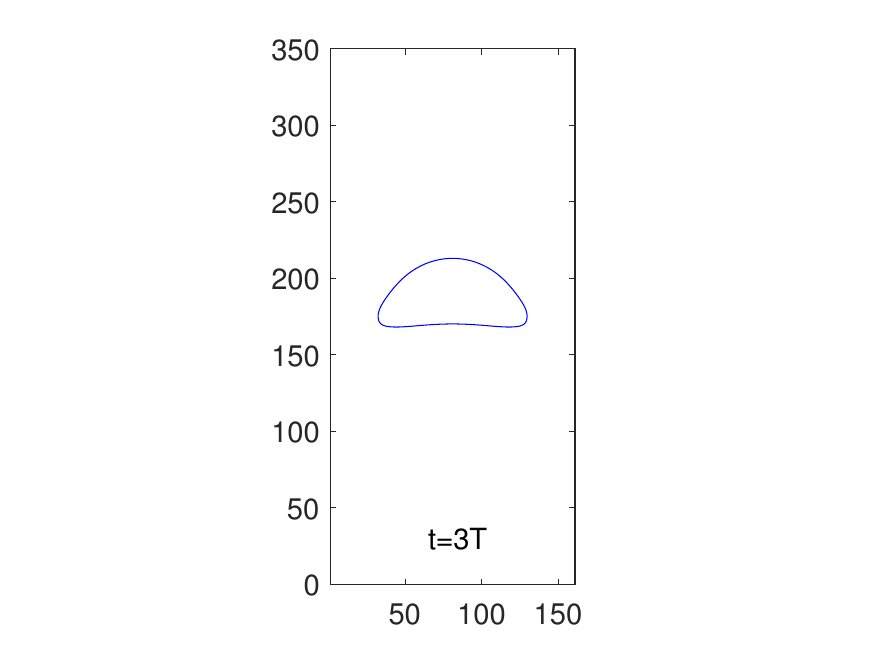}
\end{subfigure}
\begin{subfigure}{0.15\textwidth}
\includegraphics[trim = 100 10 20 15,clip, width = 40mm]{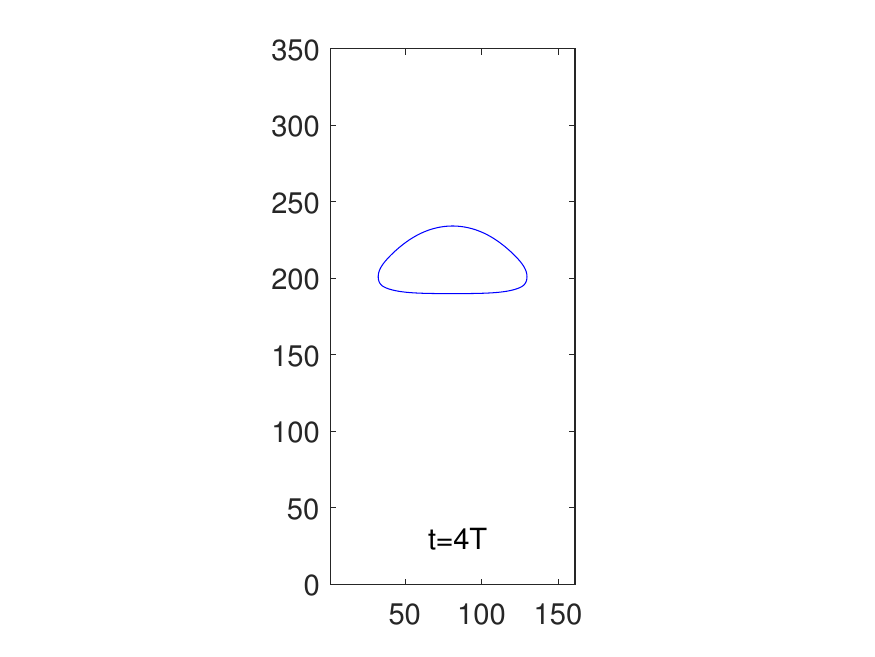}
\end{subfigure}
\begin{subfigure}{0.15\textwidth}
\includegraphics[trim = 100 10 20 15,clip, width = 40mm]{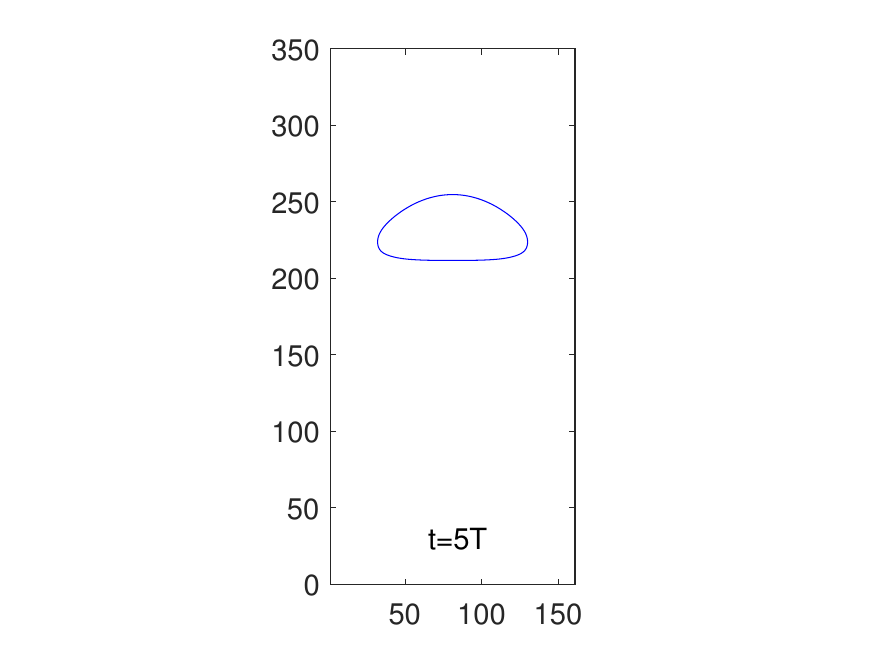}
\end{subfigure}
\text{(b) $\mbox{Eo} = 10$}
\par\bigskip

\centering
\begin{subfigure}{0.15\textwidth}
\includegraphics[trim = 100 10 20 15,clip, width = 40mm]{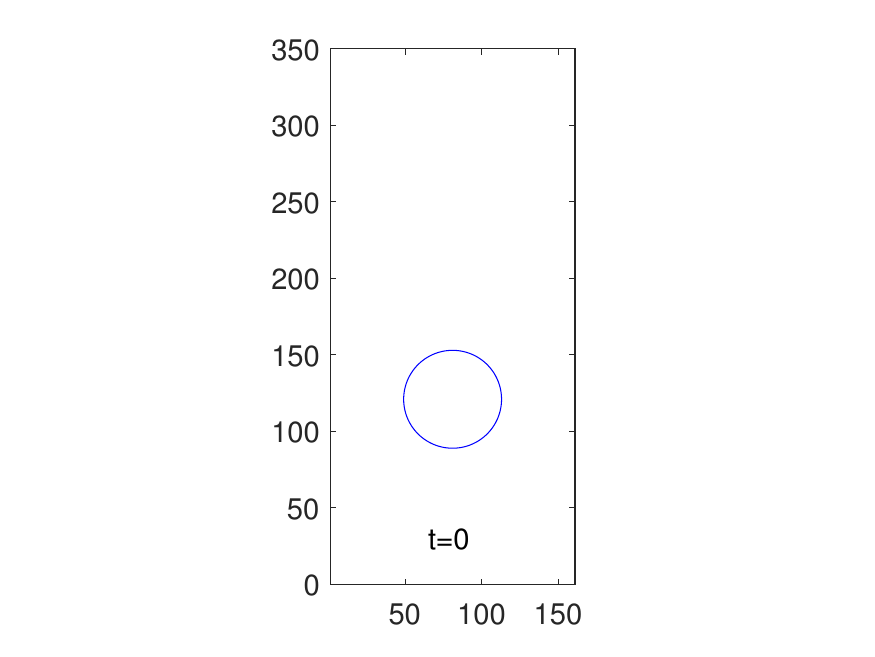}
\end{subfigure}
\begin{subfigure}{0.15\textwidth}
\includegraphics[trim = 100 10 20 15,clip, width = 40mm]{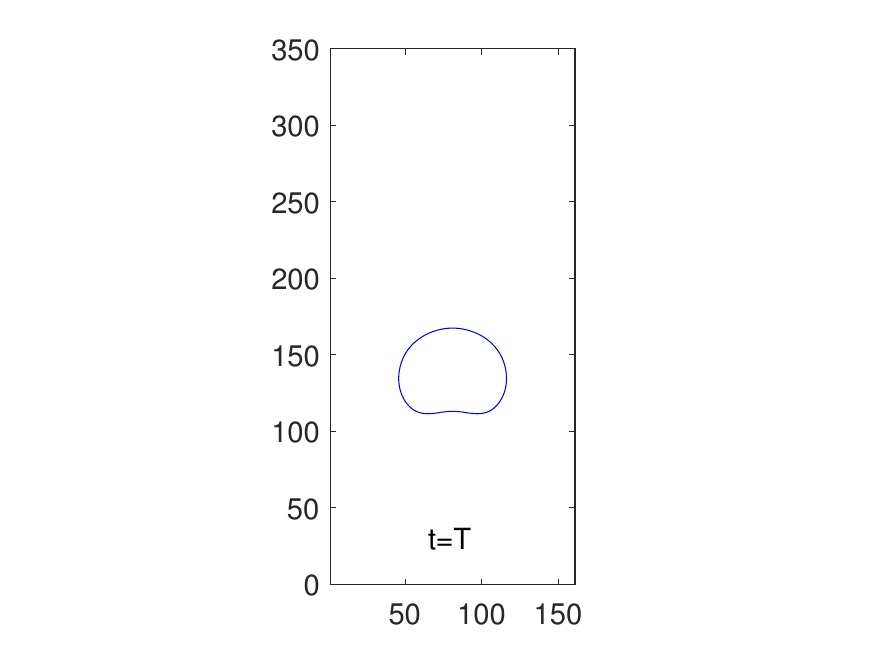}
\end{subfigure}
\begin{subfigure}{0.15\textwidth}
\includegraphics[trim = 100 10 20 15,clip, width = 40mm]{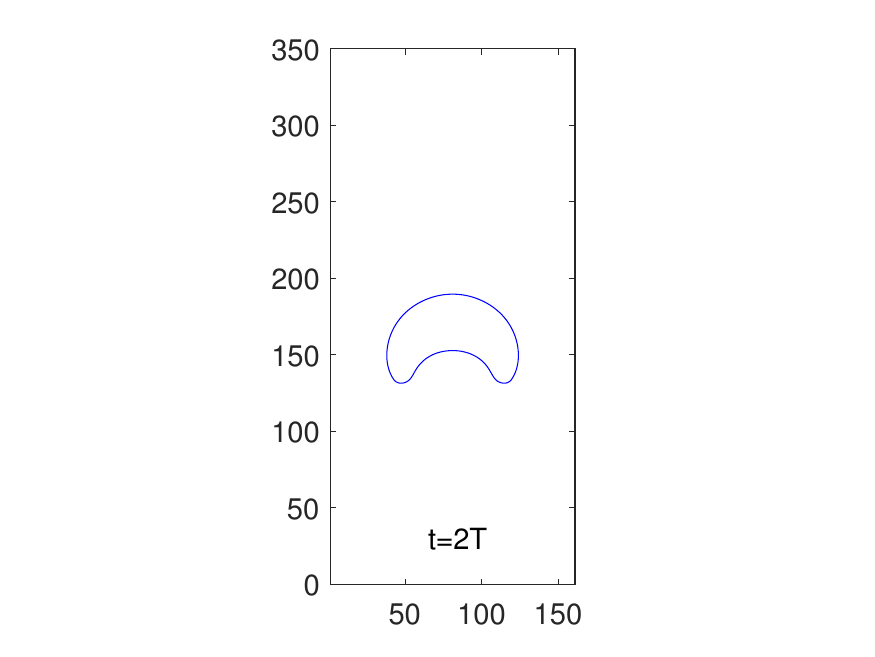}
\end{subfigure}
\begin{subfigure}{0.15\textwidth}
\includegraphics[trim = 100 10 20 15,clip, width = 40mm]{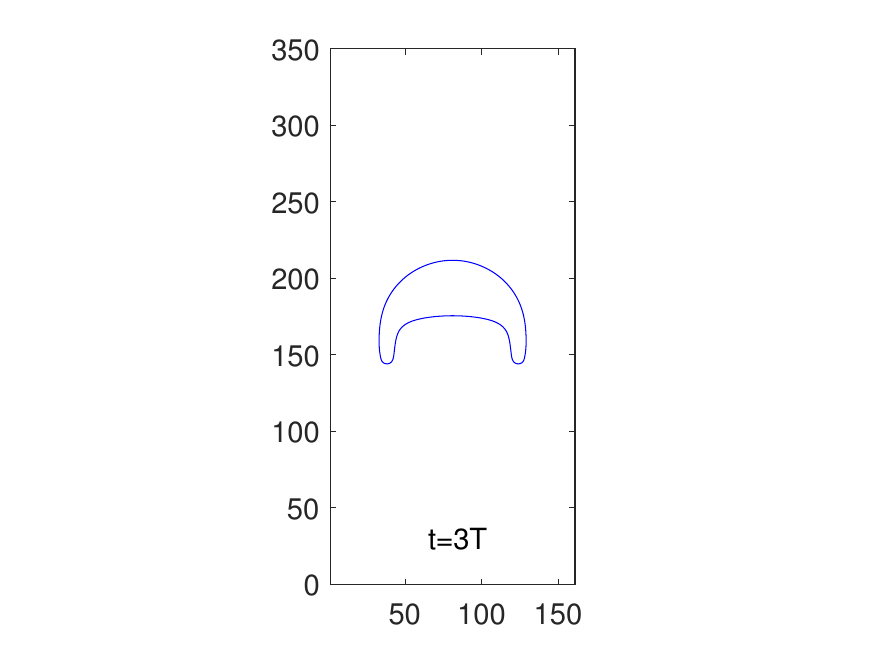}
\end{subfigure}
\begin{subfigure}{0.15\textwidth}
\includegraphics[trim = 100 10 20 15,clip, width = 40mm]{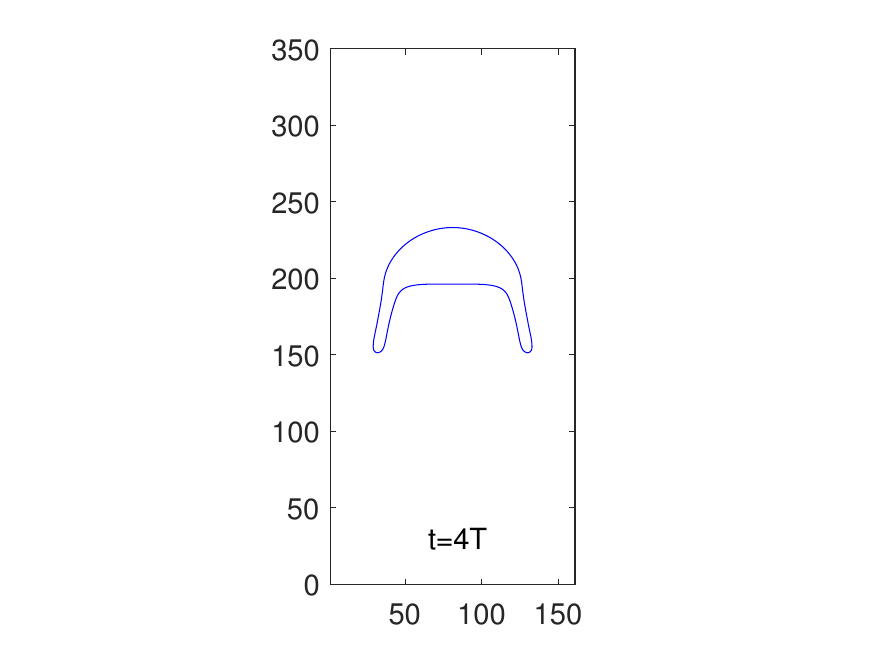}
\end{subfigure}
\begin{subfigure}{0.15\textwidth}
\includegraphics[trim = 100 10 20 15,clip, width = 40mm]{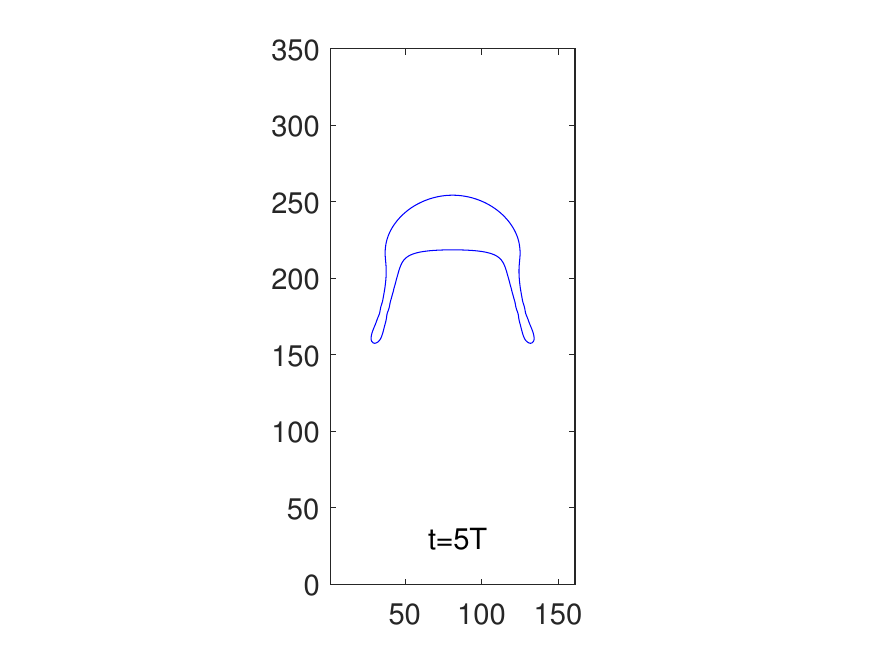}
\end{subfigure}
\text{(c) $\mbox{Eo} = 50$}
\par\bigskip

\centering
\begin{subfigure}{0.15\textwidth}
\includegraphics[trim = 100 10 20 15,clip, width = 40mm]{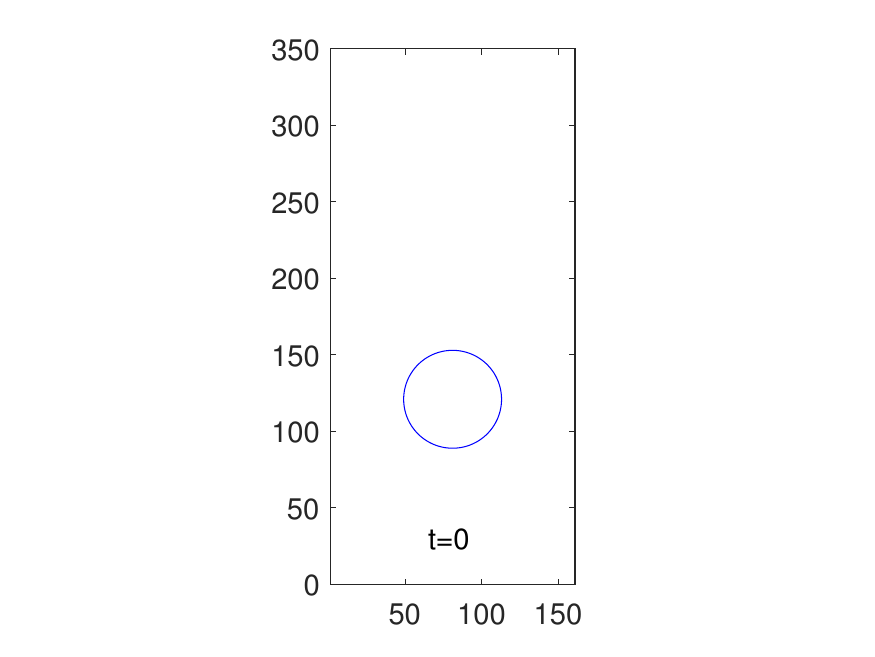}
\end{subfigure}
\begin{subfigure}{0.15\textwidth}
\includegraphics[trim = 100 10 20 15,clip, width = 40mm]{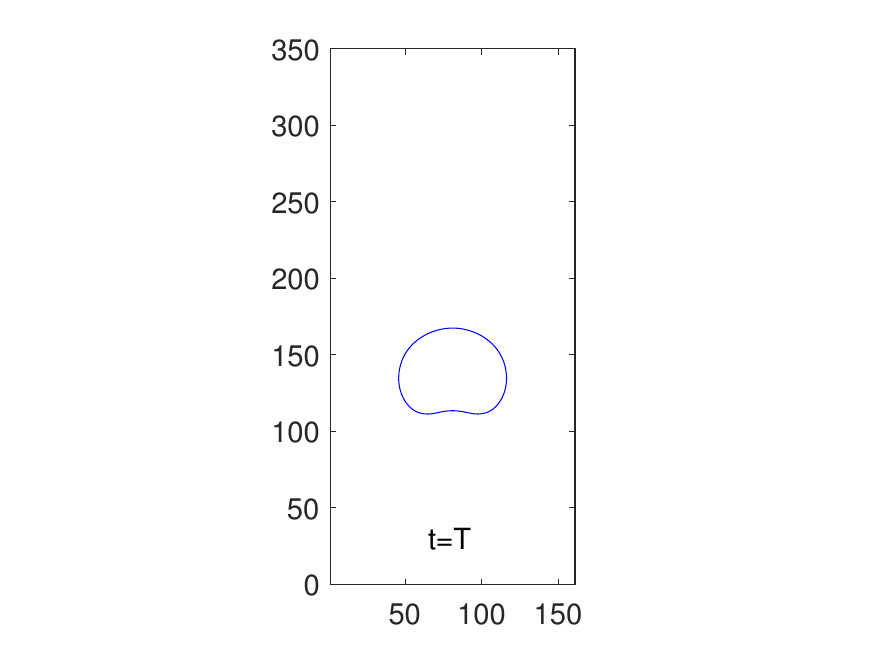}
\end{subfigure}
\begin{subfigure}{0.15\textwidth}
\includegraphics[trim = 100 10 20 15,clip, width = 40mm]{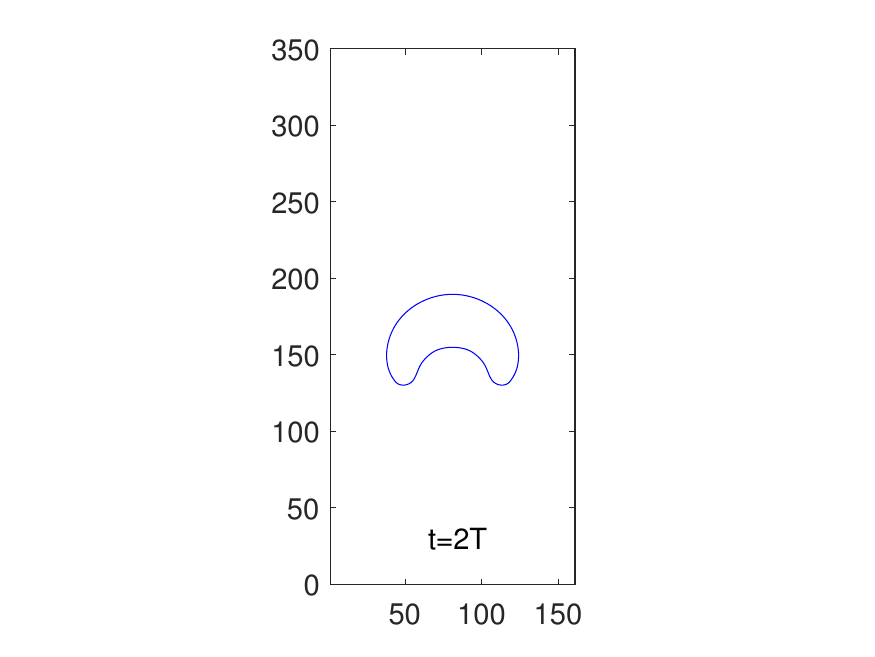}
\end{subfigure}
\begin{subfigure}{0.15\textwidth}
\includegraphics[trim = 100 10 20 15,clip, width = 40mm]{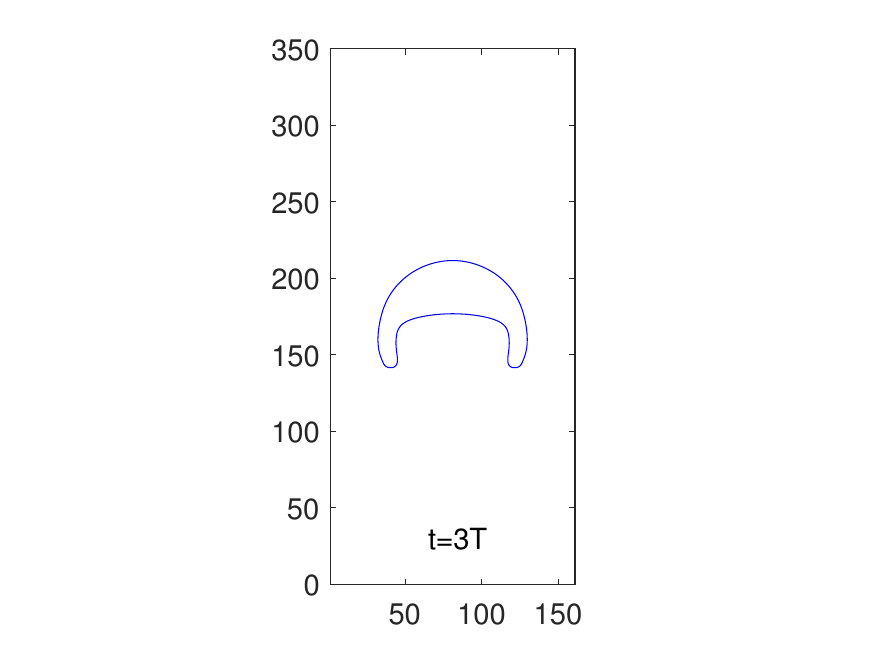}
\end{subfigure}
\begin{subfigure}{0.15\textwidth}
\includegraphics[trim = 100 10 20 15,clip, width = 40mm]{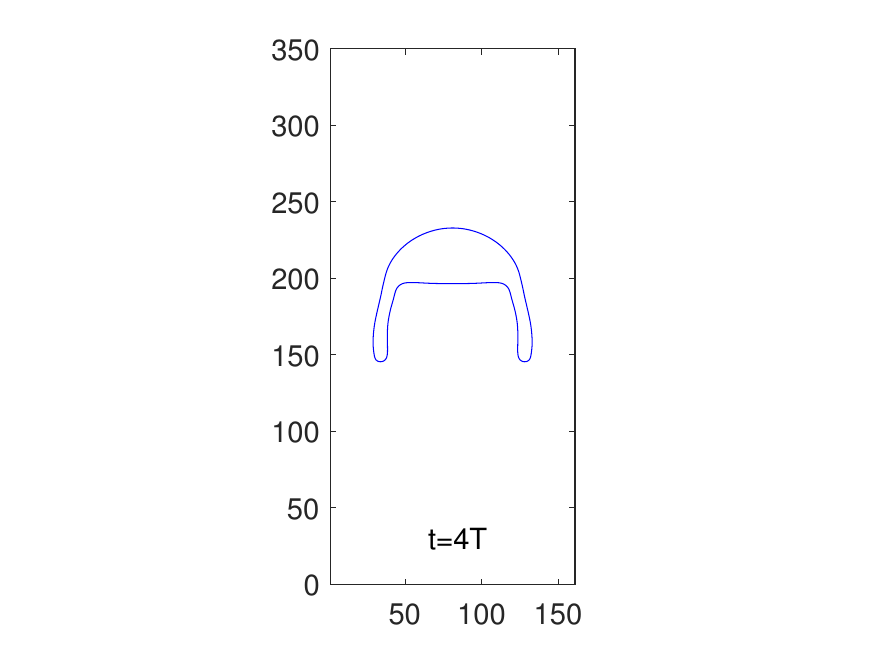}
\end{subfigure}
\begin{subfigure}{0.15\textwidth}
\includegraphics[trim = 100 10 20 15,clip, width = 40mm]{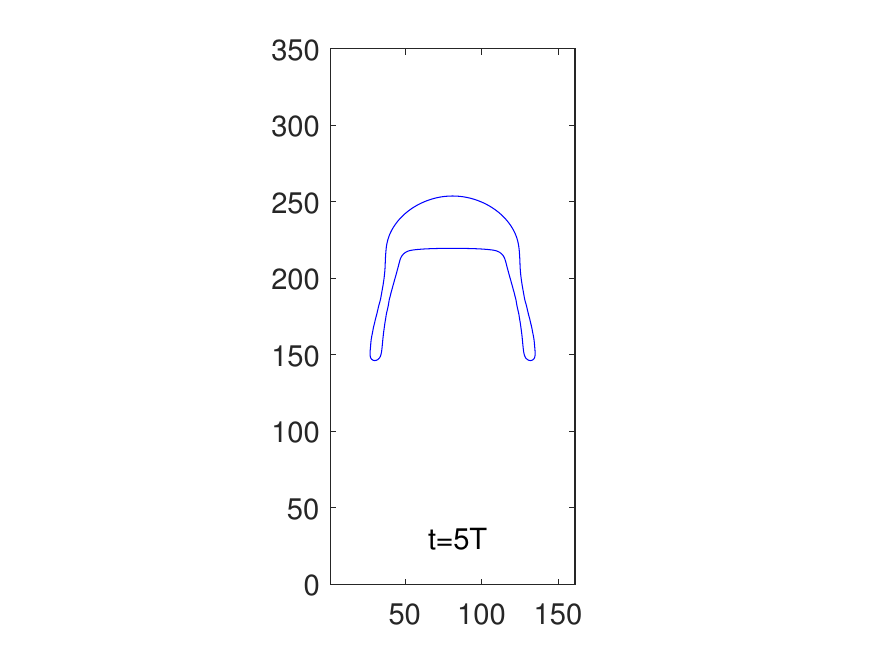}
\end{subfigure}
\text{(d) $\mbox{Eo} = 125$}
\caption{Evolution of the interface for the bubble rising at $\mbox{Re} = 35$ and $(a)$ $\mbox{Eo} = 0.1$, $(b)$ $\mbox{Eo} = 10$, $(c)$ $\mbox{Eo} = 50$, and $(d)$ $\mbox{Eo} = 125$.}
\label{Buoyancy-driven_rising_bubble}
\end{figure}

Figure~\ref{Buoyancy-driven_rising_bubble} shows the computed evolution of the interface of the rising bubble at the above four values of $\mbox{Eo}$. Due to the strong surface tension force, the form of the gas bubble remains unchanged for the rising bubble with $\mbox{Eo}=0.1$. When the surface tension force plays a relatively large role compared to the other forces, i.e., when the Eotvos number is low $(\mbox{Eo}=10)$, the bubble goes through smaller deformation that is beginning at its rear end, which then leads in a flattening of that side as the bubble rises. When $\mbox{Eo} = 50$, the driving buoyancy force exceeds the surface tension under the dominant viscous force, causing a substantially larger deformation via stretching that results in the formation of tails that extend with time. The gas bubble shape significantly changes at larger Eotvos numbers $(\mbox{Eo}=125)$ due to decreased surface tension forces, and two extended tails can be seen behind the main bubble. The findings obtained through other approaches~\cite{aland2012benchmark,wang2015mass,chen2018simplified} are quite comparable to these estimated shape changes with varying $\mbox{Eo}$ at different times.

\begin{figure}[H]
\centering
\includegraphics[trim = 0 0 0 0,clip, width = 100mm]{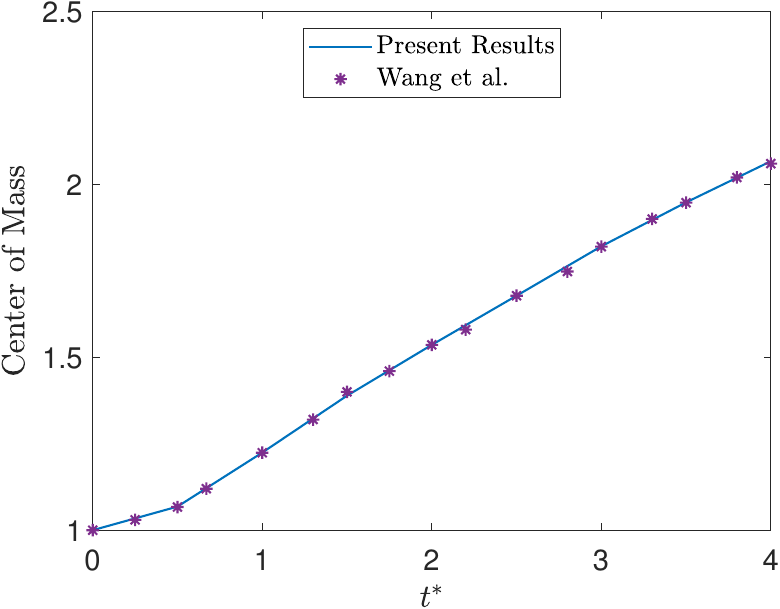}
\caption{Comparison of the non-dimensional center of mass for the rising bubble with the reference numerical results given in Wang \emph{et al.}~\cite{wang2015mass} for $\mbox{Re}=35$ and $\mbox{Eo}=125$.}
\label{mass_center_rising_bubble}
\end{figure}

In order to evaluate the current approach with more confidence, a quantitative comparison is required. The mass center of the rising bubble as its shape changes with time is shown in Fig.~\ref{mass_center_rising_bubble} for the case of $\mbox{Re} = 35$ and $\mbox{Eo}=125$. The computational results are compared with numerical results available in the literature~\cite{wang2015mass} used as a reference. The fact that the current results are in good quantitative agreement with the range of the reference data indicates that the current approach is accurate in solving this problem with a high density ratio.
Finally, we note here that the present LB numerical approach was further tested and validated against a new analytical solution for thermocapillary convection due to Marangoni stresses in two superimposed SRF liquid layers subject to a nonuniform heating in our recent study~\cite{elbousefi2023thermocapillary}.

\section{Results and Discussion} \label{Results and Discussion}
The goal of this work is to model and study the physics of the migration of surfactant-laden thermocapillary bubbles in a self-rewetting fluid (SRF) by investigating the effects of different characteristic parameters associated with this problem using the LB schemes, which were validated in the previous section. In this regard, we will study the cases where the dimensionless surface tension quadratic sensitivity coefficient is non-zero, i.e., $\sigma_{TT} \neq 0$ or $M_2 \neq 0$ to illustrate the role of self-rewetting fluids (SRFs), and also to compare and contrast the results with the normal fluids (NFs), where only the linear coefficient of the surface tension (i.e., $\sigma_{T}$ or $M_1$) exists and $M_2$ is zero. The surfactant-laden thermocapillary flow strengths are controlled by the choice of the dimensionless linear and quadratic sensitivity coefficients of the surface tension variation with temperature, i.e., $M_1$ and $M_2$, respectively, and also by the choice of the Gibbs elasticity parameter $\beta$. In this study, unlike otherwise stated, the following parameters are kept constant: $T_H=1$, $T_C=0$, $T_{ref}=0.5$, $D=0.1H$, $\tilde{\rho} = 1000$, $\tilde{\mu} = 100$, $\tilde{\alpha} = 1$, $ \mbox{Re} = 5$, and $\mbox{Ma} = 0.25$. Thus, in particular, in all the cases considered in what follows, the left end of the domain is maintained hotter at $T_H$ than the right end which is at $T_C$ (see Fig.~\ref{Model_setup} for a schematic of the problem setup). The reference surface tension for the clean interface case used in this study is assumed to be $\sigma_0=5\times 10^{-3}$. We used $W = 5$ and $M_{\phi} = 0.1$ for the model parameters associated with the conservative ACE for interface tracking. In addition, in the surfactant model, the non-dimensional numbers $\mbox{Ex}$ and $\mbox{Pi}$ have been selected to be as $\mbox{Ex}=0.16$ and $\mbox{Pi}=8.33$, similar to the values used by previous studies in Refs.~\cite{soligo2019coalescence,engblom2013diffuse} with $M_{\psi}=0.01$ and were kept fixed in all the simulations. During the simulations, we determine the migration velocity of the bubble ($\overline{U}$) by calculating the average velocity inside the bubble. The reference time is specified by the scale $T_o = R / U_o$ where $R$ is the radius of the bubble ($R = D/2$), and therefore, the dimensionless time used is $t^{*} = t / T_o$. Here, the characteristic thermocapillary velocity $U_o$ can be obtained from balancing the surface tension force for the clean interface with the viscous force, which reads
\begin{equation}
U_o= \left(\frac{\sigma_0}{\mu_a}\right)\left(\frac{R}{L}\right)\left(M_1+M_2\right). \label{Uo_eq}
\end{equation}

\subsection{Effect of Gibbs elasticity parameter $\beta$ for normal fluids at fixed $M_1$}
\begin{figure}[H]
\centering
\begin{subfigure}{0.475\textwidth}
\includegraphics[trim = 0 0 0 0,clip, width = 72mm]{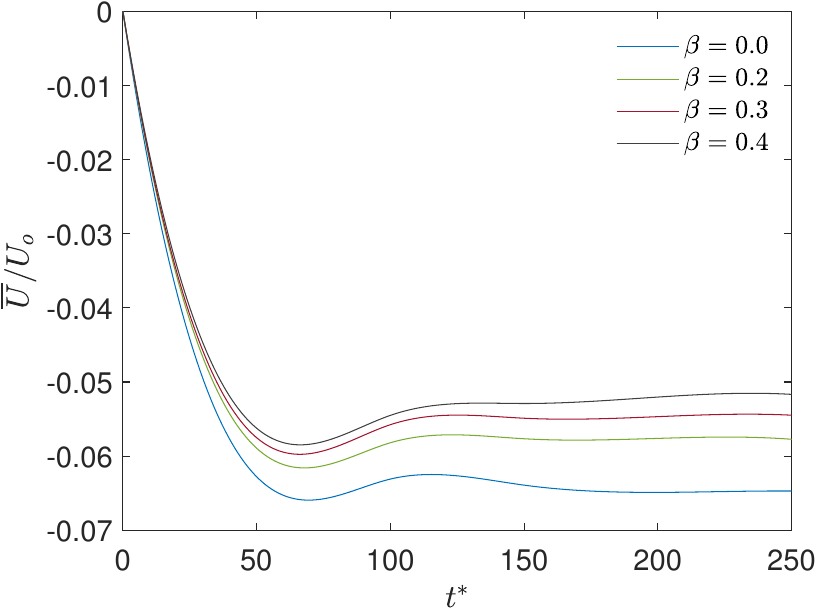}
\caption{Bubble Velocity}
\label{NF_1_a}
\end{subfigure}
\begin{subfigure}{0.475\textwidth}
\includegraphics[trim = 0 0 0 0,clip, width = 72mm]{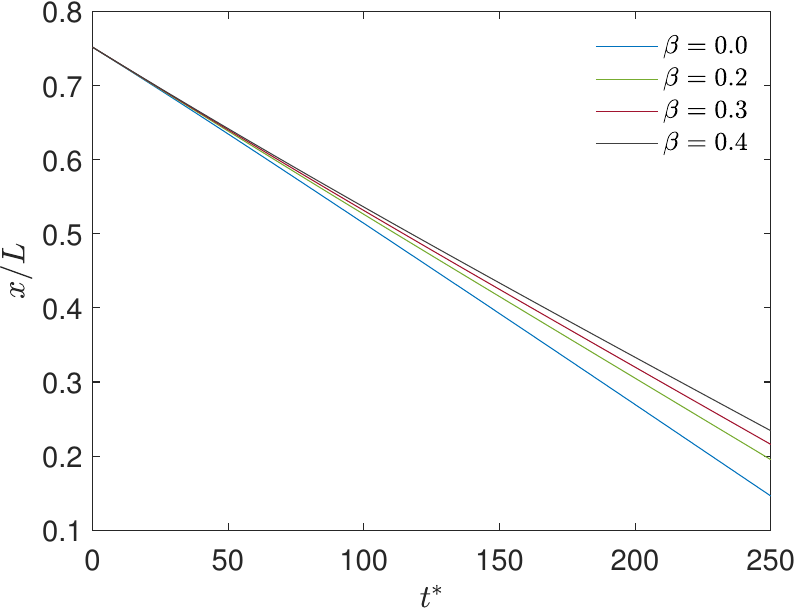}
\caption{Bubble Centroid Location}
\label{NF_1_b}
\end{subfigure}
\caption{Bubble velocity$(a)$ and its location $(b)$ at different values of the Gibbs elasticity parameter $\beta$ for surfactant-laden bubble in a normal fluid (NF). In this figure, $\mbox{Re} = 5$, $\mbox{Ma} = 0.25$, $M_1 = -2.5$, $M_2 = 0$, $D/H = 0.1$, $\psi_L = 0.1$, and $\psi_R = 0.4$.}
\label{NF_1}
\end{figure}

\begin{figure}[H]
\centering
\includegraphics[trim = 0 30 0 0,clip, width = 100mm]{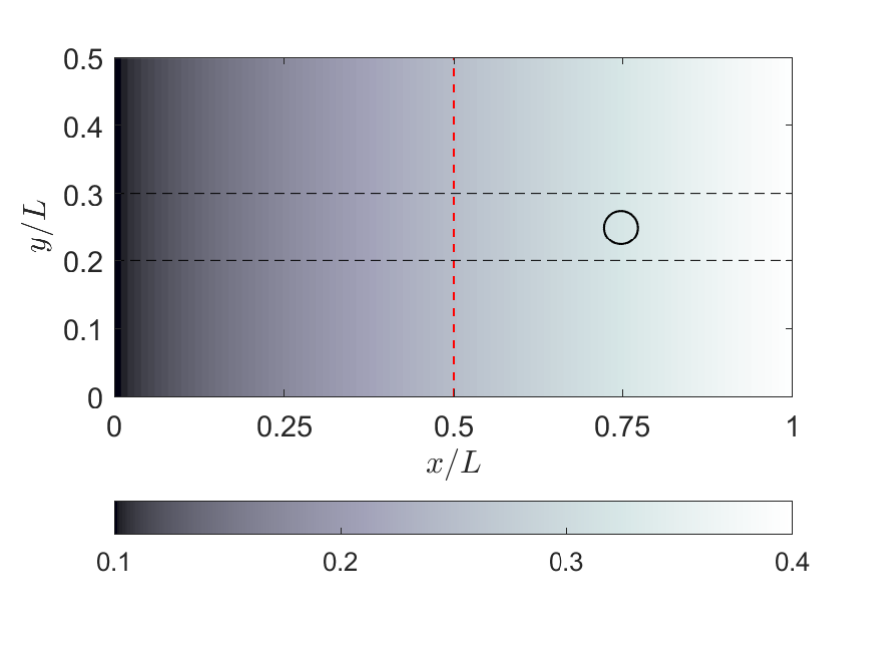}
\caption{This figure is used to indicate the horizontal slice of domain, using the black dashed lines, as seen in the time sequence snapshots in Fig.~\ref{NF_1A_8A} below. The color represents the magnitude of the normalized surfactant concentration field for the baseline case of parameters presented in Fig.~\ref{NF_1}. Similar cases are shown without color that do not have surfactant, but have the same slice of data shown.}
\label{slicefig_NF}
\end{figure}

\begin{figure}[H]
\centering
\includegraphics[trim = 0 0 0 0,clip, width = 150mm]{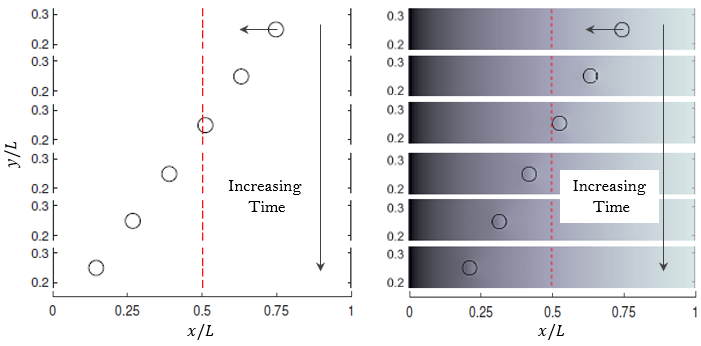}
\caption{Time sequence snapshots of bubble migration in a normal fluid with and without surfactant. The time stamps from top to bottom are $t^{*}=[0, 50, 100, 150, 200, 250]$ for $\beta=0$ (left) and $\beta=0.3$ (right), and the slice of the domain for each snapshot is indicated in Fig.~\ref{slicefig_NF}. Some relevant parameters for this problem are as follows: $\mbox{Re} = 5$, $\mbox{Ma} = 0.25$, $M_1 = -2.5$, $M_2 = 0$, $D/H = 0.1$, $\psi_L = 0.1$, and $\psi_R = 0.4$.}
\label{NF_1A_8A}
\end{figure}

To begin with, in order to establish a reference point for comparison, we will initially present results for the velocity and position of a bubble in a specific scenario involving NFs, as depicted in Fig.~\ref{NF_1}. This particular case considers $M_1 = -2.5$ and $M_2 = 0.0$. Furthermore, the selection of other fluid characteristics is such that the Gibbs elasticity parameter is varied in the range $\beta = 0, 0.2, 0.3, \text{and}\ 0.4$, while the concentration of surfactant on the left and right sides are $\psi_L = 0.1$ and $\psi_R = 0.4$, respectively. As we can see in Fig.~\ref{NF_1} for the clean interface case, i.e., without surfactants or $\beta = 0$, the bubble moves, after an initial transient phase, at a constant velocity completely to the end of the domain in one direction and adding surfactants via varying $\beta$ such that $\beta = 0.2, 0.3, \text{and}\ 0.4$ causes the rate of motion of the bubble to decrease. Figure~\ref{NF_1A_8A} shows the contours of the bubble at selected times for the case when $\beta = 0$ and $\beta = 0.3$, where, for brevity in presentation, the snapshots are taken by only focusing on the relevant section of the bubble path by slicing horizontally from the full domain presented in Fig.~\ref{slicefig_NF}. The motion of the bubble from right to left in the surfactant-laden case ($\beta \neq 0$) leads to a nonuniform distribution of the surfactant concentration field (see the contours shown in Figure~\ref{NF_1A_8A}), which in turn generates a tangential surface tension gradient or the Marangoni stress that opposes the motion of the bubble. Thus, the greater the sensitivity of surface tension on the surfactant field or $\beta$, the longer it takes for the moving bubble to cover a given distance (see Fig.~\ref{NF_1_b}), since it generates progressively greater tangential Marangoni stresses due to surfactant adsorption on its interface that retard the rate of bubble motion which manifests as a decrease in $|\bar{U}|$ as $\beta$ is increased (see Fig.~\ref{NF_1_a}). In effect, the bubble migrating in a normal fluid does not settle to an equilibrium position and monotonically and continuously moves from right to left and it attains a finite and non-zero terminal velocity $\bar{U}\neq 0$ and having a negative sign ($\bar{U} < 0$) indicating the motion is directed opposite to the positive direction of the coordinate system.

\begin{figure}[H]
\centering
\begin{subfigure}{0.475\textwidth}
\includegraphics[trim = 0 0 0 0,clip, width = 72mm]{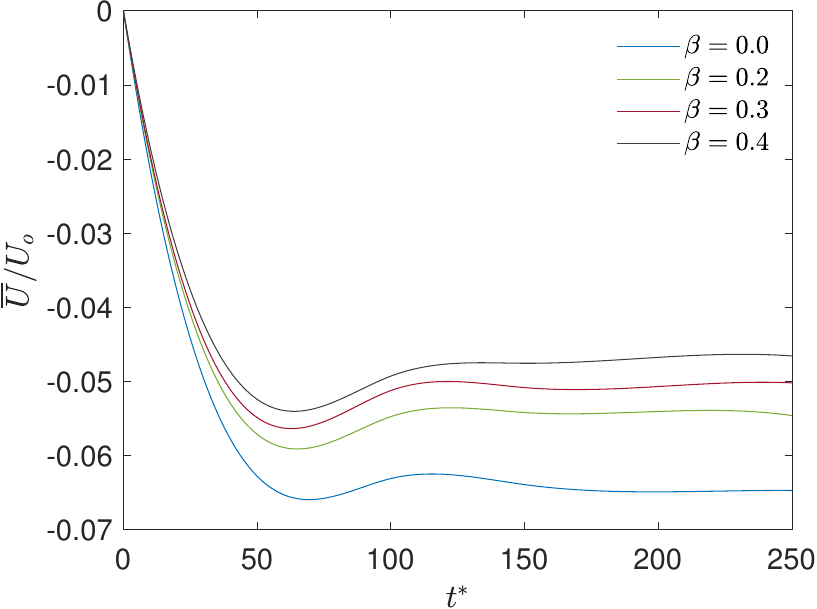}
\caption{Bubble Velocity}
\end{subfigure}
\begin{subfigure}{0.475\textwidth}
\includegraphics[trim = 0 0 0 0,clip, width = 72mm]{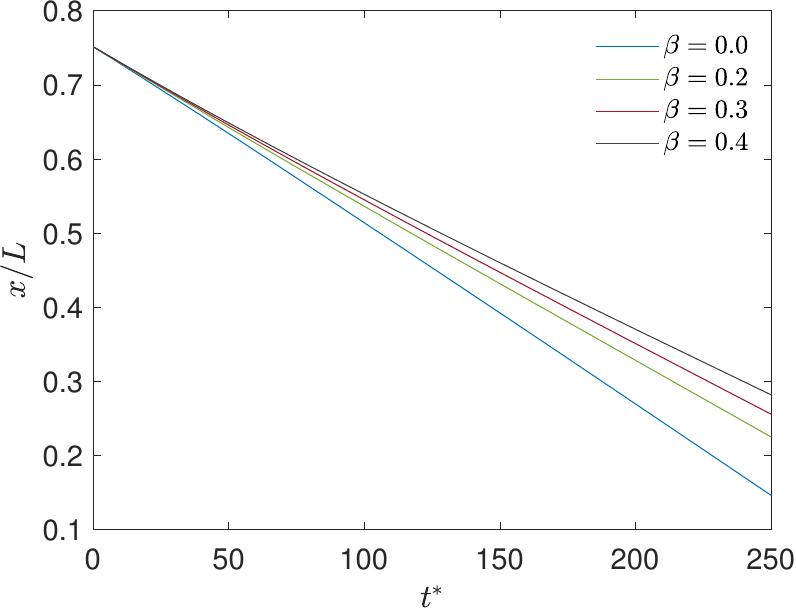}
\caption{Bubble Centroid Location}
\end{subfigure}
\caption{Bubble velocity$(a)$ and its location $(b)$ at different values of the Gibbs elasticity parameter $\beta$ for surfactant-laden bubble in a normal fluid (NF). In this figure, $\mbox{Re} = 5$, $\mbox{Ma} = 0.25$, $M_1 = -2.5$, $M_2 = 0$, $D/H = 0.1$, $\psi_L = 0.1$, and $\psi_R = 0.6$. Effective magnitude of the streamwise gradient of the surfactant concentration field $|d\psi/dx|$ is higher here when compared to Fig.~\ref{NF_1}.}
\label{NF_2}
\end{figure}

\begin{figure}[H]
\centering
\begin{subfigure}{0.475\textwidth}
\includegraphics[trim = 0 0 0 0,clip, width = 72mm]{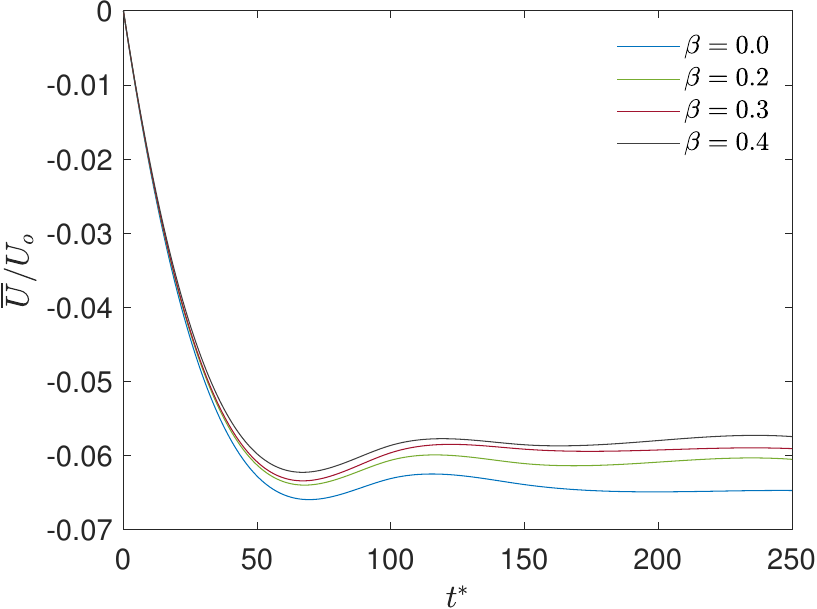}
\caption{Bubble Velocity}
\end{subfigure}
\begin{subfigure}{0.475\textwidth}
\includegraphics[trim = 0 0 0 0,clip, width = 72mm]{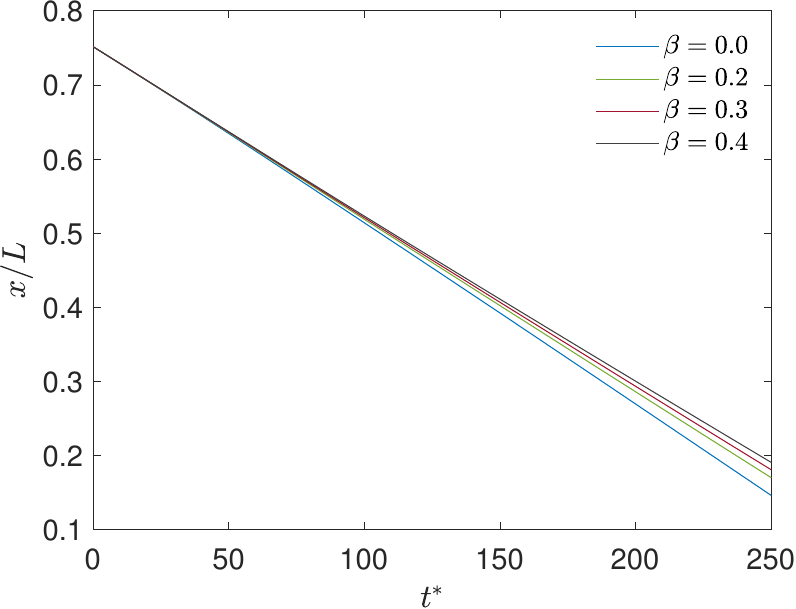}
\caption{Bubble Centroid Location}
\end{subfigure}
\caption{Bubble velocity$(a)$ and its location $(b)$ at different values of the Gibbs elasticity parameter $\beta$ for surfactant-laden bubble in a normal fluid (NF). In this figure, $\mbox{Re} = 5$, $\mbox{Ma} = 0.25$, $M_1 = -2.5$, $M_2 = 0$, $D/H = 0.1$, $\psi_L = 0.1$, and $\psi_R = 0.2$. Effective magnitude of the streamwise gradient of the surfactant concentration field $|d\psi/dx|$ is lower here when compared to Fig.~\ref{NF_1}.}
\label{NF_3}
\end{figure}

In contrast to Fig.~\ref{NF_1} where the surfactant concentration at the right and the left ends are $\psi_R = 0.4$ and $\psi_L = 0.1$, we now increase the concentration at the right end to $\psi_R = 0.6$ while keeping $\psi_L = 0.1$, as before, in Fig.~\ref{NF_2}. Effectively, this increases the streamwise gradient magnitude of the surfactant concentration $|d\psi/dx|$. As a result, comparing Figs.~\ref{NF_1} and~\ref{NF_2} for the same $\beta$, it can be seen that the magnitude of the bubble velocity decreases relative to the clean interface case ($\beta = 0$) further when $\psi_R = 0.6$ when compared to $\psi_R = 0.4$. By contrast, when the right end of the domain is loaded to $\psi_R = 0.2$, such that $|d\psi/dx|$ is lower compared to the previous two cases, the bubble travels faster for $\beta \neq 0$ (see Fig.~\ref{NF_3}). We emphasize that these observations are applicable for surfactant-laden thermocapillary bubble migration in normal fluid only.

 \subsection{Effect of Gibbs elasticity parameter $\beta$ for self-rewetting fluids at larger $M_2$}
On the other hand, when we turn off the linear coefficient of surface tension (i.e., $\sigma_T=0$ or $M_1=0$) and only retain a non-zero quadratic coefficient, i.e., $\sigma_{TT}\neq 0$, we conduct simulations to explore the migration of surfactant-laden thermocapillary bubbles in self-rewetting fluids (SRFs), while keeping the same values for the property ratios and other parameters as before. Specifically, in dimensionless form, we set $M_1=0$ and $M_2=5$ as the selected values for the SRFs' baseline case to show the effect of the Gibbs elasticity parameter, $\beta$. We impose a surfactant concentration gradient by setting $\psi_L = 0.4$ and $\psi_R = 0.1$.

\begin{figure}[H]
\centering
\begin{subfigure}{0.475\textwidth}
\includegraphics[trim = 0 0 0 0,clip, width = 72mm]{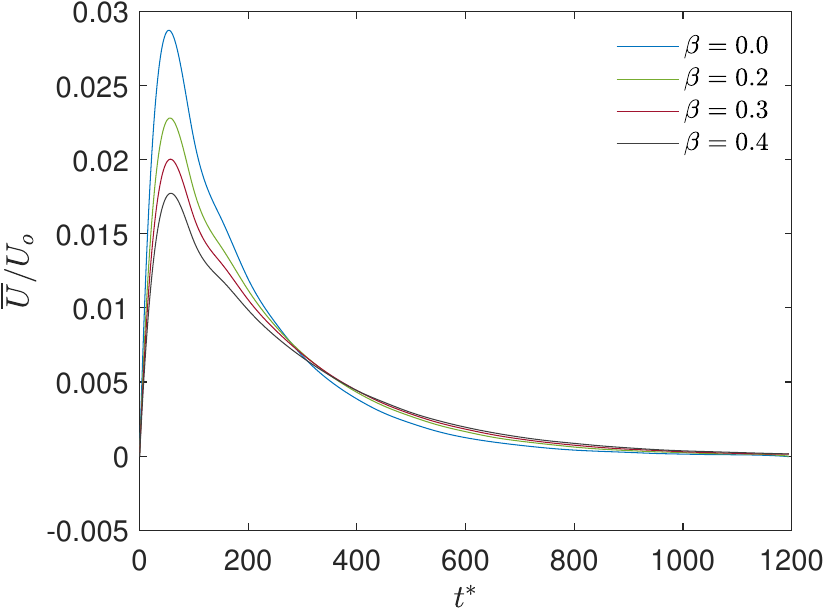}
\caption{Bubble Velocity}
\label{SRF_1_a}
\end{subfigure}
\begin{subfigure}{0.475\textwidth}
\includegraphics[trim = 0 0 0 0,clip, width = 72mm]{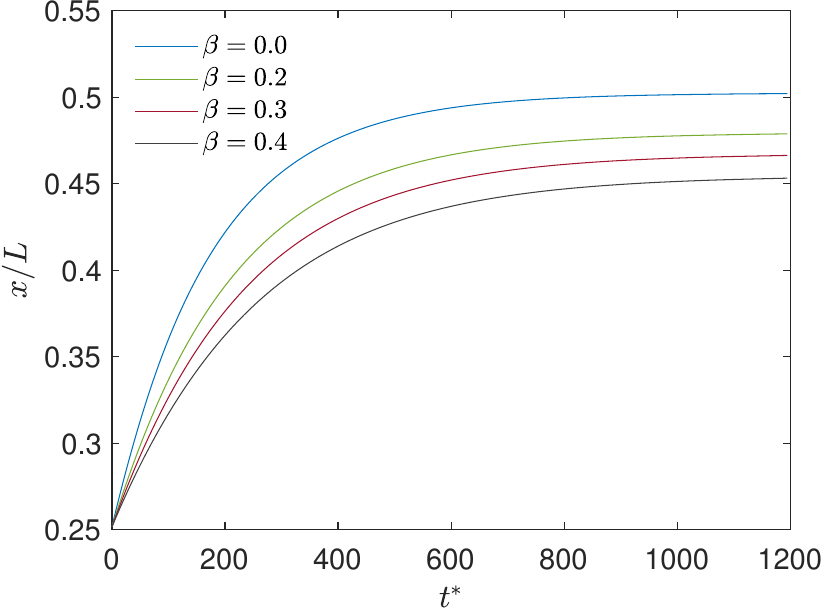}
\caption{Bubble Centroid Location}
\label{SRF_1_b}
\end{subfigure}
\caption{Bubble velocity$(a)$ and its location $(b)$ at different values of the Gibbs elasticity parameter $\beta$ for surfactant-laden bubble in a self-rewetting fluid (SRF). In this figure, $\mbox{Re} = 5$, $\mbox{Ma} = 0.25$, $M_1 = 0$, $M_2 = 5$, $D/H = 0.1$, $\psi_L = 0.4$, and $\psi_R = 0.1$.}
\label{SRF_1}
\end{figure}

The computed results, in terms of the bubble's velocity and position for different $\beta$, are depicted in Fig.~\ref{SRF_1}. First, for $\beta=0$ (i.e., with no surfactant sensitivity), the bubble initially speeds up and then slows down and continues to slow down until it reaches \emph{zero} velocity ($\bar{U}=0$) attaining an equilibrium position at the reference temperature location. This behavior is dramatically different from the normal fluid case, where, as seen earlier, there was no equilibrium position of the bubble which executes a continuous motion from the right end to the left end of the domain, i.e., from the cold to the hot sides; by contrast, the bubble in the self-rewetting fluid moves from left to right or from the hot to the cold sides eventually attaining a stationary position with $\bar{U}=0$. Moreover, observe that the presence of surfactant (i.e., $\beta \neq 0$) causes the bubble velocity to decrease for this case, since the initial bubble location $(x/L)_{initial}=0.25$ is in the higher concentration part of the surfactant gradient. During the transient phases, the greater the Gibbs elasticity parameter $\beta$, the larger is the reduction in the bubble velocity, but in all cases at long times they are come to rest, i.e., attain zero velocity (see Fig.~\ref{SRF_1_a}). Interestingly, the combined effects of the Marangoni stresses arising from the tangential gradients in the surface tension due to non uniform temperature field and the surfactant concentration field (see Eq.~(\ref{ST_SRF})), causes the bubble to attain an equilibrium position at long times that is sensitive to the presence of surfactant field and the degree of its propensity to get adsorbed in the interface. In particular, as seen in Fig.~\ref{SRF_1_b}, only when there is no surfactant effect (i.e., the clean interface case) with $\beta=0$, the equilibrium bubble position $x_{eq}$ is the same as the reference temperature location at $x= 0.5L$. In all other cases, as the surfactant's sensitivity or $\beta$ increases the equilibrium position of the bubble $x_{eq}$ is located upstream, i.e., $x_{eq}/L<0.5$; in other words, the greater the Gibbs elasticity parameter $\beta$, the further the bubble's equilibrium location is pushed upstream or towards regions with higher surfactant concentration. From a physical point of view, when the thermally-induced Marangoni stresses for SRFs (which depends on $M_2$) is equal to the counteracting surfactant-induced Marangoni stresses (which depends on $\beta$), the bubble stops moving and thereby attaining an equilibrium position that is dependent on $\beta$ for a fixed $M_2$.

\begin{figure}[H]
\centering
\begin{subfigure}{0.475\textwidth}
\includegraphics[trim = 0 0 0 0,clip, width = 72mm]{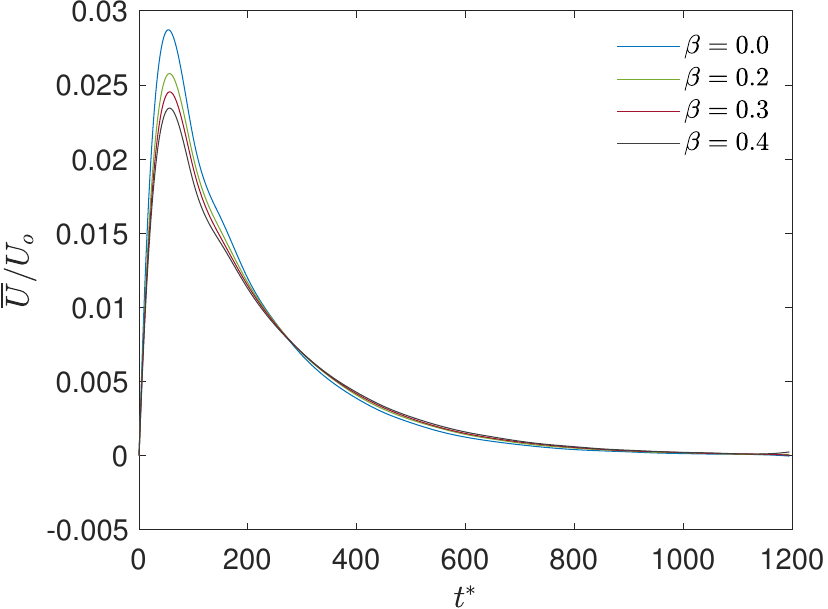}
\caption{Bubble Velocity}
\end{subfigure}
\begin{subfigure}{0.475\textwidth}
\includegraphics[trim = 0 0 0 0,clip, width = 72mm]{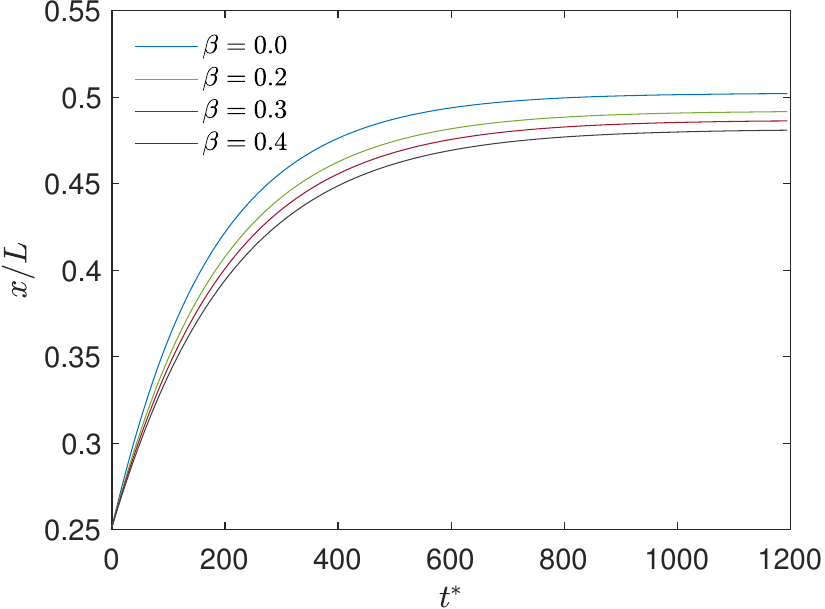}
\caption{Bubble Centroid Location}
\end{subfigure}
\caption{Bubble velocity$(a)$ and its location $(b)$ at different values of the Gibbs elasticity parameter $\beta$ for surfactant-laden bubble in a self-rewetting fluid (SRF). In this figure, $\mbox{Re} = 5$, $\mbox{Ma} = 0.25$, $M_1 = 0$, $M_2 = 5$, $D/H = 0.1$, $\psi_L = 0.2$, and $\psi_R = 0.1$. Effective magnitude of the streamwise gradient of the surfactant concentration field $|d\psi/dx|$ is lower here when compared to Fig.~\ref{SRF_1}.}
\label{SRF_4}
\end{figure}

\begin{figure}[H]
\centering
\begin{subfigure}{0.475\textwidth}
\includegraphics[trim = 0 0 0 0,clip, width = 72mm]{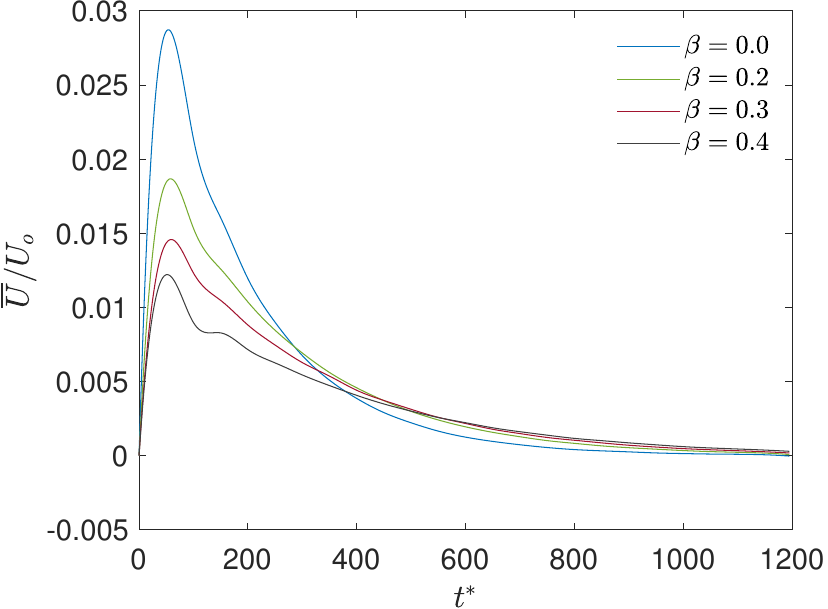}
\caption{Bubble Velocity}
\end{subfigure}
\begin{subfigure}{0.475\textwidth}
\includegraphics[trim = 0 0 0 0,clip, width = 72mm]{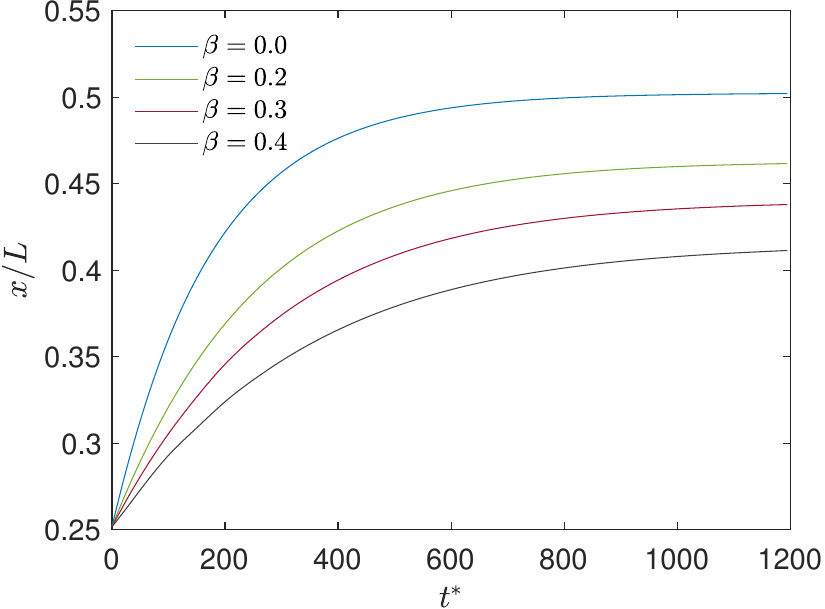}
\caption{Bubble Centroid Location}
\label{SRF_4a_b}
\end{subfigure}
\caption{Bubble velocity$(a)$ and its location $(b)$ at different values of the Gibbs elasticity parameter $\beta$ for surfactant-laden bubble in a self-rewetting fluid (SRF). In this figure, $\mbox{Re} = 5$, $\mbox{Ma} = 0.25$, $M_1 = 0$, $M_2 = 5$, $D/H = 0.1$, $\psi_L = 0.6$, and $\psi_R = 0.1$. Effective magnitude of the streamwise gradient of the surfactant concentration field $|d\psi/dx|$ is higher here when compared to Fig.~\ref{SRF_1}.}
\label{SRF_4a}
\end{figure}

Figure~\ref{SRF_1} shows these effects with the baseline case with $\psi_L = 0.4$ and $\psi_R = 0.1$ for setting up the surfactant concentration gradient along the streamwise direction. We now explore the role of changing this streamwise gradient $|d\psi/dx|$ by decreasing $\psi_L = 0.2$ while keeping $\psi_R = 0.1$ (see Fig.~\ref{SRF_4}) and the opposite case with higher $\psi_L = 0.6$ for the same $\psi_R$ (see Fig.~\ref{SRF_4a}). Clearly, when $|d\psi/dx|$ is increased as in Fig.~\ref{SRF_4a}, there is greater reduction in the bubble velocity for a specified $\beta$ when compared to the baseline case during the initial transients while the bubble halting completely at long times in each case; more importantly, comparing Figs.~\ref{SRF_1_b} and~\ref{SRF_4a_b}, it can be seen that for a specified $\beta$, the bubble equilibrates at a location further upstream for the later case with greater $|d\psi/dx|$. Thus, $(x_{eq}/L)_{G_{s,higher}} < (x_{eq}/L)_{G_{s,lower}}$, where $G_s=|d\psi|/dx$ is the magnitude of the imposed streamwise gradient of the surfactant concentration field. For example, when $\beta=0.4$, the bubble's equilibrium position ($x_{eq}/L$) is shifted from about $0.45$ to about $0.4$ when $|d\psi/dx|$ is increased via changing $\psi_L$ from $0.4$ to $0.6$. The contour plots for the bubble at selected times for the clean interface case ($\beta=0$) and the surfactant-laden case ($\beta=0.3$) presented in Fig.~\ref{SRF_1B1_1B8} corroborates these findings.


\begin{figure}[H]
\centering
\includegraphics[trim = 0 30 0 0,clip, width = 100mm]{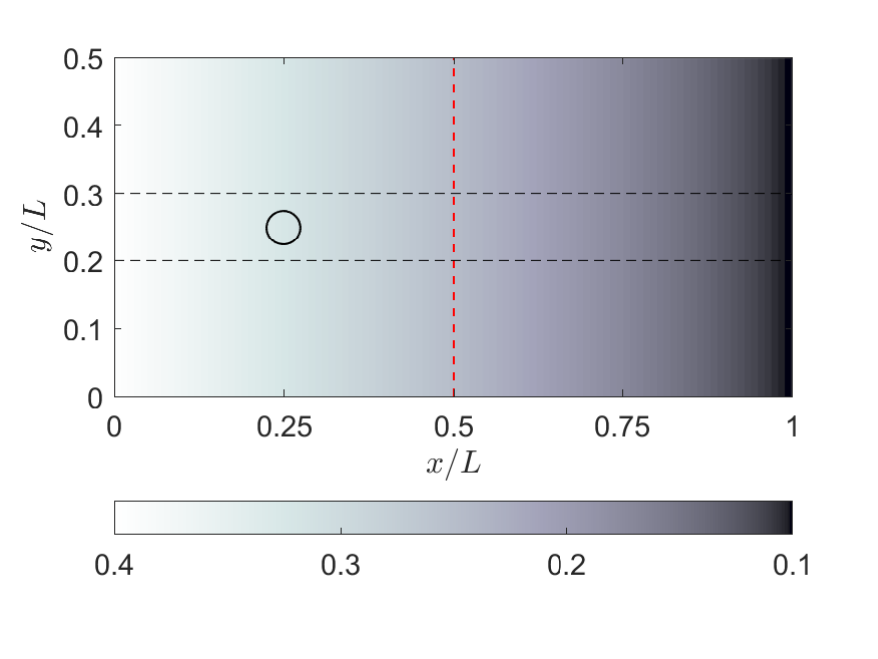}
\caption{This figure is used to indicate the horizontal slice of domain, using the black dashed lines, as seen in the time sequence snapshots in Figs.~\ref{SRF_1B1_1B8}, \ref{SRF_1a1_1a8}, and~\ref{SRF_countors_D_H_02} below. The color represents the magnitude of the normalized surfactant concentration field for the baseline case of parameters presented in Fig.~\ref{SRF_1}. Similar cases are shown without color that do not have surfactant, but have the same slice of data shown.}
\label{slicefig_SRF1}
\end{figure}

\begin{figure}[H]
\centering
\includegraphics[trim = 0 0 0 0,clip, width = 150mm]{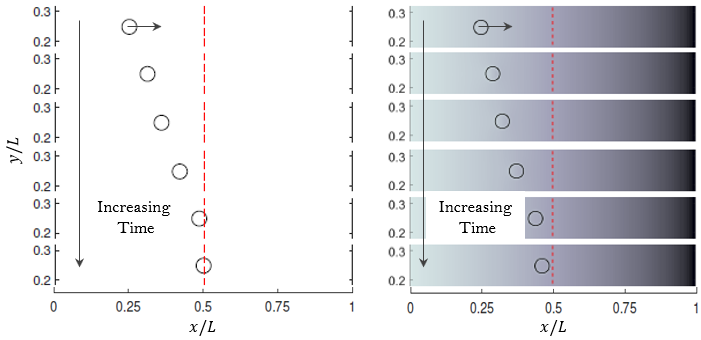}
\caption{Time sequence snapshots of bubble migration in a self-rewetting fluid with and without surfactant. The time stamps from top to bottom are $t^{*}=[0, 50, 100, 200, 500, 1200]$ for $\beta=0$ (left) and $\beta=0.3$ (right), and the slice of the domain for each snapshot is indicated in Fig.~\ref{slicefig_SRF1}. Some relevant parameters for this problem are as follows: $\mbox{Re} = 5$, $\mbox{Ma} = 0.25$, $M_1 = 0$, $M_2 = 5$, $D/H = 0.1$, $\psi_L = 0.4$, and $\psi_R = 0.1$.}
\label{SRF_1B1_1B8}
\end{figure}

\subsection{Effect of Gibbs elasticity parameter $\beta$ for self-rewetting fluids at smaller $M_2$}
\begin{figure}[H]
\centering
\begin{subfigure}{0.475\textwidth}
\includegraphics[trim = 0 0 0 0,clip, width = 72mm]{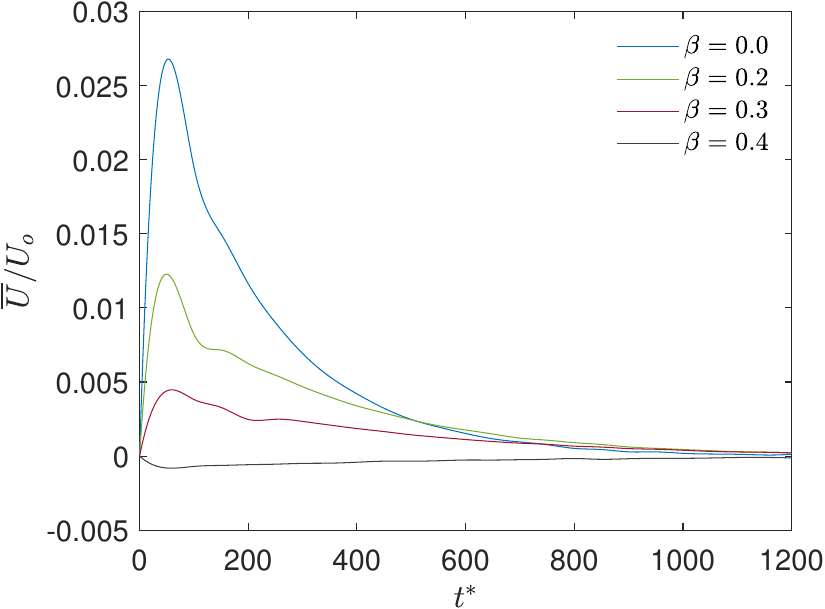}
\caption{Bubble Velocity}
\end{subfigure}
\begin{subfigure}{0.475\textwidth}
\includegraphics[trim = 0 0 0 0,clip, width = 72mm]{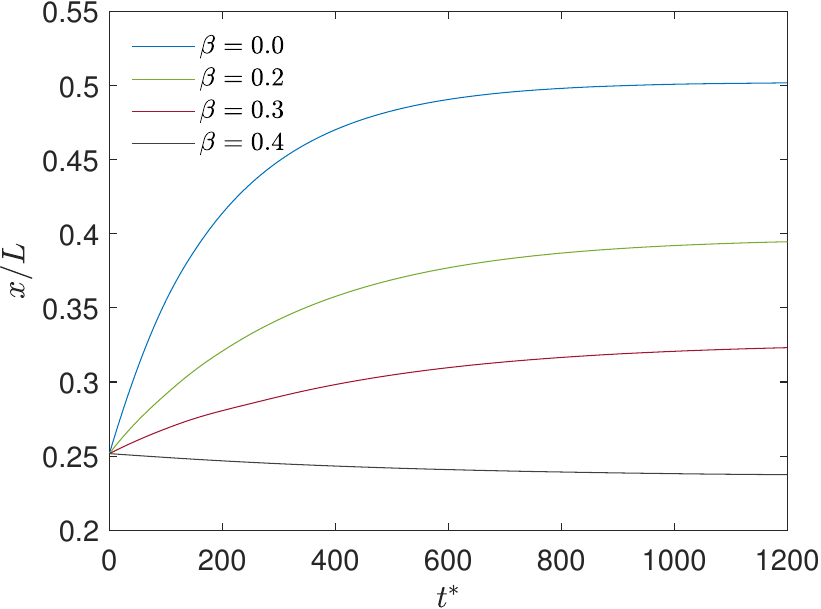}
\caption{Bubble Centroid Location}
\end{subfigure}
\caption{Bubble velocity $(a)$ and its location $(b)$ at different values of the Gibbs elasticity parameter $\beta$ for surfactant-laden bubble in a self-rewetting fluid (SRF). In this figure, $\mbox{Re} = 5$, $\mbox{Ma} = 0.25$, $M_1 = 0$, $M_2 = 1.25$, $D/H = 0.1$, $\psi_L = 0.4$, and $\psi_R = 0.1$.}
\label{SRF_smallerM2}
\end{figure}

Moving forward, we now investigate the implications of reducing the dimensionless surface tension quadratic sensitivity coefficient from $M_2 = 5.0$  for the baseline case discussed above to $M_2 = 1.25$ and its impact on the average velocity of the bubble and its centroid location for different Gibbs elasticity parameter $\beta$. The results of this simulation case are presented in Fig.~\ref{SRF_smallerM2}.
As in the higher $M_2$ case, here the bubble in the SRF moves from the left to the right sides while experiencing a
deceleration during the initial transients as $\beta$ is increased since it is launched from an initial position $(x/L)_{initial} = 0.25$ which is in the higher concentration region of the surfactant field. However, the results reveal that as $M_2$ decreases, the magnitude of the thermocapillary effects is reduced which causes the bubble to reduce its speed and, as a result, the impact of surfactants becomes even more prominent compared to the baseline case considered in the previous section. Specifically, when the bubble moves at a slower rate due to lower $M_2$, the thermally-induced contribution to the Marangoni stress is smaller and so the relative contribution of this tangential stress from surfactant's interfacial adsorption for different and higher Gibbs elasticity parameter leads to progressively larger effects.

\begin{figure}[H]
\centering
\begin{subfigure}{0.475\textwidth}
\includegraphics[trim = 0 0 0 0,clip, width = 72mm]{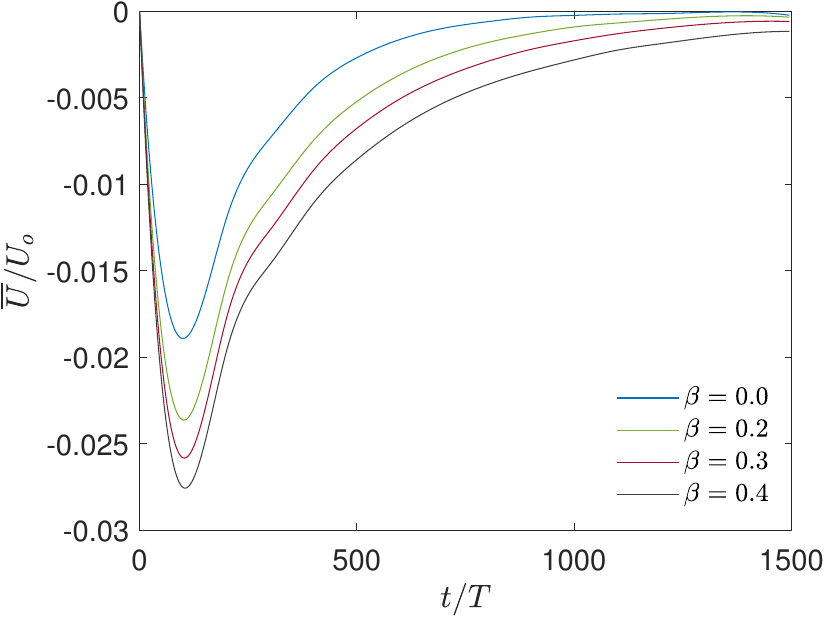}
\caption{Bubble Velocity}
\end{subfigure}
\begin{subfigure}{0.475\textwidth}
\includegraphics[trim = 0 0 0 0,clip, width = 72mm]{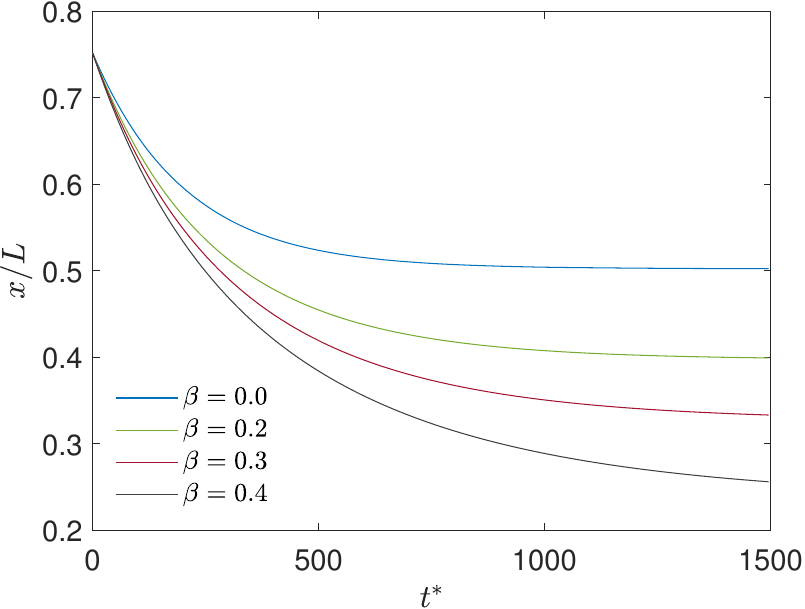}
\caption{Bubble Centroid Location}
\end{subfigure}
\caption{Bubble velocity $(a)$ and its location $(b)$ at different values of the Gibbs elasticity parameter $\beta$ for surfactant-laden bubble in a self-rewetting fluid (SRF). In this figure, $\mbox{Re} = 5$, $\mbox{Ma} = 0.25$, $M_1 = 0$, $M_2 = 1.25$, $D/H = 0.1$, $\psi_L = 0.4$, and $\psi_R = 0.1$. Here, the initial location of the bubble is at $(x/L)_{initial}=0.75$ compared that in
Fig.~\ref{SRF_smallerM2} where the bubble is launched from the initial location $(x/L)_{initial}=0.25$.}
\label{SRF_smallerM2_differentinitlocation}
\end{figure}

It would be interesting to see the bubble dynamics in SRF when it is launched from a different initial position compared to the above, i.e., from $(x/L)_{initial} = 0.75$ just like in the NF case discussed earlier. Figure~\ref{SRF_smallerM2_differentinitlocation} shows the results of the bubble velocity and its centroid location as a function of time.  Since the minimum temperature occurs at the center of the domain corresponding to the low surface tension region, the bubble seeks to move towards it. Thus, in this case, it moves from right to left. Moreover, since here the initial location of the bubble is in a lower surfactant concentration region, during the initial transients it accelerates unlike that in the case discussed above (see Fig.~\ref{SRF_smallerM2}). However, in both cases, unlike in the NF case, the bubble equilibrates, and its final position is determined by the balance of the Marangoni stresses induced by thermocapillary and surfactant effects. \emph{The previous work on bubble migration in SRF in the absence of surfactants~\cite{tripathi2015non,balla2019non,majidi2020single,mitchell2021computational} showed it always attains a final equilibrium position of $x_{eq}/L = 0.5$, where the temperature is minimum}. On the other hand, this study reveals a potentially new approach to modulate the bubble equilibrium position in SRF by adding surfactants in such a way that the Marangoni effects due to thermocapillary effects are counteracted by the solutal effects. In particular, \emph{we obtain the general result that regardless of how the bubble is initially launched, after initial transients, the bubble in SRF attains equilibrium at a location different from $x_{eq}/L = 0.5$ (corresponding to the clean interface case) that can be controlled by the Gibbs elasticity parameter $\beta$}. This can be seen by comparing Fig.~\ref{SRF_smallerM2}(b) and~\ref{SRF_smallerM2_differentinitlocation}(b).

\begin{figure}[H]
\centering
\includegraphics[trim = 0 0 0 0,clip, width = 150mm]{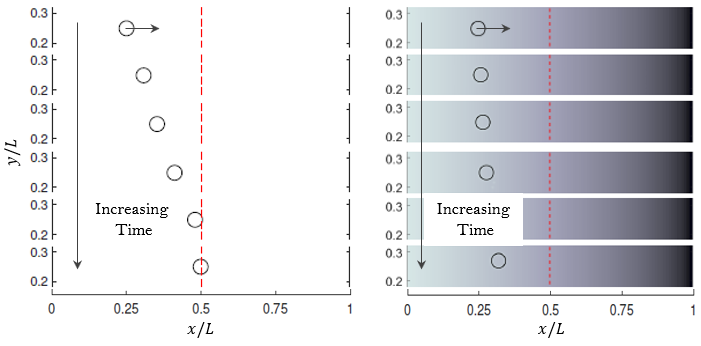}
\caption{Time sequence snapshots of bubble migration in a self-rewetting fluid with and without surfactant. The time stamps from top to bottom are $t^{*}=[0, 50, 100, 200, 500, 1200]$ for $\beta=0$ (left) and $\beta=0.3$ (right), and the slice of the domain for each snapshot is indicated in Fig.~\ref{slicefig_SRF1}. Some relevant parameters for this problem are as follows: $\mbox{Re} = 5$, $\mbox{Ma} = 0.25$, $M_1 = 0$, $M_2 = 5$, $D/H = 0.2$, $\psi_L = 0.4$, and $\psi_R = 0.1$.}
\label{SRF_1a1_1a8}
\end{figure}

These observations for the bubble dynamics in SRF in the presence of surfactants at $M_2 = 1.25$ for different $\beta$ at $\psi_L = 0.4$ and $\psi_R = 0.1$ can be clearly seen in Fig.~\ref{SRF_smallerM2} by contrasting it with Fig.~\ref{SRF_1}. In particular, the equilibrium position of the bubble with $M_2 = 1.25$ is shifted further towards the hotter, upstream side under the weakened thermally-induced Marangoni stress to a position where it balances that from the surfactant-generated Marangoni stress. For example, the bubble equilibrium position is further shifted to $x_{eq}/L=0.24$ for $\beta=0.4$ at $M_2 = 1.25$ when compared to $x_{eq}/L=0.45$ for the same $\beta$ at $M_2 = 5.0$ in Fig.~\ref{SRF_1}, which corresponds to a significant relative change in the equilibrium position by about $87.5\%$. Thus, the inclusion of surfactants provides an ability to manipulate the equilibrium position of the bubble, contingent upon the chosen value of $\beta$ related to sensitivity of surface tension on surfactant concentration, which is a material property, and the dimensionless quadratic thermocapillary coefficient $M_2$. Figure~\ref{SRF_1a1_1a8} shows the contours of a bubble during its migration in SRF from left to its equilibrium location for different dimensionless times for $\beta=0$ (left) and $\beta=0.3$ (right) and for $M_2=1.25$. We observe a behavior similar to Fig.~\ref{SRF_1B1_1B8} but with a considerably stronger effect on modulating the bubble equilibrium location where the bubble always migrates from the hot side to the cold side, which is opposite in direction with using NF, towards the reference temperature location occurring at the center of the domain based on the setup of the problem.

\begin{figure}[H]
\centering
\begin{subfigure}{0.475\textwidth}
\includegraphics[trim = 0 0 0 0,clip, width = 72mm]{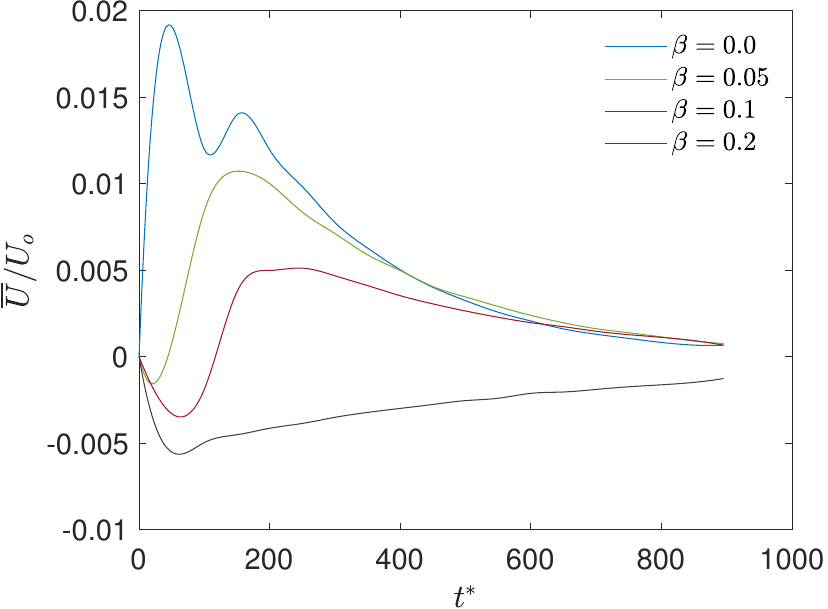}
\caption{Bubble Velocity}
\end{subfigure}
\begin{subfigure}{0.475\textwidth}
\includegraphics[trim = 0 0 0 0,clip, width = 72mm]{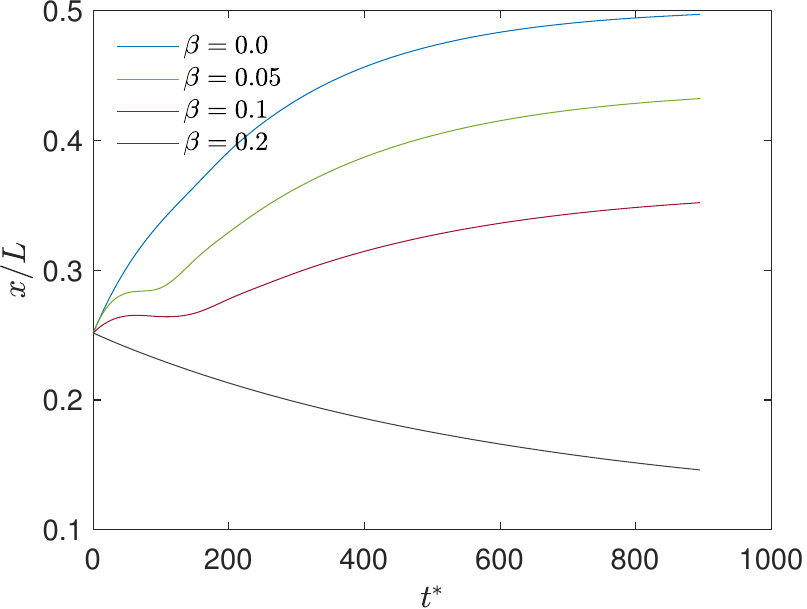}
\caption{Bubble Centroid Location}
\end{subfigure}
\caption{Bubble velocity $(a)$ and its location $(b)$ at different values of the Gibbs elasticity parameter $\beta$ for surfactant-laden bubble in a self-rewetting fluid (SRF). In this figure, $\mbox{Re} = 5$, $\mbox{Ma} = 0.25$, $M_1 = 0$, $M_2 = 0.5$, $D/H = 0.1$, $\psi_L = 0.4$, and $\psi_R = 0.1$. Here, the choice of $M_2=0.5$ represents a reduction by a factor of $2.5$ when compared to the baseline case shown in Fig.~\ref{SRF_smallerM2} with $M_2=1.25$.}
\label{SRF_smallerM2_evensmaller}
\end{figure}

In order to encompass a wider range of situations that accommodates greater role of the surfactant effects when compared to thermocapillarity in SRFs, we now reduce the quadratic sensitivity coefficient of surface tension $M_2$ even further by a factor of 2.5 when compared to the baseline case reported in Fig.~\ref{SRF_smallerM2}, which uses $M_2 = 1.25$. Thus, Fig.~\ref{SRF_smallerM2_evensmaller} shows the results of the bubble velocity and its centroid location in SRF at different values of the Gibbs elasticity parameter with $M_2 = 0.5$. It can be observed that the bubble again attains equilibrium at long times that is sensitive to the Gibbs elasticity parameter similar to the baseline case shown in Fig.~\ref{SRF_smallerM2} but with an enhanced effect due to the presence of surfactants that results in greater deviations from the equilibrium location at $x_{eq}/L=0.5$ with increase in $\beta$ when compared to the case with smaller $M_2$. As such, the general trends and findings are similar to the other cases considered earlier.

\begin{figure}[H]
\centering
\begin{subfigure}{0.475\textwidth}
\includegraphics[trim = 0 0 0 0,clip, width = 72mm]{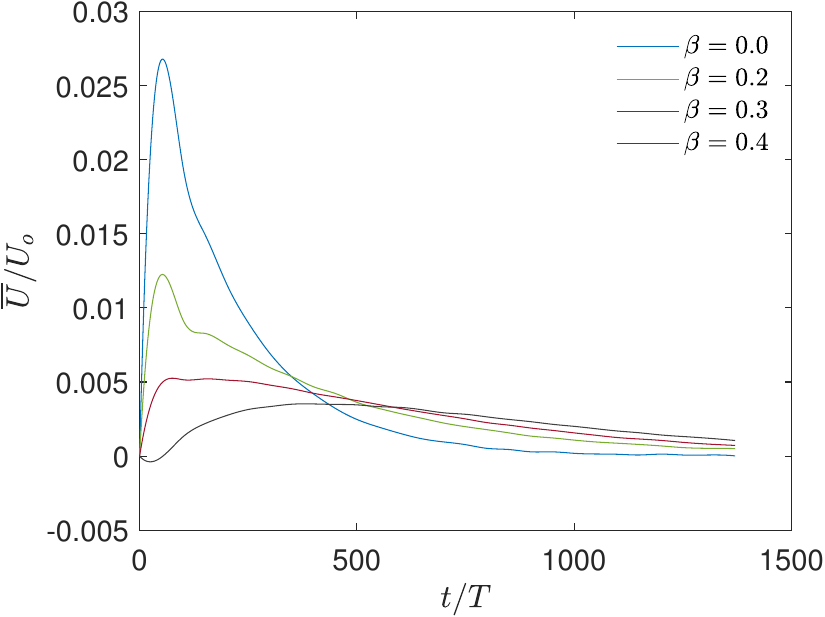}
\caption{Bubble Velocity}
\end{subfigure}
\begin{subfigure}{0.475\textwidth}
\includegraphics[trim = 0 0 0 0,clip, width = 72mm]{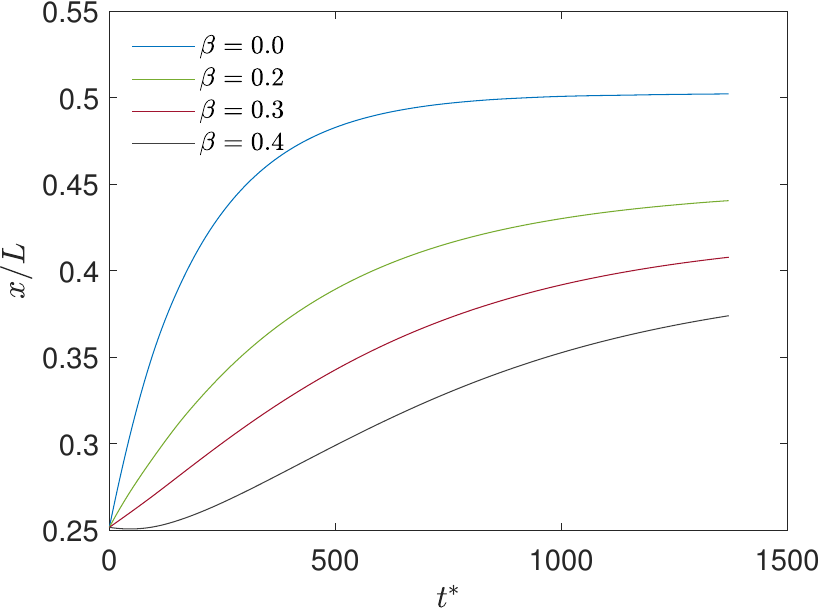}
\caption{Bubble Centroid Location}
\end{subfigure}
\caption{Bubble velocity $(a)$ and its location $(b)$ at different values of the Gibbs elasticity parameter $\beta$ for surfactant-laden bubble in a self-rewetting fluid (SRF). In this figure, $\mbox{Re} = 5$, $\mbox{Ma} = 0.25$, $M_1 = 0$, $M_2 = 1.25$, $D/H = 0.1$, $\psi_L = 0.4$, and $\psi_R = 0.1$. Here, $\psi_L$ and $\psi_R$ are set only initially and then allowed to vary through the Neumann condition compared that in Fig.~\ref{SRF_smallerM2} where $\psi_L$ and $\psi_R$ are set as boundary conditions at the respective boundaries.}
\label{SRF_smallerM2_differentBC}
\end{figure}

In the above cases, the gradient in surfactant concentration to achieve counteracting Marangoni stresses to thermocapillary effects is introduced by specifying the surfactant concentrations on the left and right sides of the domain to be at $\psi_L$ and $\psi_R$, respectively, following our previous work~\cite{premnath2018surfactant}, as representative of the presence of their sources at the respective ends of the domain. Nevertheless, we now investigate an alternative approach by specifying $\psi_L$ and $\psi_R$ as the initial condition at those locations and then allowing them to vary via the Neumann condition. The results with this approach for the baseline case with $M_2 = 1.25$ for the bubble velocity and its centroid location are shown in Fig.~\ref{SRF_smallerM2_differentBC}. Compared to the baseline case shown in Fig.~\ref{SRF_smallerM2}, the results are generally qualitatively similar with somewhat reduced effect of surfactants at higher $\beta$ in modulating the equilibrium location of the bubble. Overall, the main conclusions made earlier regarding the influence of surfactants on the bubble dynamics in SRFs are still valid.

\subsection{Effect of bubble size or diameter $D$}
Next, we explore the effect of bubble size on its dynamics when moving in a SRF for different surfactant sensitivities. In all the previous cases, we considered the bubble size $D$ to be such that $D/H = 0.1$. In this section, we double the bubble size by setting $D/H = 0.2$ and study its influence on its velocity and position. The fluid properties for this study are similar to those mentioned previously and are also shown in the figures captions. Smaller bubbles have a higher curvature, which can affect the distribution of surfactants on their surface. This can lead to more pronounced surface tension gradients, causing stronger Marangoni flows. On the other hand, larger bubbles have a smaller curvature and could result in a more uniform distribution of surfactants around the interface. The surface tension gradients are less pronounced compared to smaller bubbles, thereby resulting in weaker Marangoni-driven flows. The size of the bubble thus influences its migration velocity under Marangoni forces.

\begin{figure}[H]
\centering
\begin{subfigure}{0.475\textwidth}
\includegraphics[trim = 0 0 0 0,clip, width = 72mm]{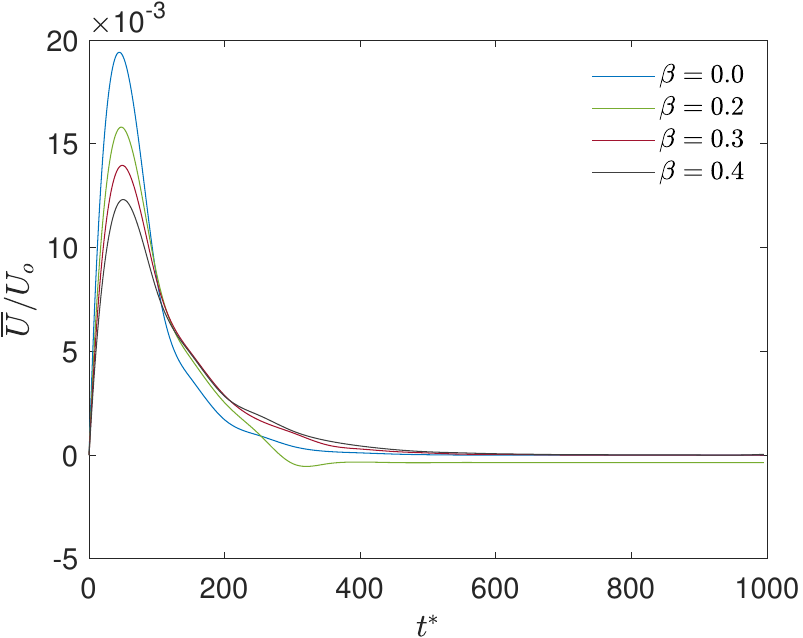}
\caption{Bubble Velocity}
\end{subfigure}
\begin{subfigure}{0.475\textwidth}
\includegraphics[trim = 0 0 0 0,clip, width = 72mm]{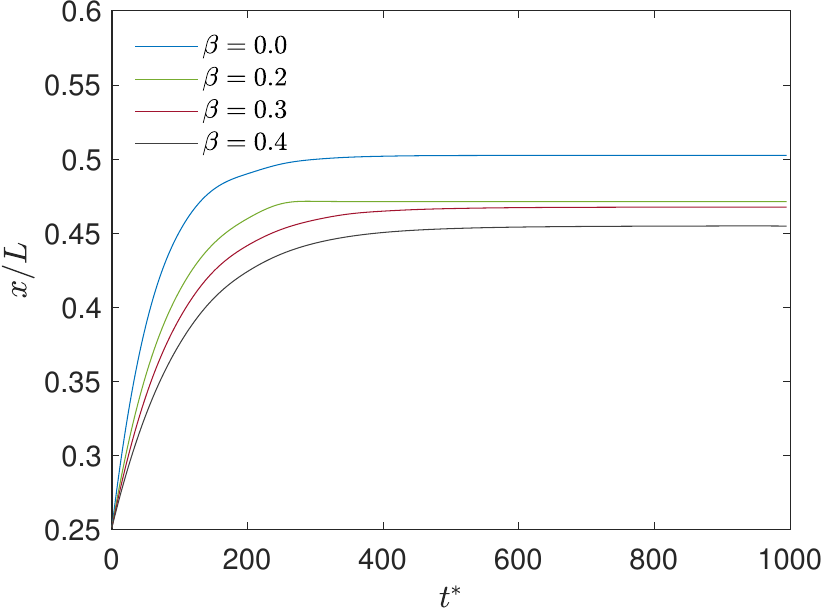}
\caption{Bubble Centroid Location}
\end{subfigure}
\caption{Bubble velocity$(a)$ and its location $(b)$ at different values of the Gibbs elasticity parameter $\beta$ for surfactant-laden bubble in a self-rewetting fluid (SRF). In this figure, $\mbox{Re} = 5$, $\mbox{Ma} = 0.25$, $M_1 = 0$, $M_2 = 5$, $D/H = 0.2$, $\psi_L = 0.4$, and $\psi_R = 0.1$.}
\label{SRF_D_H_02}
\end{figure}

\begin{figure}[H]
\centering
\includegraphics[trim = 0 0 0 0,clip, width = 150mm]{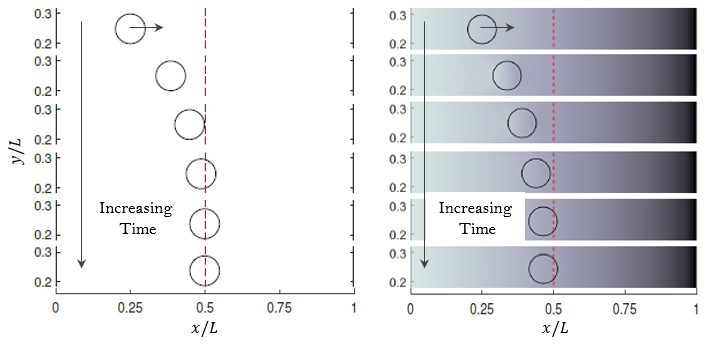}
\caption{Time sequence snapshots of bubble migration in a self-rewetting fluid with and without surfactant. The time stamps from top to bottom are $t^{*}=[0, 50, 100, 200, 500, 1000]$ for $\beta=0$ (left) and $\beta=0.3$ (right), and the slice of the domain for each snapshot is indicated in Fig.~\ref{slicefig_SRF1}. Some relevant parameters for this problem are as follows: $mbox{Re} = 5$, $\mbox{Ma} = 0.25$, $M_1 = 0$, $M_2 = 5$, $D/H = 0.2$, $\psi_L = 0.4$, and $\psi_R = 0.1$.}
\label{SRF_countors_D_H_02}
\end{figure}

Due to stronger Marangoni flows, smaller bubbles (see Fig.~\ref{SRF_1}) migrate faster and slow down more quickly compared to larger bubbles (see Fig.~\ref{SRF_D_H_02}), especially if the effect is primarily driven by surface tension gradients due to solely thermocapillary effects in the absence of surfactants in SRFs (i.e., $\beta=0$). In other words, increasing the bubble size causes a significant reduction in its velocity during the initial transients while the final equilibrium position is not significantly influenced by this change. Moreover, the bubble contours during its migration in a SRF from left to its equilibrium location for different dimensionless times for the larger bubble size presented in Fig.~\ref{SRF_countors_D_H_02}, shows that the bubble reaches its equilibrium location relatively faster than the case with the smaller bubble (see Fig.~\ref{SRF_1B1_1B8}).

\subsection{Effect of Reynolds number $\mbox{Re}$}
\begin{figure}[H]
\centering
\begin{subfigure}{0.475\textwidth}
\includegraphics[trim = 0 0 0 0,clip, width = 72mm]{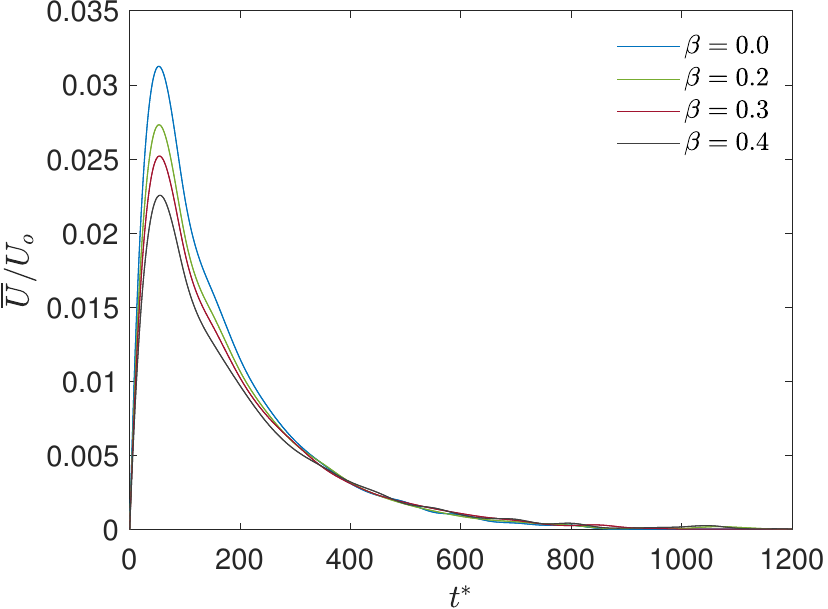}
\caption{Bubble Velocity}
\end{subfigure}
\begin{subfigure}{0.475\textwidth}
\includegraphics[trim = 0 0 0 0,clip, width = 72mm]{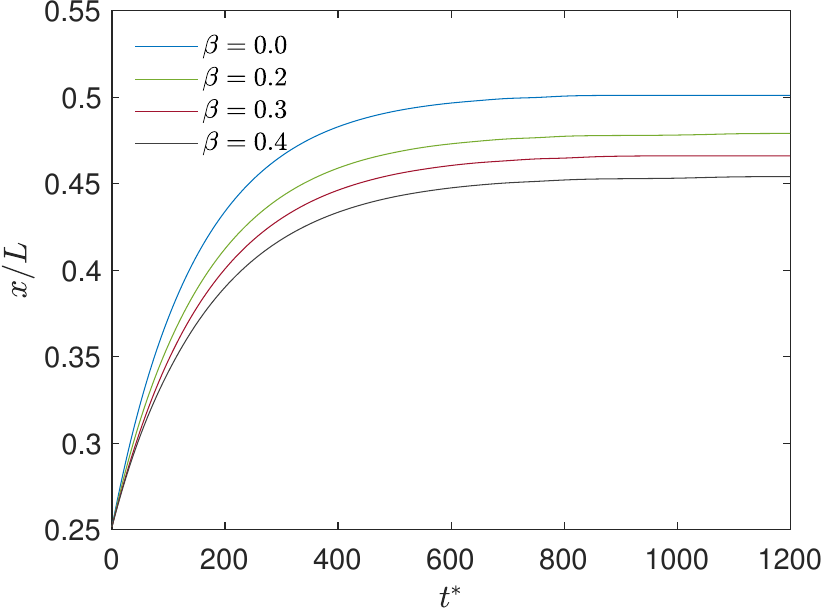}
\caption{Bubble Centroid Location}
\end{subfigure}
\caption{Bubble velocity$(a)$ and its location $(b)$ at different values of the Gibbs elasticity parameter $\beta$ for surfactant-laden bubble in a self-rewetting fluid (SRF). In this figure, $\mbox{Re} = 0.1$, $\mbox{Ma} = 0.25$, $M_1 = 0$, $M_2 = 5$, $D/H = 0.1$, $\psi_L = 0.4$, and $\psi_R = 0.1$.}
\label{SRF_Re_01}
\end{figure}

\begin{figure}[H]
\centering
\begin{subfigure}{0.475\textwidth}
\includegraphics[trim = 0 0 0 0,clip, width = 72mm]{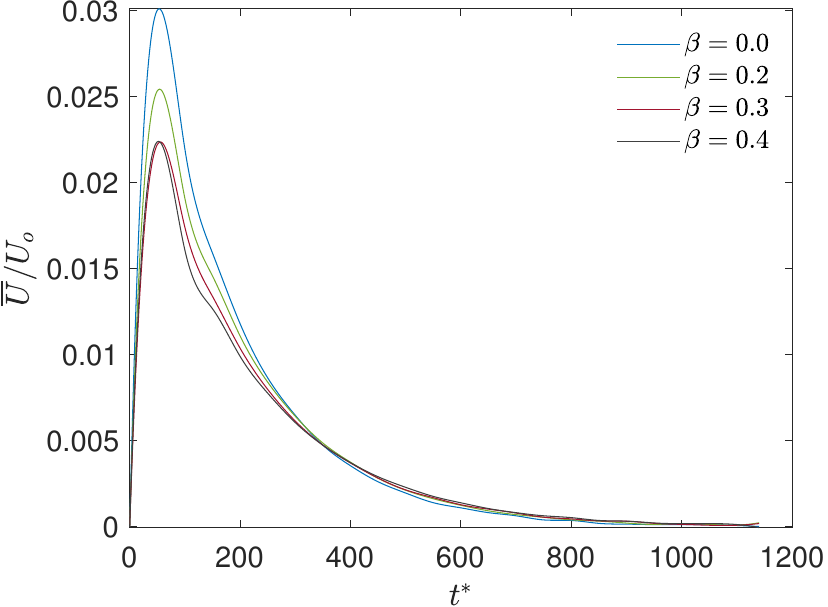}
\caption{Bubble Velocity}
\end{subfigure}
\begin{subfigure}{0.475\textwidth}
\includegraphics[trim = 0 0 0 0,clip, width = 72mm]{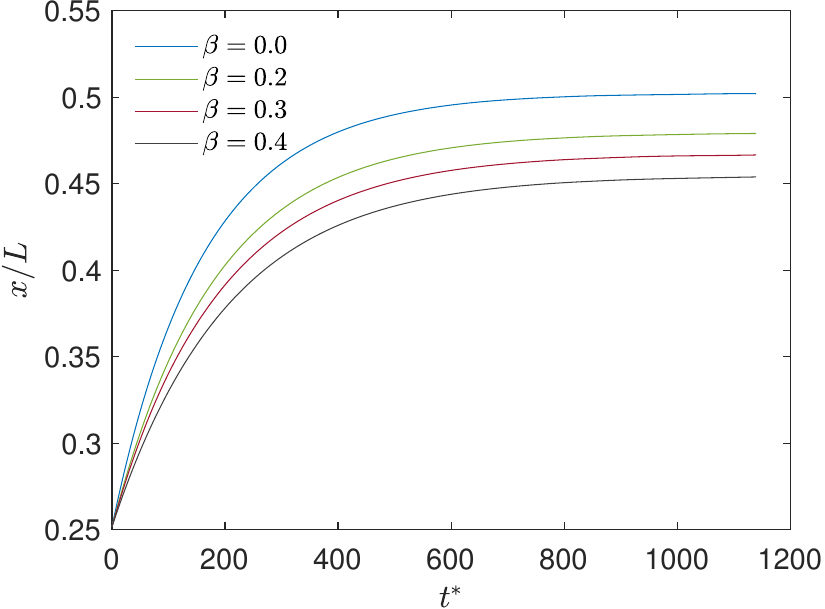}
\caption{Bubble Centroid Location}
\end{subfigure}
\caption{Bubble velocity$(a)$ and its location $(b)$ at different values of the Gibbs elasticity parameter $\beta$ $\beta$ for surfactant-laden bubble in a self-rewetting fluid (SRF). In this figure, $\mbox{Re} = 1$, $\mbox{Ma} = 0.25$, $M_1 = 0$, $M_2 = 5$, $D/H = 0.1$, $\psi_L = 0.4$, and $\psi_R = 0.1$.}
\label{SRF_Re_1}
\end{figure}

\begin{figure}[H]
\centering
\begin{subfigure}{0.475\textwidth}
\includegraphics[trim = 0 0 0 0,clip, width = 72mm]{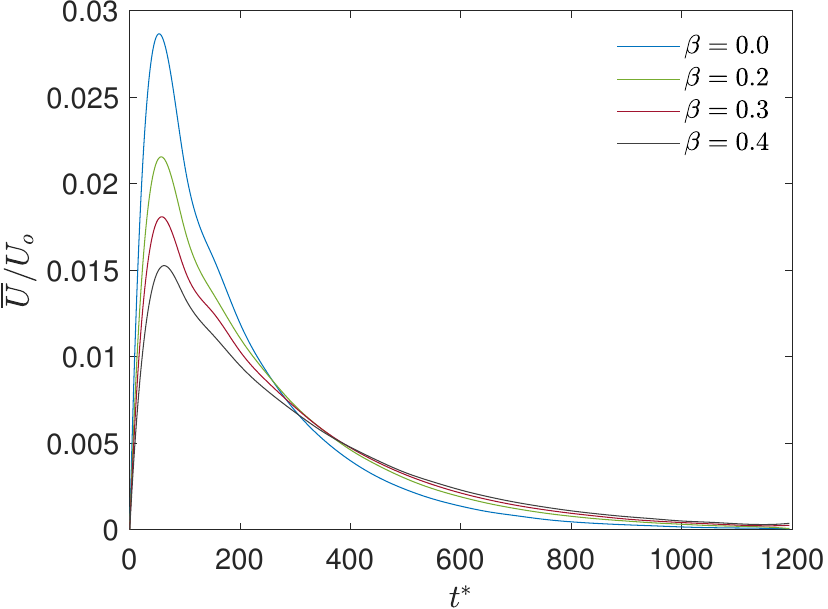}
\caption{Bubble Velocity}
\end{subfigure}
\begin{subfigure}{0.475\textwidth}
\includegraphics[trim = 0 0 0 0,clip, width = 72mm]{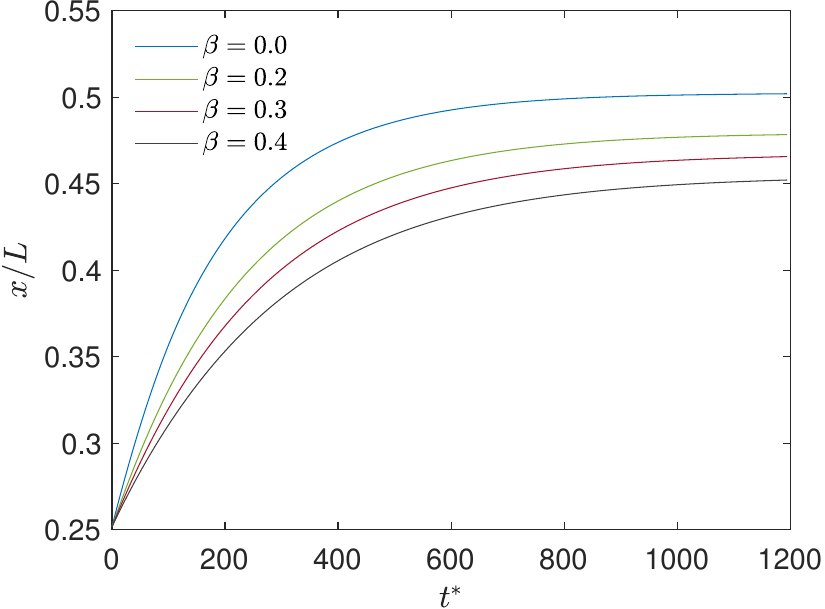}
\caption{Bubble Centroid Location}
\end{subfigure}
\caption{Bubble velocity$(a)$ and its location $(b)$ at different values of the Gibbs elasticity parameter $\beta$ for surfactant-laden bubble in a self-rewetting fluid (SRF). In this figure, $\mbox{Re} = 10$, $\mbox{Ma} = 0.25$, $M_1 = 0$, $M_2 = 5$, $D/H = 0.1$, $\psi_L = 0.4$, and $\psi_R = 0.1$.}
\label{SRF_Re_10}
\end{figure}

We now study the effect of the Reynolds number $\mbox{Re}$ on the bubble velocity and its position during its migration by varying $\mbox{Re}$ in the range ${0.1,1,5,10}$. The rest fluid properties for this study are similar to those specified in the previous case and are also shown in the captions of the associated figures. The baseline case shown in Fig.~\ref{SRF_1} already presented the case $\mbox{Re}=5$ and Figs.~\ref{SRF_Re_01}, \ref{SRF_Re_1}, and~\ref{SRF_Re_10} show the results for $\mbox{Re}=0.1,1.0, \text{and}\ 10$, respectively. As Fig.~\ref{SRF_Re_01} illustrates, bubbles with lower $\mbox{Re}$ reach a higher peak velocity during the initial transients than those with higher $\mbox{Re}$, as Fig.~\ref{SRF_Re_10} shows. The reason why bubbles in low Reynolds number environments generally move more quickly than those at higher Reynolds numbers in such thermocapillary-driven motions of the fluids and interfaces is that the greater viscous effects in low Reynolds number situations tend to transmit effects of the surface tension gradients or Marangoni stresses from the interfaces to the bulk fluid more effectively.
The presence of surfactants reduces the concentration on the trailer edge of the bubble and, in the case of a lower $\mbox{Re}$, alter its velocity during the transients the bubble velocity and shift the equilibrium location slightly upstream toward a higher surfactant gradient location. Generally, for the cases studied, it is found that the overall effect of $\mbox{Re}$ is not very significant compared to the influence of other characteristic parameters such as $M_2$ and $\beta$.

\subsection{Effect of Marangoni number $\mbox{Ma}$}
\begin{figure}[H]
\centering
\begin{subfigure}{0.475\textwidth}
\includegraphics[trim = 0 0 0 0,clip, width = 72mm]{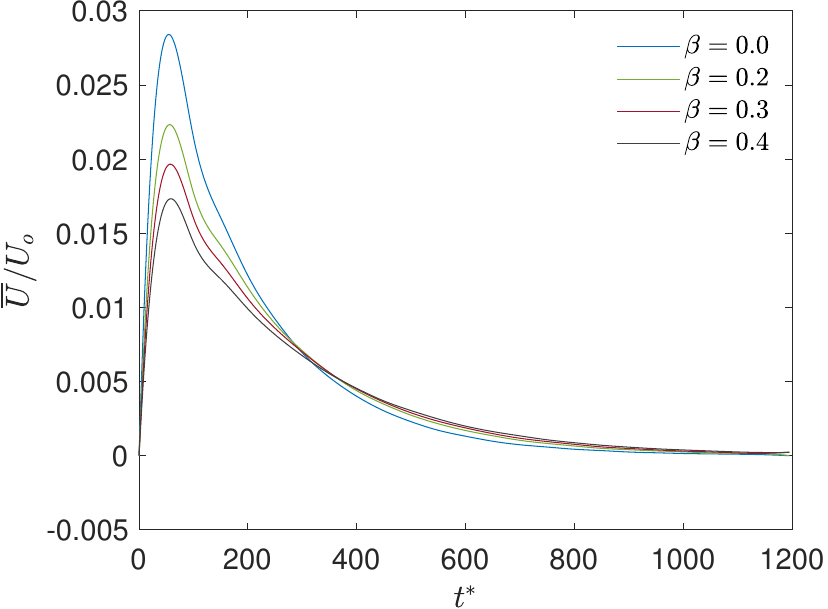}
\caption{Bubble Velocity}
\end{subfigure}
\begin{subfigure}{0.475\textwidth}
\includegraphics[trim = 0 0 0 0,clip, width = 72mm]{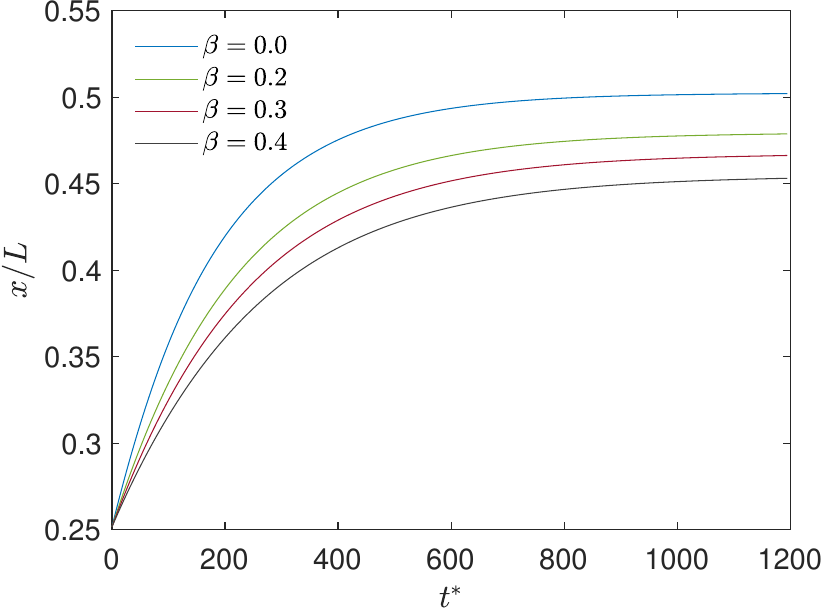}
\caption{Bubble Centroid Location}
\end{subfigure}
\caption{Bubble velocity$(a)$ and its location $(b)$ at different values of the Gibbs elasticity parameter $\beta$ for surfactant-laden bubble in a self-rewetting fluid (SRF). In this figure, $\mbox{Re} = 5$, $\mbox{Ma} = 0.1$, $M_1 = 0$, $M_2 = 5$, $D/H = 0.1$, $\psi_L = 0.4$, and $\psi_R = 0.1$.}
\label{SRF_Ma_01}
\end{figure}

\begin{figure}[H]
\centering
\begin{subfigure}{0.475\textwidth}
\includegraphics[trim = 0 0 0 0,clip, width = 72mm]{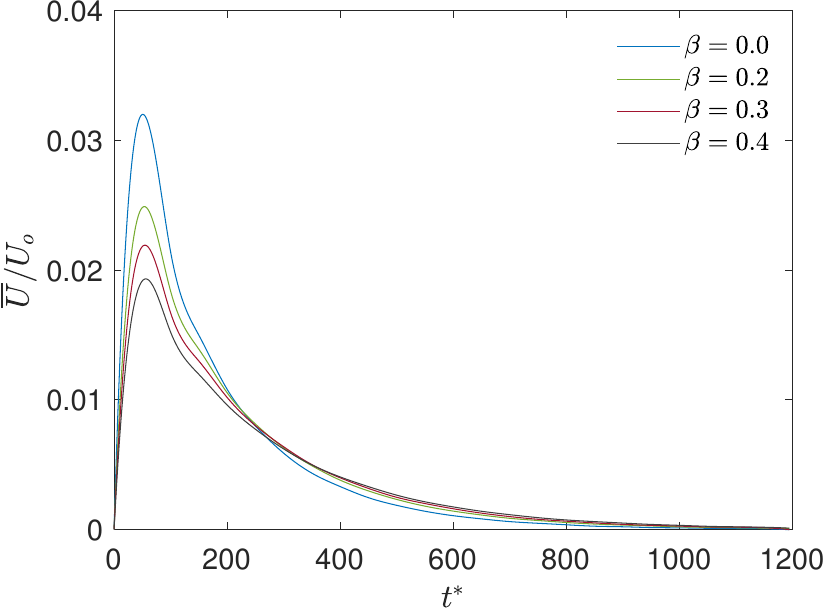}
\caption{Bubble Velocity}
\end{subfigure}
\begin{subfigure}{0.475\textwidth}
\includegraphics[trim = 0 0 0 0,clip, width = 72mm]{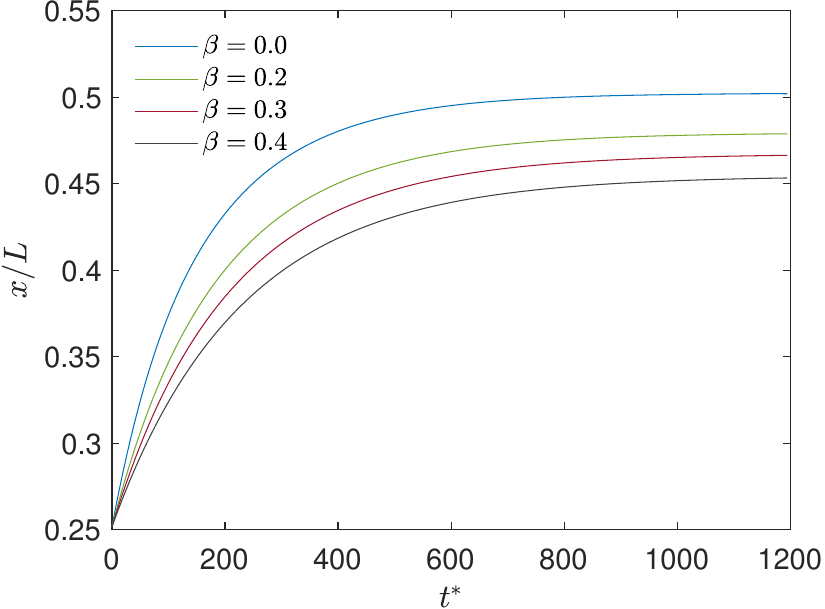}
\caption{Bubble Centroid Location}
\end{subfigure}
\caption{Bubble velocity$(a)$ and its location $(b)$ at different values of the Gibbs elasticity parameter $\beta$ for surfactant-laden bubble in a self-rewetting fluid (SRF). In this figure, $\mbox{Re} = 5$, $\mbox{Ma} = 1$, $M_1 = 0$, $M_2 = 5$, $D/H = 0.1$, $\psi_L = 0.4$, and $\psi_R = 0.1$.}
\label{SRF_Ma_1}
\end{figure}

\begin{figure}[H]
\centering
\begin{subfigure}{0.475\textwidth}
\includegraphics[trim = 0 0 0 0,clip, width = 72mm]{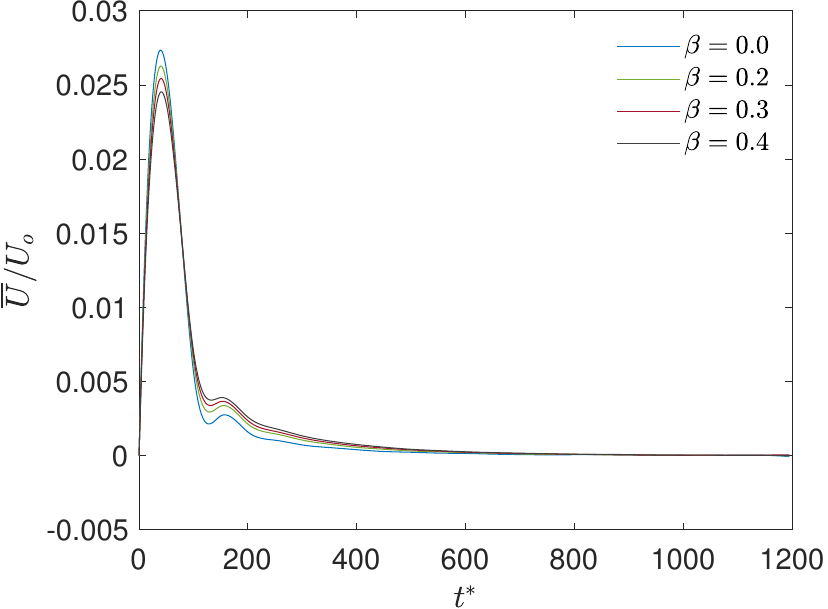}
\caption{Bubble Velocity}
\end{subfigure}
\begin{subfigure}{0.475\textwidth}
\includegraphics[trim = 0 0 0 0,clip, width = 72mm]{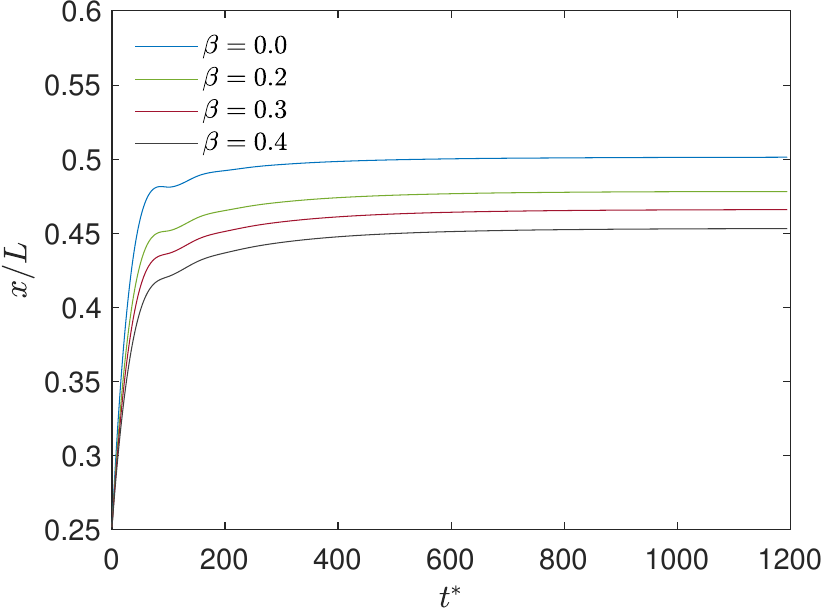}
\caption{Bubble Centroid Location}
\end{subfigure}
\caption{Bubble velocity$(a)$ and its location $(b)$ at different values of the Gibbs elasticity parameter $\beta$ for surfactant-laden bubble in a self-rewetting fluid (SRF). In this figure, $\mbox{Re} = 5$, $\mbox{Ma} = 10$, $M_1 = 0$, $M_2 = 5$, $D/H = 0.1$, $\psi_L = 0.4$, and $\psi_R = 0.1$.}
\label{SRF_Ma_10}
\end{figure}

Finally, we investigate the effect of changing the Marangoni number $\mbox{Ma}$ on the bubble migration. The baseline case (see  Fig.~\ref{SRF_1}) used $\mbox{Ma}=0.25$ and we now consider two additional cases, one with a smaller value ($\mbox{Ma}=0.1$) and two larger values ($\mbox{Ma}=1.0,10.0$) than the above reference value. All the other parameters are kept fixed as before.
According to Eq.~(\ref{ReMaCa}), the Marangoni number $\mbox{Ma}$ in the context of bubble migration is a dimensionless number that represents the effect of Marangoni convection on bubble motion relative to thermal diffusion. A larger $\mbox{Ma}$ thus means greater effect of the convective motion compared to the thermal diffusion. As a result, when comparing Fig.~\ref{SRF_1} ($\mbox{Ma}=0.25$) and Fig.~\ref{SRF_Ma_01} ($\mbox{Ma}=0.1$) with Fig.~\ref{SRF_Ma_1} ($\mbox{Ma}=1.0$) and Fig.~\ref{SRF_Ma_10} ($\mbox{Ma}=10.0$), it is seen that the bubble undergoes rapid acceleration and deceleration during the initial transients especially at higher $\mbox{Ma}$, while by contrast, the presence of surfactants or $\beta \neq 0$ has a greater role at lower $\mbox{Ma}$. However, as in the case with varying $\mbox{Re}$, changing $\mbox{Ma}$ does not significantly influence the bubble equilibrium location unlike the self-rewetting surface tension sensitivity coefficient on temperature $M_2$ or the surfactant parameters.

\section{Summary and Conclusions} \label{Sec.9}
Surface tension, a property generally influenced by temperature, plays a vital role in interfacial transport phenomena within fluids. While normal fluids (NFs) follow a linear relationship between surface tension and temperature, self-rewetting fluids (SRFs) display atypical non-linear (quadratic) dependency on temperature. This relationship is characterized by a quadratic dependence with a minimum point and a positive gradient. Consequently, SRFs possess distinctive features, such as interfacial fluid motion towards higher temperature regions. These unique characteristics make SRFs highly promising for various applications, including microfluidics, both in microgravity conditions and terrestrial settings. On the other hand, however, the inclusion of surface-active agents or surfactants result in their adsorption on fluid interfaces and can significantly alter the behavior of the latter. The combined effects of SRFs and surfactants on bubble migration were not explored previously. Thus, the primary objective of this work is to simulate the thermocapillary migration of bubbles laden with surfactants in SRFs. The computational approach based on LBM using central moment was initially validated against some benchmark problems to ensure its validity.

Simulations resulted in a number of interesting conclusions. It is found that the bubble migrates in SRF towards the minimum temperature location, and it equilibrates at the reference temperature location located at the middle of the domain for the clean interface case where the Gibbs elasticity parameter $\beta=0$. Additionally, including surfactants  ($\beta \neq 0$) is shown to slow the bubble down and move the equilibrium location upstream toward a higher surfactant gradient location. Also, the terminal bubble location in a SRF at long times can be shifted by increasing $\beta$. As a result, $\beta$ can be used to control the strength of the surfactant. The equilibrium position of the bubble is determined by the competing tangential Marangoni stresses due to thermocapillarity in SRFs and from the addition of surfactants. On the other hand, in NF, there is no equilibrium position for the bubble, which migrates completely from the cold side to the hot side.

Upon analyzing the influence of the other parameters associated with the surface tension, it was found that variations in the value of $M_2$ or the quadratic surface tension sensitivity coefficient on temperature, significantly impact the dominant thermocapillary force, leading to observable alterations in the velocity and equilibrium position of the bubble. When higher values of $M_2$ are present, the bubble experiences a more pronounced surface tension gradient along its path. As a result, the bubble undergoes greater acceleration, affecting both its velocity and final equilibrium position. Moreover, the surfactants exert greater effect in shifting the equilibrium position at lower $M_2$. Finally, the streamwise gradient in the imposed surfactant concentration field modulates the transient velocity before attaining the final resting state of the bubble. These findings provide potentially new pathways to manipulate the dynamics of bubbles in microchannels and other applications by exploiting the combined effects of surfactants and anomalous thermocapillary behaviour in SRFs.

\section*{Acknowledgements}
The authors would like to acknowledge the support of the US National Science Foundation (NSF) for research under Grant CBET-1705630. The third author (KNP) would also like to thank the NSF for support of the development of a computer cluster `Alderaan' hosted at the Center for Computational Mathematics at the University of Colorado Denver under Grant OAC-2019089 (Project ``CC* Compute: Accelerating Science and Education by Campus and Grid Computing''), which was used in performing the simulations.

\appendix

\newpage

\section{List of Symbols}\label{sec:nomenclature}
\noindent\fbox{%
    \parbox{\textwidth}{%
    \small
    \begin{multicols}{2}

\textbf{Nomenclature}\\
$L$         \quad \; domain length\\
$H$         \quad \; domain height\\
$D$          \quad \; bubble diameter \\
$T_H$      \quad   hot temperature\\
$T_C$       \quad   cold temperature\\
$T_{ref}$   \; reference temperature\\
$\Delta T = T_H-T_C$    \;    temperature difference\\
$\psi $       \;\;\;\; dimensionless surfactant concentration\\
\phantom{1 cm} $(0 \le \psi \le 1)$\\
$k$    \quad \; thermal conductivity \\
$\bm{u}$    \;\;\;\;  velocity field\\
$p$        \;\;\;\;\; pressure field\\
$T$      \quad\,  temperature field\\
$\mbox{Re} = {U R}/ {\nu_a}$              \quad\;\;  the Reynolds number\\
$\mbox{Ma} = {U R}/ {\alpha_a}$            \quad \; the Marangoni number\\
$\mbox{Ca}= {U \mu_a} /{\sigma_{0}}$         \quad \; the capillary number\\
$\mbox{Eo} = {\rho_a U_0^2 D}/ {\sigma }$    \quad  the Eotvos number\\
$\overline{U}$        \;\;\; migration velocity of the bubble\\
$U_o$         \;\;  characteristic thermocapillary velocity\\

\textbf{Greek Symbols:}\\
$\rho$             \quad fluid density\\
$\mu$           \quad   dynamic viscosity\\
$\alpha$           \quad  thermal diffusivity of the fluid\\
$\nu=\mu/\rho$          \quad \; kinematic viscosity of the fluid\\
$\tilde{\rho} =  {\rho_a}/{\rho_b}$    \;\;  density ratio \\
$\tilde{\mu} =  {\mu_a}/{\mu_b}$    \; dynamic viscosity ratio\\
$\tilde{k} = { k_a}/{k_b}$           \quad  thermal conductivity  ratio\\
$\beta$       \;\;\;  Gibbs elasticity parameter\\
$\sigma_{0}$ \;\;  the value of the surface tension at\\
\phantom{1 cm} a reference temperature $T_{ref}$ \\
$\sigma_{T}$ \;   surface tension linear sensitivity coefficient\\
$\sigma_{TT}$ surface tension quadratic sensitivity coefficient\\
$M_1 =\left({\Delta T}/{\sigma_0}\right) \sigma_T$   \quad\;\; dimensionless surface \\
\phantom{1 cm} tension linear sensitivity coefficient\\
$M_2 =\left({\Delta T^2}/{\sigma_0}\right)\sigma_{TT}$\;  dimensionless surface\\
\phantom{1 cm} tension quadratic sensitivity coefficient\\

$M_{\psi}$ \;\; local surfactant mobility\\
$m_{\psi}$ \quad the scale for the mobility parameter\\
$\mu_{\psi}$ \;\;\;  chemical potential\\
$\phi$ \quad\;  order parameter of the phase field variable\\
$\lambda$, $s$, and $w$ \quad model parameters\\
$\mbox{Pi}$ \;\;\; indicates the proportionate contribution\\
\phantom{1 cm} of surfactant diffusion\\
$\mbox{Ex}$ \quad represents the relative intensity of\\
\phantom{1 cm} adsorption and solubility effects\\
$M_{\phi}$ \;  local mobility for interface tracking\\
\phantom{1 cm} (in LBM) \\

$\bm{e}_\alpha$ \quad particle velocity directions\\
\phantom{1 cm} (in LBM) \\
$f_\alpha$ \quad distribution function for interface tracking\\
\phantom{1 cm} (in LBM) \\
$g_\alpha$ \quad distribution function for two-fluid flow\\
\phantom{1 cm} (in LBM) \\
$h_\alpha$ \quad distribution function for temperature field\\
\phantom{1 cm} (in LBM) \\
$q_\alpha$ \quad distribution function for surfactant\\
\phantom{1 cm} concentration field  (in LBM) \\

\;

\textbf{Subscripts:}\\
$A$            \quad  ambient fluid \\
$B$     \quad  bubble\\
$L$       \quad  left\\
$R$       \quad  right

   \end{multicols}
    }%
    }

\section{LBM for phase-field based interface capturing} \label{Sec.4.1}
In order to solve the conservative ACE provided in Eq.~(\ref{eqn1}), we will now discuss a central moment LB approach that involves evolving a distribution function $f_\alpha$ on the D2Q9 lattice, where $\alpha=0,1,2,\ldots,8$ denotes the discrete particle directions. During the collision, the distribution functions $\mathbf{f}=(f_0,f_1,f_2,\ldots,f_8)^\dagger$ generally rela-x to the corresponding equilibrium distribution functions given by $\mathbf{f}^{eq}=(f_0^{eq},f_1^{eq},f_2^{eq},\ldots,f_8^{eq})^\dagger$. These distribution functions must then be implemented via their central moments in the following.

First, in this regard, the following vectors, expressed in standard Dirac's bra-ket notation, may be used to describe the components of the particle velocities of this lattice as
\mathleft
\begin{subequations}
\begin{equation}     
\qquad \left| \bm{e}_x \right> = ( 0, 1, 0, -1, 0, 1, -1, -1, 0)^\dagger, \nonumber 
\end{equation}
\begin{equation}    
\qquad \left| \bm{e}_y \right> = ( 0, 0, 1, 0,-1, 1, 1, -1, -1)^\dagger.  \nonumber
\end{equation}
\end{subequations}
We also need the following 9-dimensional vector to define the zeroth moment of $f_\alpha$:
\mathleft
\begin{eqnarray} 
\qquad \left|\mathbf{1}\right> = (1,1,1,1,1,1,1,1,1)^{\dag}. \nonumber
\end{eqnarray}
In other words, the order parameter $\phi$ of the phase-field model should result from its inner product with a set of distribution functions $\left<\mathbf{f}|\mathbf{1}\right>$. Next, the central moment LB method will be built using the nine non-orthogonal basis vectors shown below (which diffrents from methodology described in~\cite{hajabdollahi2021central}):
\mathleft
\begin{gather}
\qquad \left| P_0 \right> = \left| \mathbf{1} \right>, \quad
\left| P_1 \right> = \left| \bm{e}_x \right>, \quad
\left| P_2 \right> = \left| \bm{e}_y \right>, \nonumber \\[2mm]
\qquad \left| P_3 \right> = \left| \bm{e}_x^2+\bm{e}_y^2 \right>, \quad
\left| P_4 \right> = \left| \bm{e}_x^2-\bm{e}_y^2 \right>, \quad
\left| P_5 \right> = \left| \bm{e}_x \bm{e}_y \right>,\nonumber \\[2mm]
\qquad \left| P_6 \right> = \left| \bm{e}_x^2 \bm{e}_y \right>,\quad
\left| P_7 \right> = \left| \bm{e}_x \bm{e}_y^2 \right>,\quad
\left| P_8 \right> = \left| \bm{e}_x^2 \bm{e}_y^2 \right>.  \nonumber
\end{gather}
Symbols such as $\left| e_x^2 e_y \right> =  \left| e_x e_x e_y \right>$ denote a vector that results from the element-wise vector multiplication of vectors $\left| e_x \right>$,$\left| e_x \right>$ and $\left| e_y \right>$. Using the moment basis vectors mentioned above, they may be arranged into the following matrix, which maps the distribution functions to the \emph{raw} moments:
\mathleft
\begin{equation}  \label{eqn34}
\qquad \mathbf{P} = \left[
\left<P_0\right|,
\left<P_1\right|,
\left<P_2\right|,
\left<P_3\right|,
\left<P_4\right|,
\left<P_5\right|,
\left<P_6\right|,
\left<P_7\right|,
\left<P_8\right|
\; \right].
\end{equation}
It should be emphasized that by shifting the particle velocity $\bm{e}_\alpha$ by the fluid velocity $\bm{u}$, the \emph{central} moments are obtained from the distribution moments. Using this, we can then define the raw moments of the distribution function and equilibrium $f_\alpha^{eq}$ explicitly as
\mathleft
\begin{subequations}
\begin{equation}
\qquad \left( \begin{array}{c}\kappa'_{mn}\\[2mm]   \kappa'^{\;eq}_{mn} \end{array} \right)  = \sum_{\alpha = 0}^{8} \left( \begin{array}{c}f_{\alpha} \\[2mm]   f_{\alpha}^{eq} \end{array} \right)  e_{\alpha x}^m   e_{\alpha y}^n,
\end{equation}
and the corresponding central moments as
\mathleft
\begin{equation}
\qquad \left( \begin{array}{c}\kappa_{mn} \\[2mm]   \kappa_{mn}^{eq} \end{array} \right)  = \sum_{\alpha = 0}^{8} \left( \begin{array}{c}f_{\alpha} \\[2mm]   f_{\alpha}^{eq} \end{array} \right) (e_{\alpha x}-u_x)^m  ( e_{\alpha y}-u_y)^n.
\end{equation}
\end{subequations}
The raw moment of order $(m+n)$ is thus represented by $\kappa'_{mn}$, and the corresponding central moment is $\kappa_{mn}$. The two vectors that follow can be used to easily group together all the possible raw moments and central moments for the D2Q9 lattice:
\mathleft
\begin{subequations}
\begin{eqnarray}
\qquad \bm{\kappa^{'}} \! \! \! &=& \! \! \! ( \K_{00}^{'}, \K_{10}^{'},\K_{01}^{'}, \K_{20}^{'}, \K_{02}^{'}, \K_{11}^{'},\K_{21}^{'}, \K_{12}^{'},\K_{22}^{'} ),\label{eqn:4a} \\[3mm]
\qquad \bm{\kappa} \! \! \! &=& \! \! \! ( \K_{00},\K_{10}, \K_{01}, \K_{20}, \K_{02}, \K_{11}, \K_{21}, \K_{12}, \K_{22} ).
\end{eqnarray}
\end{subequations}
It should be noted that one can readily map from the distribution functions to the raw moments via $\bm{\kappa^{'}} = \PP\mathbf{f}$, which can then be transformed into the central moments through $\bm{\kappa} = \F \bm{\kappa^{'}}$, where the frame transformation matrix $\tensr{F}$ follows readily from binomial expansions of $(e_{\alpha x}-u_x)^m  ( e_{\alpha y}-u_y)^n$ to relate to $e_{\alpha x}^m   e_{\alpha y}^n$ etc. Similarly, the inverse mappings from central moments to raw moments, from which the distribution functions can be recovered via the matrices $\tensr{F}^{-1}$ and $\tensr{P}^{-1}$, respectively. All these mapping relations are explicitly listed in Appendix~\ref{App B}.

As previously indicated, an important aspect of our approach is carrying out the collision step in a way that causes those different central moments shown above to relax to their their corresponding central moment equilibria. The discrete central moment equilibria $\Keq{mn}$ defined above can be obtained by matching them to the corresponding central moments of the continuous Maxwell distribution function after replacing the density $\rho$ with the order parameter $\phi$; Moreover, $M_{\phi}\theta n_x$ and $M_{\phi}\theta n_y$~\cite{hajabdollahi2021central,ouderji2019phdthesis} must be added to the first order central moment equilibrium components in order to account for the interface sharpening flux terms in the conservative ACE (Eq.~(\ref{eqn1})). Therefore, we have
\mathleft
\begin{gather}
\qquad \Keq{00} = \phi, \qquad
\Keq{10} = M_{\phi} \theta  n_x,\qquad
\Keq{01} = M_{\phi} \theta  n_y,\nonumber \\[2mm]
\qquad \Keq{20} = c_{s\phi}^2 \phi,\qquad
\Keq{02} = c_{s\phi}^2 \phi,\qquad
\Keq{11} = 0,\nonumber  \\[2mm]
\qquad \Keq{21} = 0,\qquad
\Keq{12} = 0,\qquad
\Keq{22} = c_{s\phi}^4 \phi,
\end{gather}
where $c_{s\phi}^2=1/3$.

We can now summarize the central moment LB algorithm for solving the conservative ACE for a time step $\Delta t$, starting from $f_\alpha=f_\alpha(\bm{x},t)$, as follows, based on the above considerations, which were inspired by the algorithmic implementation presented in~\cite{geier2015cumulant} (see also~\cite{yahia2021central,yahia2021three} for its further extensions):
\begin{itemize}
  \item Compute pre-collision raw moments from distribution functions via $\bm{\kappa^{'}} = \PP\mathbf{f}$ (see Eq.~(\ref{eq:tensorP}) in Appendix~\ref{App B} for $\tensr{P}$)
  \item Compute pre-collision central moments from raw moments via $\bm{\kappa} = \F \bm{\kappa^{'}}$ (see Eq.~(\ref{eq:tensorF}) in Appendix~\ref{App B} for $\tensr{F}$)
  \item Perform collision step via relaxation of central moments $\kappa_{mn}$ to their equilibria $\kappa_{mn}^{eq}$: \newline
        \begin{equation}\label{eq:centralmomentrelaxationCACE}
           \qquad \tilde{\kappa}_{mn} = \kappa_{mn} + \wphi{mn}  (\Keq{mn} - \kappa_{mn}),
        \end{equation}
        where $(mn)=(00),(10),(01),(20),(02),(11),(21),(12)$, and $(22)$, and $\wphi{mn}$ is the relaxation parameter for moment of order ($m+n$). Here, the implicit summation convention of repeated indices is not assumed. The relaxation parameters of the first order moments $\omega_{10}^{\phi} = \omega_{01}^{\phi} = \omega^{\phi}$ are related to the mobility coefficient $M_{\phi}$ in Eq.~(\ref{eqn1}) via $M_{\phi}= c_{s\phi}^2 \left( \frac{1}{\omega^{\phi}} - \frac{1}{2}\right)\Delta t$, and the rest of the relaxation parameters are typically set to unity, i.e., $\wphi{mn}=1.0$, where $(m+n) \geq 2$. The results of Eq.~(\ref{eq:centralmomentrelaxationCACE}) are then grouped in $\bm{\tilde{\kappa}}$.
  \item Compute post-collision raw moments from post-collision central moments via $\bm{\tilde{\kappa}^{'}} = \F^{-1} \bm{\tilde{\kappa}}$ (see Eq.~(\ref{eq:tensorFinverse}) in Appendix~\ref{App B} for $\tensr{F}^{-1}$)
  \item Compute post-collision distribution functions from post-collision raw moments via $\mathbf{\tilde{f}} = \PP^{-1}\bm{\tilde{\kappa}^{'}}$ (see Eq.~(\ref{eq:tensorPinverse}) in Appendix~\ref{App B} for $\tensr{P}^{-1}$)
  \item Perform streaming step via $f_{\alpha}(\bm{x}, t+ \Delta t) = \tilde{f}_{\alpha}(\bm{x}-\bm{e}_{\alpha} \Delta t)$, where $\alpha = 0,1,2,...,8$.
  \item Update the order parameter $\phi$ of the phase-field model for interface capturing through \newline
        \begin{equation}
           \qquad \qquad\phi = \sum_{\alpha=0}^{8} f_{\alpha}.
        \end{equation}
\end{itemize}
\section{LBM for two-fluid flow with capillary and Marangoni forces}\label{Sec.4.2}
The motion of binary fluids with interfacial forces represented in Eqs.~(\ref{eqn4})-(\ref{eqn5}) will then be solved using a central moment LBM by evolving another distribution function $g_\alpha$, where $\alpha=0,1,2,\ldots,8$. Our approach is based on discretizing the modified continuous Boltzmann equation and using a matching principle with their continuous counterparts to obtain the discrete central moment equilibria and central moments of the source terms for the body forces, as explained in Ref.~\cite{hajabdollahi2021central}. Nevertheless, in the following, we consider the simpler, non-orthogonal moment basis vectors as provided previously in Eq.~(\ref{eqn34}), in contrast to Ref.~\cite{hajabdollahi2021central,ouderji2019phdthesis}, where an orthogonal moment basis is applied resulting in the so-called cascaded LB method.

The distribution function $g_\alpha$, its equilibrium $g_\alpha^{eq}$, and the source term $S_\alpha$ are defined first, as in the previous section. The latter accounts for the surface tension, body forces, and those resulting from the application of a transformation to simulate flows at high-density ratios in the incompressible limit (see~\cite{he1999lattice,hajabdollahi2021central}).
\mathleft
\begin{subequations}
\begin{equation}
\qquad \left( \begin{array}{c} {\eta}'_{ mn}\\[1mm]   \eta'^{\;eq}_{ mn}\\[1mm]   {\sigma}'_{ mn} \end{array} \right)  =  \sum_{\alpha = 0}^{8} \left( \begin{array}{c}g_{\alpha} \\[1mm]   g_{\alpha}^{eq}\\[1mm]  S_\alpha \end{array} \right)  e_{\alpha x}^m   e_{\alpha y}^n,
\end{equation}
\mathleft
\begin{equation}
\qquad \left( \begin{array}{c}{\eta}_{ mn}\\[1mm]   \eta^{\;eq}_{ mn}\\[1mm]   {\sigma}_{ mn} \end{array} \right)  = \sum_{\alpha = 0}^{8} \left( \begin{array}{c}g_{\alpha} \\[1mm]   g_{\alpha}^{eq}\\[1mm]  S_\alpha \end{array} \right)  (e_{\alpha x}-u_x)^m  ( e_{\alpha y}-u_y)^n.
\end{equation}
\end{subequations}
For convenience, we can group the source term, equilibrium, and components of the distribution function for the D2Q9 lattice as the following vectors: $\mathbf{g}^{eq}=(g_0^{eq},g_1^{eq},g_2^{eq},\ldots,g_8^{eq})^\dagger$, $\mathbf{S}=(S_0,S_1,S_2,\ldots, g_8)^\dagger$. Additionally, we use the following to group the possible raw moments and the central moments already defined above for the D2Q9 lattice:
\mathleft
\begin{subequations}
\begin{eqnarray}
\qquad \bm{{\eta}^{'}} \! \! \! &=& \! \! \! ( {\eta}_{00}^{'}, {\eta}_{10}^{'},{\eta}_{01}^{'}, {\eta}_{20}^{'}, {\eta}_{02}^{'}, {\eta}_{11}^{'},{\eta}_{21}^{'}, {\eta}_{12}^{'},{\eta}_{22}^{'} ),\label{eqn:4a} \\[3mm]
\qquad \bm{{\eta}} \! \! \! &=& \! \! \! ( {\eta}_{00},{\eta}_{10}, {\eta}_{01}, {\eta}_{20}, {\eta}_{02}, {\eta}_{11}, {\eta}_{21}, {\eta}_{12}, {\eta}_{22} ),
\end{eqnarray}
\end{subequations}
and similarly for raw moments and the central moments the equilibrium and the source term.

The collision step will be performed such that different central moments shown above relax to their corresponding central moment equilibria, which are augmented by changes in the central moments due to the net forces; the latter is given by sum the surface tension force $\bm{F}_s=(F_{sx},F_{sy})$, which can have contributions from both the capillary and Marangoni forces as represented in Eq.~(\ref{eq:surfacetensionforcecomponents}), and any external force $\bm{F}_{ext}=(F_{ext,x},F_{ext,y})$, i.e., $\bm{F}_{t}=\bm{F}_{s}+\bm{F}_{ext}$ or $(F_{tx},F_{ty})=(F_{sx}+F_{ext,x},F_{sy}+F_{ext,y})$. Moreover, the use of an incompressible transformation as mentioned above leads to a pressure-based formulation, involving the incorporation of a net pressure force $\bm{F}_p$ arising from $\varphi(\rho)=p-\rho c_s^2$, i.e., $\bm{F}_p=-\bm{\nabla}\varphi$, or $(F_{px},F_{py})=(-\partial_x\varphi,-\partial_y\varphi)$ (see~\cite{hajabdollahi2021central} for details). Then, the discrete central moment equilibria $\eta_{mn}$ defined above can be obtained by matching them to the corresponding continuous central moments of the equilibrium that arise from the incompressible transformation, and similarly for the central moments of the source term $\sigma_{mn}$, which then results in the following expressions for the D2Q9 lattice~\cite{hajabdollahi2021central}:
\mathleft
\begin{gather}
\qquad {\eta}_{00}^{eq} = p, \quad {\eta}_{10}^{eq} = -\varphi(\rho) u_x, \quad {\eta}_{01}^{eq} = -\varphi(\rho)u_y, \quad {\eta}_{20}^{eq} =  p c_s^2 + \varphi(\rho)u_x^2,\nonumber \\
\qquad {\eta}_{02}^{eq} =  p c_s^2 + \varphi(\rho)u_y^2, \quad  {\eta}_{11}^{eq} = \varphi(\rho)u_x u_y , \quad {\eta}_{21}^{eq} = -\varphi(\rho)(u_x^2+ c_s^2) u_y,\nonumber \\
\qquad {\eta}_{12}^{eq} = -\varphi(\rho)(u_y^2+ c_s^2) u_x, \quad {\eta}_{22}^{eq} = c_s^6 \rho + \varphi(\rho)(u_x^2+ c_s^2) (u_y^2+ c_s^2).
\end{gather}
and
\begin{gather}
\qquad {\sigma}_{00}= \Gamma_{00}^p, \quad {\sigma}_{10} = c_s^2 F_{tx}-u_x{\Gamma}_{00}^p, \quad {\sigma}_{01} =  c_s^2 F_{ty}-u_y{\Gamma}_{00}^p, \nonumber \\
\qquad {\sigma}_{20} = 2c_s^2 F_{px}u_x+(u_x^2+c_s^2){\Gamma}_{00}^p,\quad {\sigma}_{02} =  2c_s^2 F_{py}u_y+(u_y^2+c_s^2){\Gamma}_{00}^p, \nonumber\\
\qquad {\sigma}_{11} = c_s^2 (F_{px}u_y+F_{py}u_x)+u_x u_y{\Gamma}_{00}^p,\quad {\sigma}_{21} = 0, \quad {\sigma}_{12} = 0, \quad {\sigma}_{22} = 0,
\end{gather}
where $\Gamma_{00}^p=(F_{px}u_x+F_{py}u_y)$.

We can now summarize the central moment LB algorithm for computing the two-fluid motion with interfacial forces for a time step $\Delta t$ starting from $g_\alpha=g_\alpha(\bm{x},t)$ using the above developments as follows:
\begin{itemize}
  \item Compute pre-collision raw moments from distribution functions via $\bm{\eta^{'}} = \PP\mathbf{g}$ (see Eq.~(\ref{eq:tensorP}) in Appendix~\ref{App B} for $\tensr{P}$)
  \item Compute pre-collision central moments from raw moments via $\bm{\eta} = \F \bm{\eta^{'}}$ (see Eq.~(\ref{eq:tensorF}) in Appendix~\ref{App B} for $\tensr{F}$)
  \item Perform collision step via relaxation of central moments $\eta_{mn}$ to their equilibria $\eta_{mn}^{eq}$ and augmented with the source terms $\sigma_{mn}$: \newline
      In order to allow for an independent specification of the shear viscosity $\nu$ from the bulk viscosity $\zeta$, the trace of the second order moments ${\eta}_{20} + {\eta}_{02}$ should be evolved independently from the other second-order moments. To accomplish this, prior to collision, we combine the diagonal parts of the second-order moments as follows (see e.g.,~\cite{geier2015cumulant,yahia2021central,yahia2021three}):
        \mathleft
        \begin{gather}
        \qquad {\eta}_{2s} = {\eta}_{20} + {\eta}_{02}, \qquad {\eta}_{2s}^{eg} = {\eta}_{20s}^{eg} + {\eta}_{02}^{eg}, \qquad {\sigma}_{2s} = {\sigma}_{20s} + {\sigma}_{02},\nonumber \\[2mm]
        \qquad {\eta}_{2d} = {\eta}_{20} -  {\eta}_{02}, \qquad {\eta}_{2d}^{eg} = {\eta}_{20s}^{eg} -  {\eta}_{02}^{eg},  \qquad {\sigma}_{2d} = {\sigma}_{20s} - {\sigma}_{02},\nonumber
        \end{gather}
        and thus ${\eta}_{2s}$ and ${\eta}_{2d}$ will be evolved independently under collision. Then, the post-collision central moments under relaxation and augmentation due to the forces can be computed via
        \mathleft
        \begin{equation} \label{eq:centralmomentrelaxationtwofluidmotion}
        \qquad  \tilde{\eta}_{ mn} =  {\eta}_{mn} + \omega_{mn} \left( {\eta}_{mn}^{eq} -{\eta}_{mn} \right)  + \left(1-\omega_{mn}/2 \right)  {\sigma}_{mn}\Delta t,
        \end{equation}
        where $\omega_{mn}$ is the relaxation time corresponding to the central moment ${\eta}_{ mn}$, and $(mn)= (00), (10), (01), (2s), (2d), (11), (21), (12), \mbox{and}, (22)$. Here, the relaxation parameter $\omega_{2s}$ is related to the bulk viscosity via $\zeta= c_s^2 \left(1/\omega_{2s}- 1/2\right)\Delta t$, while the relaxation parameters  $\omega_{2d}$ and $\omega_{11}$ are related to shear viscosity via $\nu= c_s^2 \left( 1/\omega_{ij} - 1/2\right)\Delta t$ where $(ij)=(2d),(11)$. Typically, $c_s^2=1/3$. In view of Eq.~(\ref{eqn11}) it should be noted that if the bulk fluid properties are different, the relaxation parameters $\omega_{2d}$ and $\omega_{11}$ will then vary locally across the interface. The rest of the relaxation parameters of central moments are generally set to unity, i.e., $\omega_{ij}=1.0$, where $(ij)=(00),(10),(01),(2s),(21),(12),(22)$.\newline
        Also, the combined forms of the post-collision central moments $\tilde{\eta}_{2s}$ and $\tilde{\eta}_{2d}$ are then segregated in their individual components $\tilde{\eta}_{20}$ and $\tilde{\eta}_{02}$ via
        \mathleft
        \begin{equation*}
        \qquad  \tilde{\eta}_{20} = \frac{1}{2} \left(\tilde{\eta}_{2s}+\tilde{\eta}_{2d} \right), \qquad  \tilde{\eta}_{02} = \frac{1}{2} \left(\tilde{\eta}_{2s}-\tilde{\eta}_{2d} \right).
        \end{equation*}
        Finally, the results of Eq.~(\ref{eq:centralmomentrelaxationtwofluidmotion}) by accounting for the above segregation are then grouped in $\bm{\tilde{\eta}}$.
  \item Compute post-collision raw moments from post-collision central moments via $\bm{\tilde{\eta}^{'}} = \F^{-1} \bm{\tilde{\eta}}$ (see Eq.~(\ref{eq:tensorFinverse}) in Appendix~\ref{App B} for $\tensr{F}^{-1}$)
  \item Compute post-collision distribution functions from post-collision raw moments via $\mathbf{\tilde{g}} = \PP^{-1}\bm{\tilde{\eta}^{'}}$ (see Eq.~(\ref{eq:tensorPinverse}) in Appendix~\ref{App B} for $\tensr{P}^{-1}$)
  \item Perform streaming step via $g_{\alpha}(\bm{x}, t+ \Delta t) = \tilde{g}_{\alpha}(\bm{x}-\bm{e}_{\alpha} \Delta t)$, where $\alpha = 0,1,2,...,8$.
  \item Update the pressure field $p$ and the components of the fluid velocity $\bm{u}=(u_x,u_y)$ through \newline
        \begin{equation}
           \qquad \qquad p = \sum_\alpha g_\alpha + \frac{1}{2} \bm{F}_p \cdot {\bm{u}}\Delta t,\quad \rho c_s^2 \bm{u} = \sum_\alpha g_\alpha \bm{e}_\alpha + \frac{1}{2} c_s^2 \bm{F}_{t}\Delta t.
        \end{equation}
\end{itemize}

\section{LBM for energy transport equation} \label{Sec.4.3}
The energy transport equation (Eq.~(\ref{energy eqn})) can be solved using a central moment LB technique by evolving a third distribution function $h_\alpha$, where $\alpha=0,1,2,\ldots,8$, on the D2Q9 lattice. As an advection-diffusion equation, Eq.~(\ref{energy eqn}) is constructed in a manner similar to the LB scheme for the conservative ACE previously discussed, but without including a term like the interface sharpening flux term that is included in the latter case. As previously, we first establish the following raw moments and central moments, respectively, of the distribution function $h_\alpha$, as well as its equilibrium $h_\alpha^{eq}$:
\mathleft
\begin{subequations}
\begin{equation}
\qquad \left( \begin{array}{c}\chi'_{mn}\\[2mm]   \chi'^{\;eq}_{mn} \end{array} \right)  = \sum_{\alpha = 0}^{8} \left( \begin{array}{c} h_{\alpha} \\[2mm]  h_{\alpha}^{eq} \end{array} \right)  e_{\alpha x}^m   e_{\alpha y}^n,
\end{equation}
\mathleft
\begin{equation}
\qquad \left( \begin{array}{c}\chi_{mn} \\[2mm]   \chi_{mn}^{eq} \end{array} \right)  = \sum_{\alpha = 0}^{8} \left( \begin{array}{c} h_{\alpha} \\[2mm]  h_{\alpha}^{eq} \end{array} \right) (e_{\alpha x}-u_x)^m  ( e_{\alpha y}-u_y)^n.
\end{equation}
\end{subequations}
For convenience, we list the components of the distribution function and its equilibrium, respectively, using $\mathbf{h}=(h_0,h_1,h_2,\ldots,h_8)^\dagger$ and $\mathbf{h}^{eq}=(h_0^{eq},h_1^{eq},h_2^{eq},\ldots,h_8^{eq})^\dagger$, and analogously for the raw moments and central moments via
\mathleft
\begin{subequations}
\begin{eqnarray}
\qquad \bm{\chi^{'}} \! \! \! &=& \! \! \! ( \chi_{00}^{'}, \chi_{10}^{'},\chi_{01}^{'}, \chi_{20}^{'}, \chi_{02}^{'}, \chi_{11}^{'},\chi_{21}^{'}, \chi_{12}^{'},\chi_{22}^{'} ),\label{eqn:4a} \\[3mm]
\qquad \bm{\chi} \! \! \! &=& \! \! \! ( \chi_{00},\chi_{10}, \chi_{01}, \chi_{20}, \chi_{02}, \chi_{11}, \chi_{21}, \chi_{12}, \chi_{22} ).
\end{eqnarray}
\end{subequations}
Similar to Sec.~\ref{Sec.4.1}, to construct a central moment-based collision model for solving the energy equation, we obtain the discrete equilibrium central moments from the corresponding continuous counterpart of the Maxwellian by replacing the density $\rho$ with the temperature $T$, and the results read as
\mathleft
\begin{gather}
\qquad \chi_{00}^{eq} = T, \qquad
\chi_{10}^{eq} = 0,\qquad
\chi_{01}^{eq} = 0,\nonumber \\[2mm]
\qquad \chi_{20}^{eq} = c_{sT}^2 T,\qquad
\chi_{02}^{eq} = c_{sT}^2 T,\qquad
\chi_{11}^{eq} = 0,\nonumber  \\[2mm]
\qquad \chi_{21} = 0,\qquad
\chi_{12}^{eq} = 0,\qquad
\chi_{22}^{eq} = c_{sT}^4 T,
\end{gather}
where, typically, $c_{sT}^2 = 1/3$. Then, the computational procedure for solving the energy equation for a time step $\Delta t$ starting from $h_\alpha=h_\alpha(\bm{x},t)$ can be summarized as follows:
\begin{itemize}
  \item Compute pre-collision raw moments from distribution functions via $\bm{\chi^{'}} = \PP\mathbf{h}$ (see Eq.~(\ref{eq:tensorP}) in Appendix~\ref{App B} for $\tensr{P}$)
  \item Compute pre-collision central moments from raw moments via $\bm{\chi} = \F \bm{\chi^{'}}$ (see Eq.~(\ref{eq:tensorF}) in Appendix~\ref{App B} for $\tensr{F}$)
  \item Perform collision step via relaxation of central moments $\chi_{mn}$ to their equilibria $\chi_{mn}^{eq}$: \newline
        \begin{equation}\label{eq:centralmomentrelaxationenergyequation}
           \qquad \tilde{\chi}_{mn} = \chi_{mn} + \omega^T_{mn}  (\chi_{mn}^{eq} - \chi_{mn}),
        \end{equation}
        where $(mn)=(00),(10),(01),(20),(02),(11),(21),(12)$, and $(22)$, and $\omega^T_{mn}$ is the relaxation parameter for moment of order ($m+n$). The relaxation parameters of the first order moments $\omega_{10}^T$=$\omega_{01}^T$=$ \omega^T$ are related to the thermal diffusivity $\alpha=k/(\rho c_p)$ via $\alpha = c_{sT}^2 \left(1/\omega^{T} - 1/2\right)\Delta t$, and the rest of the relaxation parameters of higher central moments are typically set to unity. The results of Eq.~(\ref{eq:centralmomentrelaxationenergyequation}) are then grouped in $\bm{\tilde{\chi}}$.
  \item Compute post-collision raw moments from post-collision central moments via $\bm{\tilde{\chi}^{'}} = \F^{-1} \bm{\tilde{\chi}}$ (see Eq.~(\ref{eq:tensorFinverse}) in Appendix~\ref{App B} for $\tensr{F}^{-1}$)
  \item Compute post-collision distribution functions from post-collision raw moments via $\mathbf{\tilde{h}} = \PP^{-1}\bm{\tilde{\chi}^{'}}$ (see Eq.~(\ref{eq:tensorPinverse}) in Appendix~\ref{App B} for $\tensr{P}^{-1}$)
  \item Perform streaming step via $h_{\alpha}(\bm{x}, t+ \Delta t) = \tilde{h}_{\alpha}(\bm{x}-\bm{e}_{\alpha} \Delta t)$, where $\alpha = 0,1,2,...,8$.
  \item Update the temperature field $T$ is obtained from \newline
        \begin{equation}
           \qquad \qquad T = \sum_{\alpha=0}^{8} h_{\alpha}.
        \end{equation}
\end{itemize}
\section{LBM for surfactant concentration equation} \label{Sec.4.4}
The surfactant concentration equation (Eq.~(\ref{SUR_mu_psi_2})), together with Eq.~(\ref{SUR_P}), is solved using a central moment LB technique by evolving a fourth distribution function $q_\alpha$, where $\alpha=0,1,2,\ldots,8$, on the D2Q9 lattice. As before, we begin by defining the distribution function $q_\alpha$ and its equilibrium $q_\alpha^{eq}$, as well as the following raw and central moments, respectively:
\mathleft
\begin{subequations}
\begin{equation}
\qquad \left( \begin{array}{c}\Upsilon'_{mn}\\[2mm]   \Upsilon'^{\;eq}_{mn} \end{array} \right)  = \sum_{\alpha = 0}^{8} \left( \begin{array}{c} q_{\alpha} \\[2mm]  q_{\alpha}^{eq} \end{array} \right)  e_{\alpha x}^m   e_{\alpha y}^n,
\end{equation}
\mathleft
\begin{equation}
\qquad \left( \begin{array}{c}\Upsilon_{mn} \\[2mm]   \Upsilon_{mn}^{eq} \end{array} \right)  = \sum_{\alpha = 0}^{8} \left( \begin{array}{c} q_{\alpha} \\[2mm]  q_{\alpha}^{eq} \end{array} \right) (e_{\alpha x}-u_x)^m  ( e_{\alpha y}-u_y)^n.
\end{equation}
\end{subequations}
For convenience, we list the components of the distribution function and its equilibrium, respectively, using $\mathbf{q}=(q_0,q_1,q_2,\ldots,q_8)^\dagger$ and $\mathbf{q}^{eq}=(q_0^{eq},q_1^{eq},q_2^{eq},\ldots,q_8^{eq})^\dagger$, and analogously for the raw moments and central moments via
\mathleft
\begin{subequations}
\begin{eqnarray}
\qquad \bm{\Upsilon^{'}} \! \! \! &=& \! \! \! ( \Upsilon_{00}^{'}, \Upsilon_{10}^{'},\Upsilon_{01}^{'}, \Upsilon_{20}^{'}, \Upsilon_{02}^{'}, \Upsilon_{11}^{'},\Upsilon_{21}^{'}, \Upsilon_{12}^{'},S_{22}^{'} ),\label{eqn:4a} \\[3mm]
\qquad \bm{\Upsilon} \! \! \! &=& \! \! \! ( \Upsilon_{00},\Upsilon_{10}, \Upsilon_{01}, \Upsilon_{20}, \Upsilon_{02}, \Upsilon_{11}, \Upsilon_{21}, \Upsilon_{12}, \Upsilon_{22} ).
\end{eqnarray}
\end{subequations}
Similar to Section.~\ref{Sec.4.1}, we obtain the discrete equilibrium central moments from the corresponding continuous counterpart of the Maxwellian by substituting the surfactant concentration $\psi$ for the density $\rho$. This allows us to build a central moment-based collision model for solving the advection and diffusion parts of the surfactant concentration equation given in Eq.~(\ref{SUR_mu_psi_2}); moreover, we account for the fluxes due to $\bm{R}=(R_x, R_y)$ via $m_{\psi} \psi (1-\psi) \bm{R}$ appearing in Eq.~(\ref{SUR_mu_psi_2}) by means of further corrections to the first order equilibrium moments. The outcomes read as follows.
\mathleft
\begin{gather}
\qquad \Upsilon_{00}^{eq} = \psi, \qquad
\Upsilon_{10}^{eq} = -M_\psi \psi (1-\psi) R_x, \nonumber \\[2mm]
\qquad \Upsilon_{01}^{eq} = -M_\psi \psi (1-\psi) R_y,\qquad
\Upsilon_{20}^{eq} = c_{s_\psi}^2 \psi,\qquad
\Upsilon_{02}^{eq} = c_{s_\psi}^2 \psi,\nonumber  \\[2mm]
\qquad \Upsilon_{11}^{eq} = 0,\qquad
\Upsilon_{21} = 0,\qquad
\Upsilon_{12}^{eq} = 0,\qquad
\Upsilon_{22}^{eq} = c_{s_\psi}^4 \psi,
\end{gather}
where $c_{s_\psi}^2 = 1/3$ and $\bm{R}=(R_x, R_y)$ is defined in Eq.~(\ref{SUR_P}). Then, the computational procedure for solving the surfactant concentration equation for a time step $\Delta t$ starting from $q_\alpha=q_\alpha(\bm{x},t)$ can be summarized as follows:
\begin{itemize}
  \item Compute pre-collision raw moments from distribution functions via $\bm{\Upsilon^{'}} = \PP\mathbf{q}$ (see Eq.~(\ref{eq:tensorP}) in Appendix~\ref{App B} for $\tensr{P}$)
  \item Compute pre-collision central moments from raw moments via $\bm{\Upsilon} = \F \bm{\Upsilon^{'}}$ (see Eq.~(\ref{eq:tensorF}) in Appendix~\ref{App B} for $\tensr{F}$)
  \item Perform collision step via relaxation of central moments $\Upsilon_{mn}$ to their equilibria $\Upsilon_{mn}^{eq}$:
        \begin{equation}\label{eq:centralmomentrelaxationsurfactantconcentrationequation}
           \qquad \tilde{\Upsilon}_{mn} = \Upsilon_{mn} + \omega^{\psi}_{mn}  (\Upsilon_{mn}^{eq} - \Upsilon_{mn}),
        \end{equation}
where $(mn)=(00),(10),(01),(20),(02),(11),(21),(12)$, and $(22)$, and $\omega^{\psi}_{mn}$ is the relaxation parameter for moment of order ($m+n$). The relaxation parameters of the first order moments $\omega_{10}^{\psi}$=$\omega_{01}^{\psi}$=$ \omega^{\psi}$ are related to the local surfactant mobility $M_{\psi} = m_{\psi} \psi (1-\psi)$ via $m_{\psi} = c_{s_\psi}^2 \left(1/\omega^{\psi} - 1/2\right)\Delta t$, and the rest of the relaxation parameters of higher central moments are typically set to unity. The results of Eq.~(\ref{eq:centralmomentrelaxationsurfactantconcentrationequation}) are then grouped in $\bm{\tilde{\Upsilon}}$.
  \item Compute post-collision raw moments from post-collision central moments via $\bm{\tilde{\Upsilon}^{'}} = \F^{-1} \bm{\tilde{\Upsilon}}$ (see Eq.~(\ref{eq:tensorFinverse}) in Appendix~\ref{App B} for $\tensr{F}^{-1}$)
  \item Compute post-collision distribution functions from post-collision raw moments via $\mathbf{\tilde{q}} = \PP^{-1}\bm{\tilde{\Upsilon}^{'}}$ (see Eq.~(\ref{eq:tensorPinverse}) in Appendix~\ref{App B} for $\tensr{P}^{-1}$)
  \item Perform streaming step via $q_{\alpha}(\bm{x}, t+ \Delta t) = \tilde{q}_{\alpha}(\bm{x}-\bm{e}_{\alpha} \Delta t)$, where $\alpha = 0,1,2,...,8$.
  \item Finally, update the surfactant concentration field $\psi$ which is obtained from
\begin{equation}
\qquad \qquad \psi = \sum_{\alpha=0}^{8} q_{\alpha}.
\end{equation}
\end{itemize}

\section{Transformation Matrices for Central Moment LB Schemes on a D2Q9 lattice} \label{App B}
Here, we summarize the various mapping relations that are needed prior to and following the collision step, where different central moments are relaxed to their equilibria, in the central moment LB scheme on the D2Q9 lattice. See Appendix~\ref{Sec.4.1} for related discussion and the notations adopted.

The transformation matrix $\tensr{P}$ mapping a vector of distribution functions $\mathbf{f}$ to a vector of raw moments $\bm{\kappa^{'}}$ is given by
\begin{equation}\label{eq:tensorP}
\tensr{P} = \begin{bmatrix}
     1  &\quad    1  &\quad    1  &\quad      1  &\quad    1  &\quad     1 &\quad    1  &\quad     1  &\quad     1 \\[10pt]
     0  &\quad    1  &\quad     0  &\quad    \um1  &\quad     0  &\quad    1 &\quad   \um1 &\quad   \um1  &\quad     1 \\[10pt]
     0  &\quad    0  &\quad     1  &\quad     0  &\quad   \um1  &\quad    1  &\quad     1  &\quad    \um1  &\quad  \um1 \\[10pt]
     0  &\quad    1  &\quad     0  &\quad     1  &\quad     0  &\quad    1  &\quad     1  &\quad     1  &\quad     1 \\[10pt]
     0  &\quad    0  &\quad     1  &\quad     0  &\quad     1  &\quad    1  & \quad    1  &\quad     1  &\quad     1 \\[10pt]
     0  &\quad    0  &\quad     0  &\quad     0  &\quad     0  &\quad    1  &\quad    \um1  &\quad     1  &\quad   \um1 \\[10pt]
     0  &\quad    0  &\quad     0  &\quad     0  &\quad     0  &\quad    1  &\quad     1  &\quad  \um1  &\quad  \um1 \\[10pt]
     0  &\quad    0  &\quad     0  &\quad     0  &\quad     0  &\quad    1  &\quad    \um1  &\quad  \um1  &\quad    1 \\[10pt]
     0  &\quad    0  &\quad     0  & \quad    0  &\quad     0  & \quad   1  &\quad     1  &\quad     1  &\quad    1
\end{bmatrix}
\end{equation}
Next, the transformation matrix $\tensr{F}$ mapping a vector of raw moments $\bm{\kappa^{'}}$ to a vector of central moments $\bm{\kappa}$ reads as
\begin{equation}\label{eq:tensorF}
\tensr{F}=
\begin{bmatrix}
      1  &    0  &    0  &     0  &    0  &     0 &    0  &     0  &     0 \\[10pt]

     \um u_x  &   1  &    0  &     0  &    0  &     0 &    0  &     0  &     0 \\[10pt]

      \um u_y  &    0  &   1  &     0  &    0  &     0 &    0  &     0  &     0 \\[10pt]

      u_x^2  &   \um 2u_x  &    0  &     1  &    0  &     0 &    0  &     0  &     0 \\[10pt]

      u_y^2  &    0  &    \um 2u_y  &     0  &    1  &     0 &    0  &     0  &     0 \\[10pt]

     u_x u_y  &   \um u_y  &    \um u_x  &     0  &    0  &     1 &    0  &     0  &     0 \\[10pt]

      \um u_x^2 u_y  &   2u_x u_y  &   u_x^2  &     \um u_y  &    0  &     \um 2u_x &    1  &     0  &     0 \\[10pt]

      \um u_x u_y^2 &    u_y^2  &    2u_x u_y  &     0  &    \um u_x  &    \um 2u_y &    0  &     1  &     0 \\[10pt]

      u_x^2 u_y^2  &    \um u_x u_y^2  &    \um u_x^2 u_y  &    u_y^2  &   u_x^2  &    4u_x u_y &    \um 2 u_y  &     \um 2  u_x  &     1 \\
\end{bmatrix}
\end{equation}
Then, the transformation matrix $\tensr{F}^{-1}$ mapping a vector of (post-collision) central moments $\bm{\tilde{\kappa}}$ to a vector of (post-collision) raw moments $\bm{\tilde{\kappa}}^{'}$ can be written as
\begin{equation}\label{eq:tensorFinverse}
\tensr{F}^{-1}=
\begin{bmatrix}
      1  &    0  &    0  &     0  &    0  &     0 &    0  &     0  &     0 \\[10pt]

      u_x  &   1  &    0  &     0  &    0  &     0 &    0  &     0  &     0 \\[10pt]

      u_y  &    0  &   1  &     0  &    0  &     0 &    0  &     0  &     0 \\[10pt]

      u_x^2  &   2u_x  &    0  &     1  &    0  &     0 &    0  &     0  &     0 \\[10pt]

      u_y^2  &    0  &    2u_y  &     0  &    1  &     0 &    0  &     0  &     0 \\[10pt]

     u_x u_y  &   u_y  &    u_x  &     0  &    0  &     1 &    0  &     0  &     0 \\[10pt]

      u_x^2 u_y  &   2u_x u_y  &   u_x^2  &     u_y  &    0  &     2u_x &    1  &     0  &     0 \\[10pt]

      u_x u_y^2 &    u_y^2  &    2u_x u_y  &     0  &    u_x  &    2u_y &    0  &     1  &     0 \\[10pt]

      u_x^2 u_y^2  &    u_x u_y^2  &    u_x^2 u_y  &    u_y^2  &   u_x^2  &    4u_x u_y &    2 u_y  &     2  u_x  &     1 \\
\end{bmatrix}
\end{equation}
It may be noted that if $\tensr{F}=\tensr{F}(u_x,u_y)$, then $\tensr{F}^{-1}=\tensr{F}(-u_x,-u_y)$ (see~\cite{yahia2021central}).

Finally, we express the transformation matrix $\tensr{P}^{-1}$ mapping a vector of (post-collision) raw moments $\bm{\tilde{\kappa}^{'}}$ to a vector of
(post-collision) distribution functions $\mathbf{\tilde{f}}$ as
\begin{equation}\label{eq:tensorPinverse}
\tensr{P}^{-1} =
\begin{bmatrix}
     1  &\quad    0  &\quad    0  &\quad    \um 1  &\quad   \um 1  &\quad     0 &\quad    0  &\quad     0  &\quad     1 \\[10pt]
     0  &\quad    \frac{1}{2}  &\quad     0  &\quad    \frac{1}{2}  &\quad     0  &\quad    0 &\quad   0 &\quad   \um \frac{1}{2}  &\quad     \um \frac{1}{2} \\[10pt]
     0  &\quad    0  &\quad     \frac{1}{2}  &\quad     0  &\quad   \frac{1}{2}  &\quad    0  &\quad     \um \frac{1}{2}  &    0  &\quad   \um \frac{1}{2} \\[10pt]
     0  &\quad    \um \frac{1}{2}  &\quad     0  &\quad     \frac{1}{2}  &\quad     0  &\quad   0 &\quad     0  &\quad     \frac{1}{2}  &\quad    \um \frac{1}{2} \\[10pt]
     0  &\quad    0  &\quad     \um \frac{1}{2}  &\quad     0  &\quad     \frac{1}{2}  &\quad    0  &\quad     \frac{1}{2} &     0  &\quad     \um \frac{1}{2} \\[10pt]
     0  &\quad    0  &\quad     0  &\quad     0  &\quad     0  &\quad    \frac{1}{4}  &\quad    \frac{1}{4}  &\quad     \frac{1}{4}  &\quad   \frac{1}{4} \\[10pt]
     0  &\quad    0  &\quad     0  &\quad     0  &\quad     0  &\quad    \um \frac{1}{4}  &\quad     \frac{1}{4}  &\quad  \um \frac{1}{4}  &\quad \frac{1}{4} \\[10pt]
     0  &\quad    0  &\quad     0  &\quad     0  &\quad     0  &\quad    \frac{1}{4}  &\quad    \um \frac{1}{4} &\quad  \um \frac{1}{4}  &\quad    \frac{1}{4} \\[10pt]
     0  &\quad    0  &\quad     0  &\quad     0  &\quad     0  &\quad    \um \frac{1}{4}  &\quad     \um \frac{1}{4}  &\quad     \frac{1}{4}  &\quad    \frac{1}{4}
\end{bmatrix}
\end{equation}


\newpage
\bibliographystyle{unsrt}

\end{document}